\documentclass[journal=mamobx,manuscript=article,layout=twocolumn]{achemso}
\usepackage[version=3]{mhchem}
\usepackage{amssymb}
\usepackage{amsmath}
\usepackage[normalem]{ulem}
\usepackage{relsize}
\usepackage{color}
\usepackage{float}

\let\oldmaketitle\maketitle
\let\maketitle\relax

\SectionNumbersOn

\author{Vanessa Weith}
\author{Alexei Krekhov}
\author{Walter Zimmermann}
\email{walter.zimmermann@uni-bayreuth.de}
\phone{+49 (0)921 553181}
\fax{+49 (0)921 555820}
\affiliation[Universit\"at Bayreuth]
{Theoretische Physik I, Universit\"at Bayreuth, D-95440 Bayreuth}

\title{Stability and Orientation of Lamellae in Diblock Copolymer Films}

\begin{document}

\twocolumn[
\begin{@twocolumnfalse}
\oldmaketitle
\centering\date{\today}
\begin{abstract}
The dynamics of microphase separation and the orientation of lamellae 
in diblock copolymers
is investigated in terms of a mean-field model. 
The formation of lamellar structures
and their stable states are explored and it is
shown that lamellae are stable not only for
the period of the structure corresponding to the minimum of the free energy. 
The range of wavelengths of stable lamellae is determined 
by a functional approach, introduced with this work, which
is in agreement with the results of a linear stability analysis.
The effects of the interaction of block copolymers with confining plane 
boundaries on the lamellae orientation 
are studied by an extensive analysis of the free energy.
By changing the surface property at one boundary,  
a transition from a preferentially perpendicular to
a parallel lamellar orientation {\it with respect to the boundaries} is found, which is rather
independent of the distance between the boundaries.
Computer simulations reveal, that the time scale 
of the lamellar orientational order dynamics, which is
quantitatively characterized in terms of an
orientational order parameter and the structure factor, 
depends significantly on the
properties of the confining boundaries as well as on the quench depth.
\end{abstract}
\end{@twocolumnfalse}]
\section{Introduction}
\label{intro}
During microphase separation diblock copolymers
can form various nanoscopic structures like lamellae, cylinders, spheres
or bicontinuous gyroids, depending on their composition 
\cite{Bates:1994.1,Bates:1994.2,Matsen:1996.1,Bates:99.1,Fasolka:01.1}.
These self-organized periodic nanoscale patterns in bulk materials as well as in 
block copolymer (BCP) 
films attract great attention because of interesting
phenomena in these systems and promising applications in nanofabrication, see, e.g., 
the reviews~\cite{Segalman:2005.1,Register:2010.1,Russell:2009.1,Hamley:09.1}.

For lamellar structures the typical width of lamellae is of the order
of the length of two covalently bonded polymers and
range from $10\cdots 100$ nm. They are readily tunable
by varying the molecular weights of both blocks of the copolymer. 
In bulk materials lamellae are locally ordered 
but on larger length scales one finds  
a random orientational order \cite{Gido:93.1}. 
Near a substrate lamellae are oriented
parallel or perpendicularly onto it.
The orientation is a direct result of a surface
and interfacial energy minimization \cite{Fredrickson:87.1}. When coating a substrate with
the same material as one block of a diblock copolymer, 
this block is selected and lamellae orient parallel to the substrate. 
Covering the substrate with a thin film of
a equimolar random copolymer, its interaction
with the two dissimilar BCP blocks 
can be balanced, so that the substrate
behaves neutrally \cite{Kellogg:96.1,Fasolka:01.1}.
In this case the lamellae orient
perpendicularly near the substrate 
but the lamellae orientation in the
plane of the substrate is disordered at large length scales.

Various strategies are investigated 
to achieve a long-range orientational order
of the lamellae in BCP films by the application of 
electric fields \cite{Amundson:91.1, Thurn-Albrecht:02.1, Boker:02.1, Xu:07.1}, 
shear flow \cite{Albalak:93.1, Wiesner:97.1, Angelescu:04.1}, 
directional solidification \cite{Hashimoto:99.1, DeRosa:00.1, Yoon:06.1, Berry:07.1},
use of topographically
\cite{Segalman:01.1, Cheng:06.1,Bita:2008.1} or chemically \cite{Kim:03.1,Segalman:2005.1} patterned
substrates.

By chemical patterning of 
substrates with a periodicity close to the  lamellae width, 
a long range orientational order of lamellae can be induced \cite{Kim:03.1}.
Such long range order can be achieved even in the case of
small mismatches between both periodicities. This
raises the question, whether straight lamellae 
are also stable at a wavelength apart from the optimal one
at the minimum of the free energy, $\lambda_e$. 
Does for lamellae in BCPs also exist a continuous wave number band 
around $2\pi/\lambda_e$, similar as  
in other common pattern forming systems
\cite{Cross:93.1,Cross:09.1}, even further below
the critical temperature of microphase separation? 
This question is investigated in Sec.~\ref{statstab}, where we introduce
a general method for the determination of stable wave number bands 
in  pattern forming systems with a potential, like
BCPs. We determine by this method 
the stable wave number range of lamellae
and we find also perfect agreement
with the results of a standard linear stability analysis of straight lamellae
(see also Appendix \ref{app_numstab} and \ref{app_stab}).
In Sec.~\ref{statstab} we calculate in addition analytical solutions
for the lamellar structure in the weak
as well as in the strong segregation limit and present
also analytical results for the stability range in the weak segregation limit
in part \ref{stab_weak}.

To induce a long range orientational lamellar order in the plane of BCP films with 
the lamellae perpendicular to the substrate, 
the film may be confined
between two lateral boundaries at small and medium distances 
as in Refs.~\cite{Park:07.1,Nealey:2010.1}. It depends again on the surface
preparation of the lateral boundaries, whether lamellae become oriented 
by energetic reasons either parallel or perpendicularly with respect to them.
The homogeneous lamellar structures in such quasi two-dimensional systems confined between 
two boundaries are analysed in Sec.~\ref{confined} and Appendix \ref{analconf}.
We determine for various selectivities at
the boundaries the lamellar orientation corresponding 
to the lowest free energy.
For nonsymmetric boundary conditions we find in two dimensions
as a function of the surface selectivity of one boundary  
a transition from a preferentially perpendicular to a parallel
orientation to the boundaries, which is rather
independent of the distance between the boundaries.
The dynamical evolution, including the coarsening and
the development of the orientational order 
of lamellae between two
boundaries is investigated in Sec.~\protect\ref{Dynphasesep}.
The final Sec.~\ref{conclusion} includes besides a 
discussion also experimental suggestions.
%

\section{Model equation}
\label{model}
Microphase separation in an incompressible $AB$-diblock copolymer melt
is described in terms of a  time-dependent Ginzburg-Landau model
for the conserved mean-field order parameter 
$\psi(\mathbf r,t)      \sim \phi_A(\mathbf r,t) - \phi_B(\mathbf r,t)$, with
the local concentrations $\phi_{A,B}$ of the components $A$ and $B$. 

Spatial variations of the order parameter involve a spatial
dependence of the chemical potential $\mu(\mathbf r,t)$
and the mass current ${\mathbf j}(\mathbf r,t)$, which determine 
the dynamics of $\psi(\mathbf r,t)$ via 
the continuity equation 
\begin{equation}
\label{eq:mass_conserv}
\frac{\partial \psi(\mathbf r,t)}{\partial t} = -\nabla \mathbf j(\mathbf r,t) \;.
\end{equation}
Gradients of the chemical potential
drive the mass current 
\begin{equation}
\label{eq:current}
\mathbf j(\mathbf r,t) = -M \nabla \mu(\mathbf r,t) \;,
\end{equation}
with the Onsager coefficient $M (>0)$ 
\cite{Gunton:83.1} that describes the mobility of the monomers $A$ with respect to $B$.
The  functional derivative of a free energy functional $F\lbrace\psi\rbrace$
with respect to  the order parameter determines the chemical potential

\begin{equation}
\label{eq:mu}
\mu(\mathbf r,t) = \frac{\delta F\lbrace\psi\rbrace}{\delta \psi} \;,
\end{equation}
and via Eq.~(\ref{eq:mu}) and Eq.~(\ref{eq:current})
 also the dynamics of $ \psi(\mathbf r,t)$:
\begin{eqnarray}
\label{eq:bcp}
\frac{\partial \psi(\mathbf r,t)}{\partial t} =
M \nabla^2 \frac{\delta F\lbrace\psi\rbrace}{\delta \psi} \;.
\end{eqnarray}
The free energy $F\lbrace \psi \rbrace$ acts as a global Lyapunov functional
and is always decreasing with time towards
its minimum:
\begin{eqnarray}
\label{eq:lyapunov}
\frac{\partial F\lbrace \psi \rbrace}{\partial t} &=& 
\int\limits_{V} \left( \frac{\delta F}{\delta \psi} \frac{\partial \psi}{\partial t} \right) d\mathbf{r} = 
M \int\limits_{V} \left( \frac{\delta F}{\delta \psi} \nabla^2 \frac{\delta F}{\delta \psi} \right) d\mathbf{r} 
\nonumber \\
&=& - M \int\limits_{V} \left( \nabla \frac{\delta F}{\delta \psi} \right)^2 d\mathbf{r} \le 0 \;.
\end{eqnarray}

Employing a generalized random phase approximation, the bulk 
free energy functional for diblock copolymers
was derived by Leibler \cite{Leibler:80.1} in the weak 
segregation limit, which is applicable to the slightly quenched regime
of   microphase separation. Here we use the extended free energy 
introduced by Ohta and Kawasaki \cite{Ohta:86.1},
\begin{eqnarray}
\label{free}
\hspace*{-0.7cm}&& \frac{F_{b}\lbrace \psi \rbrace}{k_B T} = \int\limits_V\left[-\frac{b}{2}\psi^2+\frac{u}{4}\psi^4+\frac{K}{2}(\nabla \psi)^2\right] d\mathbf{r} \qquad\qquad\,\,\,\,
\nonumber \\
\hspace*{-0.7cm}&& \; +\frac{D}{2}\int\limits_V\int\limits_V G(\mathbf r, \mathbf r') [\psi(\mathbf r,t)-\bar \psi] [\psi(\mathbf r',t)-\bar\psi] d\mathbf{r} d\mathbf{r}' ,
\end{eqnarray}
which also  includes  the strong segregation  limit, 
applicable to the deeply quenched regime.
The functional comprises
the temperature independent phenomenological constants $u$, $K$, 
 $D$  and  $\bar \psi = \langle\psi\rangle$ is the spatial average 
of the order parameter $\psi(\mathbf r,t)$.
$b$ is the temperature dependent {\it control parameter} of the model
and for $b>b_c$ microphase separation sets in. 
The second term in Eq.~(\ref{free}) with the double integral 
covers the long-range interaction due to the connectivity of the subchains
and the Green's function $G(\mathbf r, \mathbf r')$ satisfies Laplace's equation 
$\nabla^2 G(\mathbf r, \mathbf r')=-\delta(\mathbf r-\mathbf r')\label{greens}$.

The coefficients of the mean field
free energy in Eq.~(\ref{free}) can be related to  microscopic models
under the following assumptions: All chains have the same index of polymerization $N$, 
 are composed of by the same number $N_A(N_B)$ 
of monomers of type $A$ ($B$), with  $N=N_A + N_B$, 
and have therefore the same composition $f=N_A/N$. Both
blocks have the same Kuhn statistical
segment length $l$. 

The control parameter $b$ in Eq.~(\ref{free}) is related
to the Flory-Huggins interaction parameter $\chi$ 
as follows \cite{Ohta:86.1},
\begin{eqnarray}
\label{eq:b_micro}
b = 2 \chi - \frac{s(f)}{2 N f^2 (1-f)^2} \;,
\end{eqnarray}
where  $s(f)$ is of order unity and 
depends on approximations \cite{Fredrickson:87.1,Leibler:80.1}.
Microphase separation occurs for $\chi > \chi_c$ ($T<T_c$)
and the temperature dependence of the Flory-Huggins interaction parameter $\chi$
is taken as 
\begin{eqnarray}
\label{eq:chi}
\chi = A + B/T \;,
\end{eqnarray}
where the coefficients $A$ and $B$ are determined from experiments \cite{Hashimoto:83.1}.
The parameter $u >0$ in Eq.~(\ref{free}) remains undetermined in the strong segregation limit \cite{Ohta:86.1}.
For the lamellar structure in the weak segregation limit $u$ 
can be identified with the vertex
function $u = \Gamma_4(0,0)$, 
which depends on the composition $f$ and the polymerization index $N$ \cite{Leibler:80.1,Fredrickson:87.1}.
The parameter $K >0$ describes the interfacial energy  between $A$ and $B$ 
domains and it depends on the composition $f$ as follows \cite{Ohta:86.1}:
\begin{equation}
\label{eq:K_micro}
K = \frac{l^2}{4 f (1-f)}\,.
\end{equation}
The parameter
\begin{equation}
\label{eq:D_micro}
D = \frac{3}{l^2 N^2 f^2 (1-f)^2} \;
\end{equation}
is positive and decays
with the polymerization degree $N$ \cite{Ohta:86.1}.
Similar functional dependencies of $K$ and $D$ 
have been obtained in Refs.~\cite{Fredrickson:87.1,Choksi:03.1} 
except the difference in numerical factors due to the use of different 
models for the polymer chain (and therefore different expressions for the radius of gyration).
Besides the derivation of the interaction parameters from microscopic models it is
also possible to determine them by fitting data from scattering experiments obtained
immediately after a quench 
(see, e.g., \cite{Hashimoto:83.1,Bates:85.1,Sakamoto:95.1} and references therein).

With the  functional given by  Eq.~(\ref{free})
and  Eq.~(\ref{eq:bcp}) the nonlinear evolution equation of  the
order parameter $\psi({\bf r},t)$ follows:
\begin{eqnarray}
\label{modeleq}
\partial_t \psi &=& M k_B T \left[ \nabla^2(-b \psi+u\psi^3-K\nabla^2\psi)\right.\nonumber \\
&& \qquad \qquad\left.- D(\psi- \bar \psi) \right] \;.
\end{eqnarray}
With the length scale  $\xi=\sqrt{K}$ and
the time scale $\tau=\xi^2/(M k_B T)$ one may introduce
with $\mathbf r=\xi\mathbf r'$ and $t=\tau t'$ 
the dimensionless  variables  $\mathbf r'$ and $t'$. Using 
in addition the  rescaled order parameter $\psi'=\sqrt{u}\psi$,
one obtains the dimensionless form of the equation 
(primes are omitted)
\begin{equation}
\label{dynglei}
\partial_t \psi = \nabla^2(-\varepsilon \psi+\psi^3-\nabla^2\psi)-\alpha(\psi-\beta) \;,
\end{equation}
with the dimensionless parameters: 
\begin{equation}
\varepsilon=b \,,\,\,\,\, \alpha=DK \,,\,\,\,\, \beta=\sqrt{u}\bar \psi\,\,.
\end{equation}
According to Eq.~(\ref{eq:K_micro}) and Eq.~(\ref{eq:D_micro}) one has the
scaling $\alpha \propto N^{-2}$ and 
the limiting case $\alpha =0$ corresponds to the Cahn-Hilliard equation \cite{Cahn:58.1}
describing, e.g., phase separation in polymer blends.

The bulk free energy in dimensionless form is given by
\begin{eqnarray}
\label{eq:Fbulk}
F_b\lbrace \psi \rbrace &=& \int\limits_{V}
\left[ -\frac{\varepsilon}{2} \psi^2 + \frac{1}{4} \psi^4 + \frac{1}{2} (\nabla \psi)^2 \right.
\nonumber \\
&& \quad \left. + \frac{\alpha}{2} h(\mathbf{r}) (\psi-\beta) \right] d\mathbf{r} \;,
\end{eqnarray}
where we have introduced for  practical reasons the auxiliary function
\begin{equation}
\label{fhelp}
h(\mathbf{r}) = \int\limits_{V} G(\mathbf r, \mathbf r')\,\left[\psi(\mathbf r')-\beta\right] \, d\mathbf r' \;.
\end{equation}
The function $h(\mathbf{r})$ fulfills  Poisson's equation
\begin{equation}
\label{fhelp2}
\nabla^2 h(\mathbf{r}) = -(\psi(\mathbf{r}) -\beta)\,,
\end{equation}
with the boundary condition 
along the direction $\mathbf n$  normal to the surface
of the volume $V$,
\begin{equation}
\mathbf n\cdot \nabla h(\mathbf r) = 0 \;,
\end{equation}
that follows from the conservation 
of the order parameter: $\int (\psi-\beta) \, d\mathbf{r} =0$.
%

\subsection{Effects of boundaries}
To study {\em unconfined} systems periodic boundary conditions 
of the order parameter $\psi$ can be applied
in each spatial direction.
For {\em confined} systems one needs for the fourth order Eq.~(\ref{dynglei})
two boundary conditions.
The first boundary condition follows from the conservation 
of the order parameter $\int (\psi-\beta)\,d\mathbf{r} =0$ 
corresponding to a zero flux at the boundary:
\begin{equation}
\label{eq:bc1}
\mathbf{n} \cdot \nabla\left( -\varepsilon \psi+\psi^3-\nabla^2\psi \right) = 0 \;.
\end{equation}
The free energy in a confined system includes
a surface contribution $F_s$,
 which can be written in dimensionless form,
\begin{eqnarray}
\label{eq:Fsurf}
F_{s} = \frac{1}{2} \int\limits_S g(\psi-\psi_S)^2\,dS \;,
\end{eqnarray}
with two phenomenological parameters $g$ and $\psi_S$.
$g >0$ is a measure of the strength of the interaction of the
block copolymer with the surface and $\psi_S$ is the
preferred difference between the concentrations of $A$ and $B$ at the surface.
The case $g = const.$ and $\psi_S=const.$ corresponds to a homogeneous surface
and non-constant $g = g(S)$ and $\psi(S)$ model patterned surfaces
as e.g. described in Ref.~\cite{Chen:98.1}.
The second boundary condition is derived from the 
local equilibrium condition of the total
free energy $F=F_b+F_s$ 
at the surface 
\begin{eqnarray}
 \frac{\delta F}{\delta \psi} = \mu +
\left[{\bf n} \cdot \nabla \psi+ g(\psi -\psi_s) \right]_S\,.
\end{eqnarray}
Since the bulk is in equilibrium
with the surface one has the second boundary condition:
\begin{eqnarray}
\label{eq:bc2}
\mathbf{n} \cdot \nabla\psi + g(\psi-\psi_S) = 0 \;.
\end{eqnarray}
Note that the surface energy in Eq.~(\ref{eq:Fsurf})
is equivalent to the expression proposed in Refs.~\cite{Binder:83.1,Fredrickson:87.1},
\begin{eqnarray}
F_{s} = \int\limits_S \left( -H_1 \psi + \frac{a_1}{2} \psi^2 \right)\,dS \;,
\end{eqnarray}
where the ``field'' $H_1$ is related to the difference of chemical potential between 
$A$- and $B$ blocks at the surface and the parameter $a_1$ is related to the 
so-called ``extrapolation length'', $a_1 \sim \delta^{-1}$, that describes the 
ability of the surface to modify the local interaction parameter $\chi$ \cite{Binder:83.1}.
In our notation one has $H_1 = g \psi_S$ and $a_1 = g$.
The usual situation for  diblock copolymers is the so-called ``ordinary transition'' with
 $\delta >0$ and therefore $a_1 >0$ (or $g >0$).
The surface modifies the local monomer interactions only within a thin 
surface layer of thickness $\mathcal{O}(l)$ 
with $l$ as the Kuhn statistical segment length  \cite{Binder:83.1}.
In this situation the local interaction parameter $\chi$ at the surface
is smaller than
  in the bulk and there is no ordering transition in the range $T >T_c$ for $H_1 =0$.
For $H_1 \neq 0$ finite values of the order parameter $\psi\not =0$
are already induced beyond the critical temperature,
 $T>T_c$. 
Walls with  the property $H_1 \neq 0$ are so-called {\it selective boundaries} and
with  $H_1 = 0$ so-called {\it neutral boundaries}.

Although we restrict our investigations to the most relevant ordinary 
transition we shortly mention 
for completeness also another  type of transition.
The so-called ``surface transition'' occurs for $a_1 <0$, which corresponds to $\delta <0$.
In this case the local interaction parameter $\chi$ is greater 
than in the bulk and even for $H_1=0$ an ordering transition may be observed at $T>T_c$ near the surface.
%

\section{Unconfined system: Periodic solutions}
\label{statstab}
Above the onset of microphase separation
a perfect lamellar order of block  copolymers is described
by  periodic solutions of  Eq.~(\ref{dynglei}) and their
properties in unconfined systems 
are investigated in this section 
for symmetric diblock copolymers, i.e.
 $\beta=0$.
An analytical approximation of the 
 amplitude of the spatially periodic solution 
immediately above onset is  given in Sec.~\ref{weaknon}. 
A method 
for the determination of the stability boundaries of
periodic solutions, which works close to
and  even far beyond onset of
microphase separation is presented 
 in Sec.~\ref{stab_func}.
It is based on an analysis of the free energy functional evaluated for 
periodic solutions. A conventional linear stability 
analysis of nonlinear periodic solutions, cf. \cite{Cross:09.1},
is presented for the present system in Appendix \ref{app_numstab}
and \ref{app_stab}.
%

\subsection{Onset of microphase separation}
\label{onset}
The  homogeneous phase of a symmetric diblock copolymer melt
is described by a vanishing order parameter: $\psi=0$.
This basic state becomes unstable with respect to small 
perturbations $\psi(\mathbf{r},t) \sim e^{\sigma t} e^{i \mathbf{k} \cdot \mathbf{r}}$
when the control parameter $\varepsilon$ is raised beyond its critical 
value $\varepsilon_c$, corresponding 
to a quench below the critical temperature $T_c$
of the diblock copolymer melt.
In this case the growth rate $\sigma$ of the perturbations 
becomes positive and microphase separation sets in.
Since Eq.~(\ref{dynglei}) is isotropic in space, $\sigma$ 
depends only on the modulus of the wave vector, $k = |\mathbf{k}|$, 
which is determined by the linear part of Eq.~(\ref{dynglei}):
\begin{eqnarray}
\label{eq:sigma_k}
\sigma(k) = k^2 (\varepsilon - k^2) - \alpha \;.
\end{eqnarray}
The neutral stability condition $\sigma(k) =0$ yields the {\it neutral curve}
\begin{eqnarray}
\label{eq:eps_N}
\varepsilon_N(k) = k^2 + \alpha/k^2 \;
\end{eqnarray}
and the basic state, $\psi =0$, is unstable above  $\varepsilon_N(k)$ with $\sigma(k) >0$.
The critical value of the control parameter, $\varepsilon_c$,  
and the critical wave number, $k_c$,
are both obtained from the extremal condition $d \varepsilon_N(k) / d k =0$ 
at the minimum of the neutral curve:
\begin{eqnarray}
\label{eq:crit}
\varepsilon_c = 2 \sqrt{\alpha} \;, \,\,\qquad
k_c = \alpha^{1/4} \;.
\end{eqnarray}
The wave numbers along the left and right part of the
neutral curve are given
as a function of the control parameter $\varepsilon$ 
by the expression
\begin{eqnarray}
\label{eq:k_N}
k_N^2(\varepsilon) = 
\frac{1}{2} \left( \varepsilon \mp \sqrt{\varepsilon^2 - \varepsilon_c^2} \right) \;\,
\end{eqnarray}
and the wave number $k_m$ at the maximum of the growth rate $\sigma(k)$
increases with $\varepsilon>\varepsilon_c$ as follows:
\begin{eqnarray}
\label{maxgrowth}
 k_m = \sqrt{\frac{\varepsilon}{2}}\,\,\,.
\end{eqnarray}
For further discussions also the reduced control parameter 
$r = \varepsilon/\varepsilon_c - 1$ and the reduced wave number $\tilde{k} = k/k_c$ are useful.
Then the rescaled neutral curve takes the following form
\begin{eqnarray}
\label{eq:r_N}
r_N(\tilde{k}) = \frac{(\tilde{k}^2 - 1)^2}{2 \tilde{k}^2},  
\end{eqnarray}
and the critical 
values of the control parameter and the
wave number are given by
\begin{eqnarray}
\label{eq:rk_c}
r_c = 0 \;, \qquad
\tilde{k}_c = 1 \;.
\end{eqnarray}
The reduced wave numbers along  $r_N(\tilde{k})$ are
\begin{eqnarray}
\label{eq:k1k2}
\tilde{k}_N^2 = 1 + r \mp \sqrt{r^2 + 2 r} \;,
\end{eqnarray}
and the reduced wave number at the maximum 
of the growth rate $\sigma(\tilde{k})$ is $\tilde{k}_m = \sqrt{1 + r}$.

Microphase separation and spatially periodic solutions of Eq.~(\ref{dynglei}) 
develop in the range $r > 0$ and  $\varepsilon > \varepsilon_c$, respectively. 
%

\subsection{Amplitude equation}
\label{weaknon}
The basic Eq.~(\ref{dynglei}) takes in terms of the reduced control parameter 
$r = \varepsilon/\varepsilon_c - 1$ the following form:
\begin{eqnarray} 
\label{eq:alt}
\partial_t \psi = \nabla^2 [-2 k_c^2 (r+1) \psi +\psi^3  - \nabla^2 \psi ] 
- k_c^4 \psi \;.
\end{eqnarray}
Equation (\ref{eq:alt})
is rotationally invariant and therefore the wave vector 
of a periodic solution can be chosen parallel to the $x$-axis with 
${\bf k}=(k,0)$.

In the range of small values of $r \gtrsim 0$ the neutral curve 
is still narrow around $k_c$
and nonlinear periodic solutions exist only for $k$ rather close 
to $k_c$. 
Small deviations of the wave vector ${\bf k}$ from 
the critical one, $(k_c,0)$, and therefore
long wavelength (slow) modulations of the periodic
solution $\propto \exp(i {\bf k}_c \cdot {\bf r})$
can be taken into account by a spatially 
dependent amplitude (envelope),
\begin{eqnarray}
\label{eq:psi_sp}
\psi = A(x,y,t) e^{i k_c x} + \textrm{c.c.}\,,
\end{eqnarray}
with $ A(x,y,t)$ slowly varying on the
scale  $2 \pi /k_c$. Such a separation
into a slowly varying amplitude and 
a fast varying periodic part is successfully used
in a broad class of pattern forming 
systems, as described for instance
in  Refs.~\cite{Newell:69.1,Cross:93.1,Newell:1993.1,Cross:09.1} .
By this separation into short and long length scales 
near threshold, $r \gtrsim 0$,
a further reduction of the basic equation (\ref{eq:alt})
to a universal equation for the envelope 
$A(x,y,t)$ is possible\cite{Newell:69.1,Cross:93.1,Newell:1993.1,Cross:09.1}
and allows  in the weak segregation regime 
further analytical progress, as described 
in the following.

The partial differential equation describing the 
dynamics of $A(x,y,t)$, the well-known Newell-Whitehead-Segel 
amplitude equation \cite{Newell:69.1} for the envelope 
of periodic solutions in isotropic systems,
can be derived by a 
multiple scale analysis~\cite{Cross:93.1,Shiwa:97.1,Cross:09.1},
\begin{align}
\label{eq:A}
\tau_0 \partial_t A =  r A + \xi_0^2 \left( \partial_x - \frac{i}{2 k_c} \partial_y^2 \right)^2 A
- g_0 |A|^2 A \;, 
\end{align}
\begin{align}
\label{eq:Ak}
\hspace{-0.7cm}
\mbox{with}\quad \tau_0= \frac{1}{2k_c^4}\,, \quad \xi_0^2=\frac{2}{k_c^2}\,, \quad g_0=\frac{3}{2k_c^2} \,.
\end{align}
Eq.~(\ref{eq:A}) has in the range $r>0$ periodic solutions
\begin{eqnarray}
\label{eq:A0}
A = A_0 e^{i (Q x+Py)} \;
\end{eqnarray}
in terms of the deviations  $Q=k-k_c$ and $P << k_c$
from the critical wave vector $(k_c,0)$.

The rotational invariance of the system allows to choose for stationary solutions 
$P=0$, i.e. 
the stationary amplitude $A_0$ can be expressed 
in terms of $Q$ or $\tilde{k} =k/k_c$:
\begin{eqnarray}
\label{eq:A0_2}
A_0^2 = \frac{2 k_c^2}{3} \left( r -  \frac{2Q^2}{ k_c^2} \right) 
      =  \frac{2 k_c^2}{3} \left[ r - 2 (\tilde{k} - 1)^2 \right]\;.
\end{eqnarray}
The amplitude $A_0$ vanishes along the neutral curve
\begin{eqnarray}
\label{eq:r_NA}
\hat{r}_N = 2 Q^2/k_c^2 = 2 (\tilde{k} - 1)^2\,,
\end{eqnarray}
which is equivalent to Eq.~(\ref{eq:r_N}) for
 $r \gtrsim 0$ and $|Q| << 1$.

The amplitude equation (\ref{eq:A}) has also non-periodic, 
inhomogeneous solutions,
as for instance described in Refs.~\cite{Kramer:84.1,Kramer:85.1,Cross:93.1,Cross:09.1} . 
$A(x,y,t)$ varies in these cases 
on length scales along the $x$ and the $y$ direction,
which are larger than $2\pi/k_c$. These length scales $\xi_1$
along the $x$ direction and $\xi_2$ along the $y$ direction
are: 
\begin{eqnarray}
\label{scalexy}
\displaystyle
\xi_1= \frac{1}{k_c} \left(\frac{2}{r}\right)^{1/2}\,, \qquad 
\xi_2 = \frac{1}{k_c} \left(\frac{1}{2r}\right)^{1/4}\,.
\end{eqnarray}
Accordingly, the envelope $A(x,y,t)$  varies 
perpendicular to the lamellae (along the $x$ direction) 
on a different length scale than 
parallel to the lamellae (along the  $y$ direction), 
when for instance the envelope decays from its bulk value 
$A \propto \sqrt{r}$ beyond threshold ($r>0$) to a small
value at the boundary. 
The ratio between the 
two length scales is
$\xi_2/\xi_1= (r/8)^{1/4}$ and therefore 
in the weak segregation regime $r \gtrsim 0$ the
length $\xi_2$ is 
always considerably smaller than the length $\xi_1$.
This difference has a strong influence on the orientation of 
lamellae near boundaries as discussed in Sec.~\ref{confined} .

\subsection{Nonlinear solutions}
\label{one-mode}
In extended systems with periodic boundary conditions
spatially periodic solutions of the nonlinear equation 
(\ref{dynglei}) of wave number $k$
 can be represented by a Fourier series, 
\begin{equation}
\label{Fourieransatz}
\psi_k(x) = \sum\limits_{j=-M}^M A_j ~e^{i k x j} \;, \qquad
A_j=A_{-j}^\ast \;,
\end{equation}
where the coefficients of 
this series are determined numerically, as described in Appendix \ref{app_numstab}.
For a truncated ansatz with one mode,
\begin{eqnarray}
\label{eq:onemode}
\psi =  a_0 \cos(k x) \;,
\end{eqnarray}
one obtains for the amplitude $a_0$:
\begin{eqnarray}
\label{eq:a0_onemode}
a_0 
&=& \pm 2\sqrt{ \frac{2 k_c^2}{3} \left[r - r_N(\tilde k) \right] ~}
 \;,
\end{eqnarray}
which becomes in the range $|Q| << 1$ identical to the expression
in Eq.~(\ref{eq:A0_2}).
Again $a_0$ only exists beyond the neutral curve $r>r_N(\tilde k)$
[resp. $\varepsilon >\varepsilon_N(k)$].
%
%

\subsection{Wave number bands of stable periodic solutions}
\label{stab_func}
Spatially periodic solutions in extended pattern 
forming systems are stable only in a  
subrange of the wave number band beyond a neutral curve as given for example
by Eq.~(\ref{eq:eps_N}) \cite{Cross:93.1}.

Stationary, spatially periodic solutions 
may be destabilized, for instance,
by small perturbations 
with a wave vector parallel to that of the nonlinear periodic pattern, 
if the so-called {\it Eckhaus stability boundary} is crossed.
Or, a periodic solution may be destabilized by perturbations with 
the wave 
vector perpendicular to that of the pattern ({\it zig-zag instability}), 
or by a combination of both types of destabilizing modes ({\it skewed varicose}) \cite{Cross:93.1}.
Such stability boundaries are determined 
by the condition, that the growth rate of
small perturbations with respect to
nonlinear periodic solutions vanishes, 
as described in more detail in Appendix \ref{app_numstab}
and \ref{app_stab}.
In systems where the dynamic equation of the
field $\psi$ can be derived from a functional
$F\lbrace \psi \rbrace$, 
the Eckhaus stability boundary and the zig-zag stability boundary 
can be determined by an analysis of the 
functional $F$
in terms of the periodic solutions $\psi({\mathbf r})$. 
The idea of this method was indicated 
earlier\cite{Kramer:85.1} and it is described below.
%

\subsubsection{Eckhaus stability boundary}
\label{stab_func_Eckhaus}

By crossing the Eckhaus boundary, nonlinear periodic solutions
become unstable with respect to small longitudinal
perturbations with a wave vector parallel to  $\mathbf{k}$
of the unperturbed pattern\cite{Eckhaus:65.1,Kramer:85.1,Cross:09.1}.
The wave vector $\mathbf{k} = (k,0)$ is chosen along the $x$-axis
and in order 
to determine the Eckhaus boundary we calculate the free energy $F$
for  a slightly perturbed solution, which has
a slightly ''compressed'' periodicity of wave number
 $k+\Delta k_x$ and free energy $F(k+\Delta k_x)/2$
in one half of the system and  a slightly 
''dilated'' periodicity of wave  number $k-\Delta k_x$
and free energy $F(k-\Delta k_x)/2$
in the other half.
The free energy of the perturbed periodic solution is then
as follows:
\begin{equation}
\label{eq:F_avr}
\bar{F}(k) = \frac{1}{2}[F(k+\Delta k_x)+F(k-\Delta k_x)] \;.
\end{equation}
Such deformations ensure that the mean value of the wave
number $\bar{k} = k$ in the whole system and  also 
the number of periodic units remains unchanged.
For small values of $\Delta k_x$ the expression at 
the right hand side of Eq.~(\ref{eq:F_avr}) can be expanded in terms of a 
Taylor series and one obtains at leading order in $\Delta k_x$:
\begin{eqnarray}
\bar{F}(k) = F(k) + \frac{1}{2}\frac{d^2 F(k)}{dk^2} (\Delta k_x)^2 + \cdots \;.
\end{eqnarray}
It depends therefore 
on the sign of the second derivative $d^2 F(k)/dk^2$, whether the slight deformation of the periodic solution 
leads to a reduction or an enhancement of the free energy $\bar{F}(k)$ with respect to $F(k)$.
In the case of $d^2 F(k)/dk^2 <0$ a simultaneous small dilation and compression of the periodic 
solution of wave number $k$ in neighboring ranges
leads to a reduction of the free energy with $\bar{F}(k) < F(k)$. 
I.e. for such parameter combinations $(\varepsilon, k)$  periodic solutions are unstable.
In the opposite case with  $d^2 F(k)/dk^2 >0$ a periodic solution with wave number $k$ is
stable with respect to longitudinal perturbations.
Therefore, the curve separating in the $\varepsilon-k$ plane the range of stable from unstable solutions is determined via the condition
\begin{equation}
\label{potEck}
\frac{d^2F(\varepsilon, k)}{dk^2} = 0 \;.
\end{equation}
%

\subsubsection{Zig-zag stability boundary}
 \label{stab_func_ZZ}
The zig-zag instability of a periodic solution of
wave vector $\mathbf{k}$ is induced by perturbations with a wave vector perpendicular 
to $\mathbf{k}$.
Utilizing that the free energy of the perturbed solution with the wave vector 
$\mathbf{k} = (k, \Delta k_y)$ only depends on the value of the
wave number
$|\mathbf{k}| = \sqrt{k^2 + (\Delta k_y)^2}$, one finds for small 
values of $\Delta k_y$ at leading order
\begin{eqnarray}
F(|\mathbf{k}|) = F(k) + \frac{1}{2 k}\frac{d F(k)}{dk} (\Delta k_y)^2 + \cdots \;.
\end{eqnarray}
In the range of $k$ with $d F(k)/dk <0$ small transversal perturbations 
of the periodic solution reduce the free energy of the system and thus the periodic solutions are unstable.
A periodic solution of wave number $k$ with $d F(k)/dk >0$ is stable
with respect to transversal perturbations.
Therefore the zig-zag stability boundary in the $\varepsilon-k$ plane is determined by the condition
\begin{eqnarray}
\label{potZZ}
\frac{d F(\varepsilon, k)}{dk} = 0 \;,
\end{eqnarray}
which corresponds for periodic solutions also to 
the minimum condition of free energy functional with respect to $k$.
The conditions for the Eckhaus boundary given by Eq.~(\ref{potEck}) and 
the zig-zag line by Eq.~(\ref{potZZ}) are valid for any two-dimensional isotropic system 
with a dynamics governed by a functional.
{\it Recipe} for a determination of the stability boundaries
of stationary periodic patterns in systems with
a potential: In a first step the nonlinear periodic solution of
wave number $k$ is determined either analytically 
(e.g. by a one-mode approximation) or numerically.
In a second step the free energy functional $F(\varepsilon, k)$ is determined 
for the periodic solution and in a third step the location of the Eckhaus stability boundary,
$\varepsilon_E(k)$, is determined via the condition 
given by  Eq.~(\ref{potEck}) and the zig-zag stability boundary by
Eq.~(\ref{potZZ}).
%

\subsection{Stability in the weak segregation regime}
\label{stab_weak}
With the one-mode approximation given by  Eq.~(\ref{eq:onemode})
the functional in Eq.~(\ref{eq:Fbulk}) can be easily evaluated and
the resulting
free energy per period $\lambda = 2\pi/k$ is given by:
\begin{eqnarray}
\label{eq:F_k}
{\cal F}(k) \equiv \frac{F_b(k)}{\lambda} = 
 - \frac{3}{32} a_0^4 \;.
\end{eqnarray}
In this case the condition (\ref{potEck}) 
leads to the following Eckhaus stability boundary:
\begin{eqnarray}
\label{eq:eps_E}
\varepsilon_E(k) = \frac{3 k^8 + 5 \alpha^2}{k^2 (k^4 + 3 \alpha)} \;.
\end{eqnarray}
This formula reads in terms of the reduced control parameter $r$ and the reduced wave number $\tilde{k}$ as follows:
\begin{eqnarray}
\label{eq:r_E}
r_E = 
\frac{3 \tilde{k}^8 - 2 \tilde{k}^6 - 6 \tilde{k}^2 + 5}{2 \tilde{k}^2 (\tilde{k}^4 + 3)} \;.
\end{eqnarray}
The zig-zag instability condition (\ref{potZZ}) 
provides the wave number $k_{ZZ}$ (resp. $\tilde{k}_{ZZ}$)
at the zig-zag stability boundary:
\begin{eqnarray}
\label{eq:zz_line}
k_{ZZ} = \alpha^{1/4} \;  \qquad ( \textrm{or} \;\; \tilde{k}_{ZZ} = 1) \;.
\end{eqnarray}
This wave number does not depend on the control parameter $\varepsilon$, respectively
$r$.
The lamella period $\lambda_e = 2\pi/k_{ZZ}$ at the minimum of the free energy 
and the corresponding free energy per period ${\cal F}_e = {\cal F}(k_{ZZ})$ are given by
\begin{eqnarray}
\label{eq:lmbdeFe}
\lambda_e = \frac{2\pi}{\alpha^{1/4}} \; \;\;\;\,\, \mbox{and}\,\,\,\,\,
{\cal F}_e = - \frac{(\varepsilon - 2\sqrt{\alpha})^2}{6} \;.
\end{eqnarray}

These results are obtained for a one-mode approximation 
of the nonlinear periodic solution  
and an analysis of the related free 
energy agrees with a conventional stability 
analysis  in terms of amplitude equations \cite{Cross:93.1},
as described for the present system 
in more detail in Appendix \ref{app_stab}.

\begin{figure}[ht]
\begin{center}
\includegraphics[width=8.25cm]{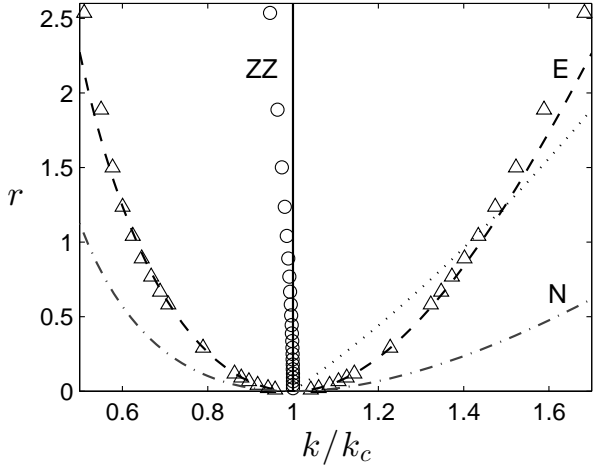} 
\end{center}
\caption{\label{stab_diagram_komplett}
The stability diagram of the periodic solutions of Eq.~(\ref{dynglei}) 
shows the neutral curve (N, dot-dashed line) according to Eq.~(\ref{eq:r_N}), 
the Eckhaus stability boundary obtained for the one-mode solution via Eq.~(\ref{eq:r_E}) (E, dashed line) 
and numerically (triangles), as well as the zig-zag stability boundary for 
the one-mode approximation given by  Eq.~(\ref{eq:zz_line}) (ZZ, solid line) and as obtained numerically (circles). 
The wave number $\tilde{k}_m = \sqrt{1 + r}$  at the maximum of the growth rate $\sigma$ is given by the dotted line.}
\end{figure}

The stability properties of periodic solutions of Eq.~(\ref{dynglei})
beyond threshold  are summarized in  Fig.~\ref{stab_diagram_komplett}.
The dot-dashed line represents the neutral 
curve $r_N(\tilde{k})$ as described by Eq.~(\ref{eq:r_N}).
In the range beyond $r_N(\tilde{k})$ periodic solutions are in
one spatial dimension only 
stable within the Eckhaus stability boundary, 
which is marked in Fig.~\ref{stab_diagram_komplett}
by triangles (full numerical analysis). For comparison
the Eckhaus-stability boundary in terms of
a one-mode approximation is given by the 
dashed line, cf. Eq.~(\ref{eq:r_E}).
The zig-zag stability boundary obtained 
by full numerical analysis is marked by the
open circles in 
Fig.~\ref{stab_diagram_komplett}
and for the one-mode approximation in Eq.~(\ref{eq:zz_line})
by the solid line.
To the left of the zig-zag boundary and to the right
of the right Eckhaus boundary, spatially periodic solutions are 
unstable in two spatial dimensions.
In a full numerical analysis
the nonlinear periodic solution
is determined numerically, as described in more detail in
Appendix \ref{app_numstab}.
Then the stability boundaries  
are determined by the conditions given in Eq.~(\ref{potEck}) and in Eq.~(\ref{potZZ}),
similar as for the
one mode approximation above.

Alternatively, the stability of
nonlinear periodic solutions is determined by a linear stability
analysis, as described in Appendix \ref{app_numstab} and in Appendix \ref{app_stab}
analytically in the range $r \gtrsim 0$. 
The analysis of the functional and the linear stability analysis 
give identical stability boundaries.

The reduced control parameter range, $0 < r < 1$, corresponds 
to a moderately deep quench of the diblock copolymer melt
and belongs to the  so-called weak segregation regime.
In this range the Eckhaus and zig-zag boundary, as  
obtained by the one-mode approximation in  Eq.~(\ref{eq:onemode}), 
coincide well with the numerical stability analysis, where several modes 
of the Fourier expansion in Eq.~(\ref{Fourieransatz}) are taken into account.
In the range $r > 1$ the spatial shape of the nonlinear 
periodic solutions becomes increasingly anharmonic and 
deviates from the one-mode solution given by Eq.~(\ref{eq:onemode}).
Accordingly, 
the results of the one-mode approximation, 
as given by Eq.~(\ref{eq:r_E}) and Eq.~(\ref{eq:zz_line}), start to 
deviate from the full numerical results 
obtained for the Eckhaus boundary (triangles) and the zig-zag line (circles).
The dotted line in Fig.~\ref{stab_diagram_komplett} shows the control parameter 
dependence of the wave number $\tilde{k}_m$ of the 
fastest growing perturbation with respect to the homogeneous state.
The curve $\tilde{k}_m(r)$ crosses the 
Eckhaus boundary at $\tilde{k} = 5^{1/4} \approx 1.5$, corresponding to $r = \sqrt{5} - 1 \approx 1.24$.
Therefore, after a  deep quench with $r > \sqrt{5} - 1$,
the wave number of lamellar structures developing during the early 
stages of microphase separation may lie in the unstable
range to the right of the Eckhaus 
stability boundary.
The processes required to relax the wave number of 
the periodic solution back to the stable wave number band leads to the appearance 
of many defects as discussed further
in Sec.~\ref{Dynphasesep}.

\subsection{Strong segregation regime}
\label{stab_strong}
\begin{figure}[ht]
\begin{center}
\includegraphics[width=8.25cm]{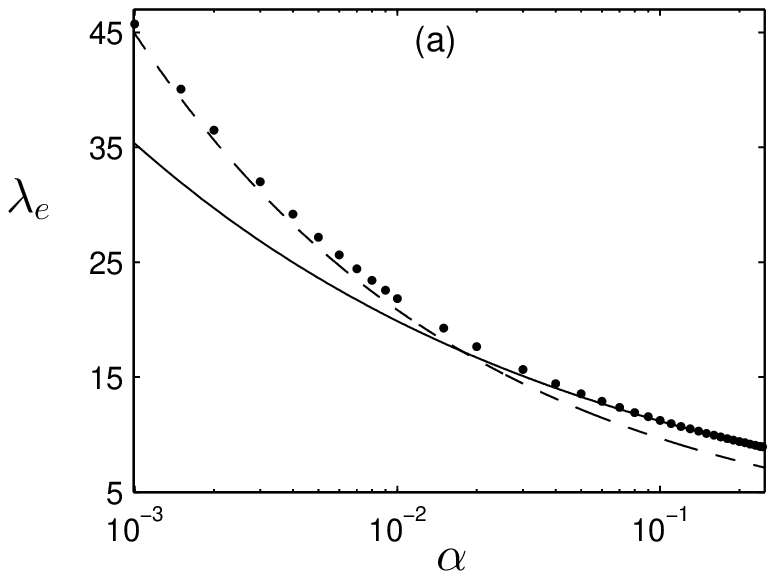}
\includegraphics[width=8.25cm]{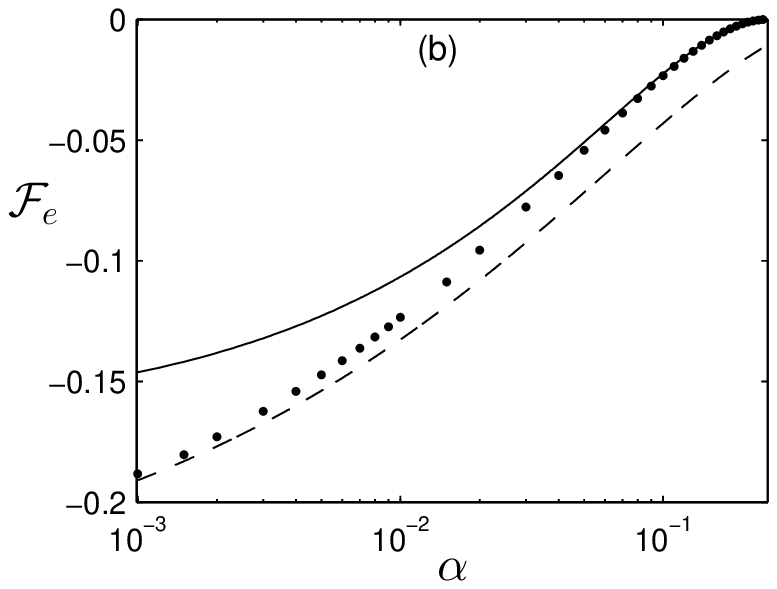}
\caption{\label{free3} The lamella period $\lambda_e$  at the minimum of
${\cal F}(k)$ is shown in  (a)
as a function of $\alpha$ for  $\varepsilon=1$
and in (b) the corresponding free energy ${\cal F}_e={\cal F}(\lambda_e)$. 
The solid lines are obtained by Eq.~(\ref{eq:lmbdeFe}) (weak segregation), 
the dashed lines by Eq.~(\ref{eq:lmbdeFe_step}) (strong segregation). 
The bullets are the results of the full numerical calculations.
 }
\end{center}
\end{figure}

With increasing of the control parameter $r$ (resp. $\varepsilon$) 
the spatially  
periodic solutions of Eq.~(\ref{dynglei}) become rather anharmonic
and for large values of $r$ 
they can be approximated by a square wave of the form
\begin{eqnarray}
\label{eq:psi_step}
  \hspace{-1mm} \psi(x) =a_0 [1 - 2\Theta(x-\lambda/4) + 2\Theta (x-3\lambda/4)], 
\end{eqnarray}
with $0 < x < \lambda$ and the Heaviside step function $\Theta(x)$.
In this limit the Green's function in Eq.~(\ref{eq:Fbulk}) is  given by $G(x,x') = |x-x'|/2$.
The gradient square term in the free energy 
Eq.~(\ref{eq:Fbulk}) can be calculated 
by using the  hyperbolic tangent profile $\psi = \mp a_0 \tanh[(x-x_i)/\xi]$ with a small 
width $\xi$ of the interfaces at $x_1=\lambda/4$ and $x_2=3\lambda/4$
as an approximation of the step function.
Within these approximations one obtains the following expression
for the free energy per period $\lambda$:
\begin{eqnarray}
\label{eq:F_step}
{\cal F} = -\frac{a_0^2}{2} \left( \varepsilon - \frac{a_0^2}{2} 
 - \frac{16}{3\lambda \xi} - \frac{\alpha \lambda^2}{48} \right) \;.
\end{eqnarray}
A minimization of this expression with respect to $a_0$ gives
\begin{eqnarray}
\label{eq:a0_step}
a_0^2 = \varepsilon - \frac{16}{3 \lambda \xi} - \frac{\alpha \lambda^2}{48} \;,
\end{eqnarray}
and with this amplitude the corresponding free energy can be further simplified to
\begin{eqnarray}
\label{eq:Flmbd_step}
{\cal F} = -\frac{1}{4} a_0^4 \;.
\end{eqnarray}
The equilibrium period of lamellae $\lambda_e$ corresponding to the minimum 
of the free energy is found by a minimization of Eq.~(\ref{eq:Flmbd_step}) with respect to $\lambda$:
\begin{eqnarray}
\label{eq:lmbde}
\lambda_e = 4 \left( \frac{2}{\alpha \xi} \right)^{1/3} \;.
\end{eqnarray}
In the strong segregation regime, i.e.
 $\varepsilon \gg \varepsilon_c$ (large $r$), or equivalently for $\alpha \to 0$ at a fixed $\varepsilon$, the width of the interface $\xi \to \sqrt{2/\varepsilon}$ becomes small \cite{Weith:2009.1}.
In this case $\lambda_e$ at the minimum of ${\cal F}$ 
and the corresponding free energy per period ${\cal F}_e$ are given by
\begin{eqnarray}
\label{eq:lmbdeFe_step}
\lambda_e &=& 4 (2 \varepsilon)^{1/6} \alpha^{-1/3} \;, \nonumber \\
{\cal F}_e &=& -\frac{1}{4} \left[ \varepsilon - (2 \varepsilon \alpha)^{1/3} \right]^2 \;.
\end{eqnarray}
The scaling $\lambda_e \sim \alpha^{-1/3}$ has to be compared with 
the scaling in the weak segregation limit
$\lambda_e \sim \alpha^{-1/4}$ according to Eq.~(\ref{eq:lmbdeFe}).
The results for $\lambda_e$ and ${\cal F}_e$ obtained in weak and strong 
segregation regime, which are given by Eq.~(\ref{eq:lmbdeFe}) and Eq.~(\ref{eq:lmbdeFe_step}), respectively, 
are in each regime in  good agreement with the results according to the full numerical solutions, 
as can be seen in Fig.~\ref{free3}.
The transition from the weak segregation regime in the range $\alpha > 0.02$ to the strong segregation regime 
in the range $\alpha < 0.02$ is clearly visible in Fig.~\ref{free3}.
The scaling exponents agree with those obtained earlier from variational calculations \cite{Ohta:86.1} and numerical simulations \cite{Oono:88.1,Liu:89.1}.
%

\section{Orientation of lamellae between substrates}
\label{confined}

The free energy of 
block copolymer films
between two confining substrates depends
on the orientation of lamellae with respect to the substrates, on the distance between the confining substrates
and on the surface properties of the bounding substrates.
In order to model for instance a BCP film with the lamellae perpendicular to a substrate,  
which is in addition laterally confined by parallel side walls as in Refs.~\cite{Park:07.1, Nealey:2010.1},
we consider in this section Eq.~(\ref{dynglei}) in two spatial dimensions in the $x-y$ plane with boundaries at $y=0$ and $y=L_y$.  
This analysis applies as well to BCP confined between two extended plane parallel substrates.  

The film thickness $L_y$ is given in terms
of the dimensionless number $d$ and the wavelength
 $\lambda_e$:
\begin{equation}
L_y = d \ \lambda_e \;.
\end{equation}

{\it Boundary conditions.} 
At plane substrates the boundary conditions for $\psi$
given by Eq.~(\ref{eq:bc1}) and  Eq.~(\ref{eq:bc2})
take the following form,
\begin{subequations}
 \label{boundall}
\begin{eqnarray}
\label{bounda1}
&&\left. \frac{\partial}{\partial y} \left( -\varepsilon \psi+\psi^3-\nabla^2\psi \right)
 \right|_{y=0,\,L_y} = 0\,, \qquad \\
\label{bounda2}
&& \qquad\, \left. \frac{\partial \psi}{\partial y} \right|_{y=0} = g(\psi-\psi_0) \vert_{y=0} \;,
\\
\label{bounda3}
&& \qquad\, \left. \frac{\partial \psi}{\partial y} \right|_{y=L_y} = g(\psi_{L_y}-\psi) \vert_{y=L_y} \;,
\end{eqnarray}
\end{subequations}
with $\psi_0=\psi_S(y=0)$ and $\psi_{L_y}=\psi_S(y=L_y)$.

Substrates preferentially wetted by one block
of an $AB$ copolymer are described by
finite values of  $\psi_0$ and $\psi_{L_y}$, 
corresponding to so-called {\it selective} 
boundary conditions. We consider either
symmetric selective boundary conditions
at the two confining substrates,  $\psi_0=\psi_{L_y} \not =0$, 
or antisymmetric ones, $\psi_0=-\psi_{L_y} \not =0$. 
Substrates being equally wetted by 
the $A$- and the $B$ block
of a copolymer correspond to {\it neutral} 
boundary conditions,
$\psi_0=\psi_{L_y} =0$.
As a third example of confined copolymer films
we investigate also {\it mixed} boundary conditions,
when one substrate acts like a selective boundary 
and the opposite one like a neutral boundary.

With the wave vector  
${\bf k}_{\perp}=(k,0)$ we describe the periodic order of lamellae
perpendicular and with
${\bf k}_{\parallel}=(0,k)$ the periodic order of lamellae parallel 
to the substrates.

{\it  Numerical method:}
To find stationary solutions of
Eq.~(\ref{dynglei}) with
the boundary conditions Eq.~(\ref{boundall})
a central difference approximation
of the  spatial derivatives is used.
In the case of  an orientation of lamellae parallel 
to the substrates  one has to consider only the $y$ 
dependence of Eq.~(\ref{dynglei}) and  
Newton's iteration method is used for its solution.
For lamellae perpendicularly oriented 
to the substrates two-dimensional simulations 
of Eq.~(\ref{dynglei}) are required. 
In this case we
use a simple relaxation method 
with the width of one period
$L_x = \lambda_e$ along the $x$ direction.

For a given solution $\psi({\bf r})$ the total free energy $F=F_b+F_s$
is calculated by integrating Eq.~(\ref{eq:Fbulk}) and Eq.~(\ref{eq:Fsurf}) numerically.
In order to determine the last term in Eq.~(\ref{eq:Fbulk}), 
Poisson's equation for the auxiliary function in
Eq.~(\ref{fhelp2}) is solved numerically by a relaxation method.
The spatial discretization was chosen to be $\delta x = \delta y =0.5$ for most of 
the calculations, which provide a relative error of the free energy less than $1$\%.
For a transition range,
where the free energies of lamellae parallel and perpendicular 
to the substrates become comparable,
the discretization was decreased to $\delta x = \delta y =0.25$  for
the purpose of a precision higher than $0.2$\%.
Note, that the values for the total free energy $F$ presented 
in the following are divided by the system 
size $A=L_x \times L_y$, i.e. we use the free energy per unit system size.
%

\subsection{Selective boundary conditions}
\label{selectbound}
In the case of homogeneous, selective  boundary conditions
with $\psi_0,\psi_{L_y} \not =0$ at the substrates,
lamellae parallel to the boundaries have a lower 
free energy than perpendicularly oriented ones, as shown in
this section. If the values $\psi_{0},\psi_{L_y}$ 
have a magnitude similar to the maximum of the amplitude of
$\psi(y)$ in the bulk, then the envelope of $\psi(y)$
is only slightly deformed near the boundaries. 

In the case $\psi_{0}$ and $\psi_{L_y}$ 
agree with the extrema of $\psi(y)$ in the bulk, then 
also the boundary condition $\left. \partial_y \psi \right|_{y=0,L_y}\approx0$ can be
fulfilled at an extremum of a periodic function $\psi(y)$
without any deformation.

\begin{figure}[ht]
\begin{center}
\includegraphics[width=4.2cm]{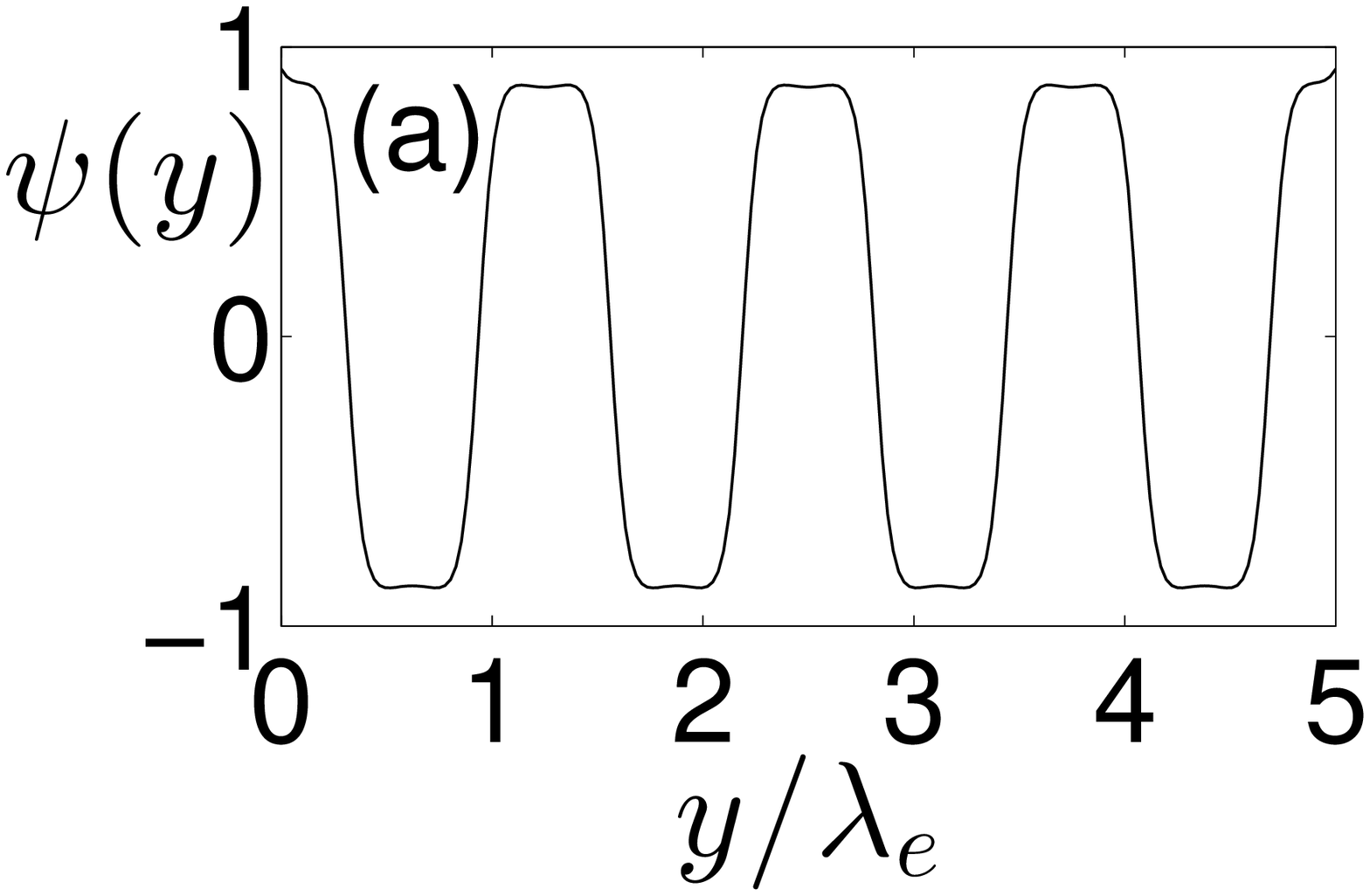} 
\hspace{-0.2cm}
\includegraphics[width=4.2cm]{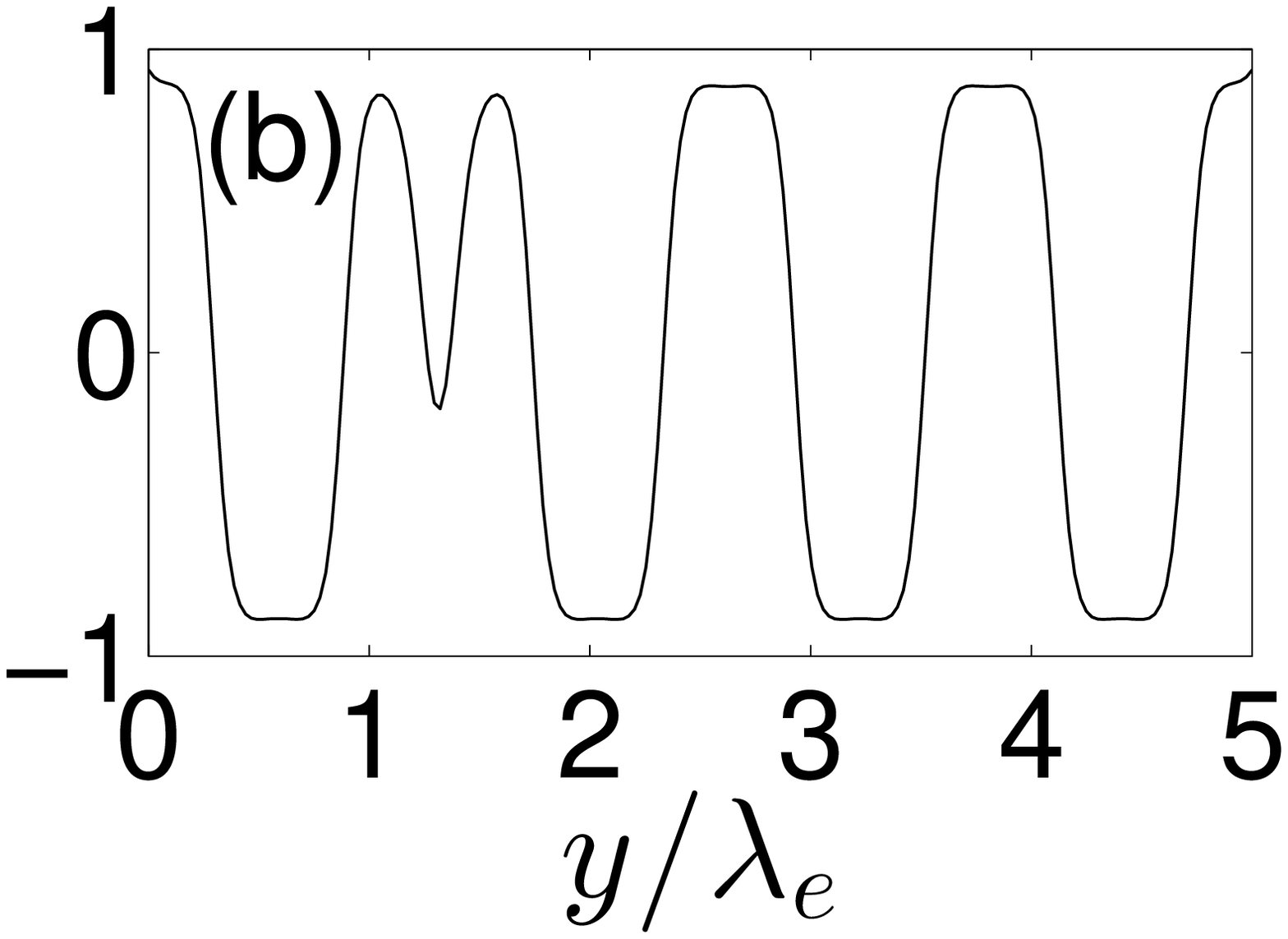} 
\includegraphics[width=4.2cm]{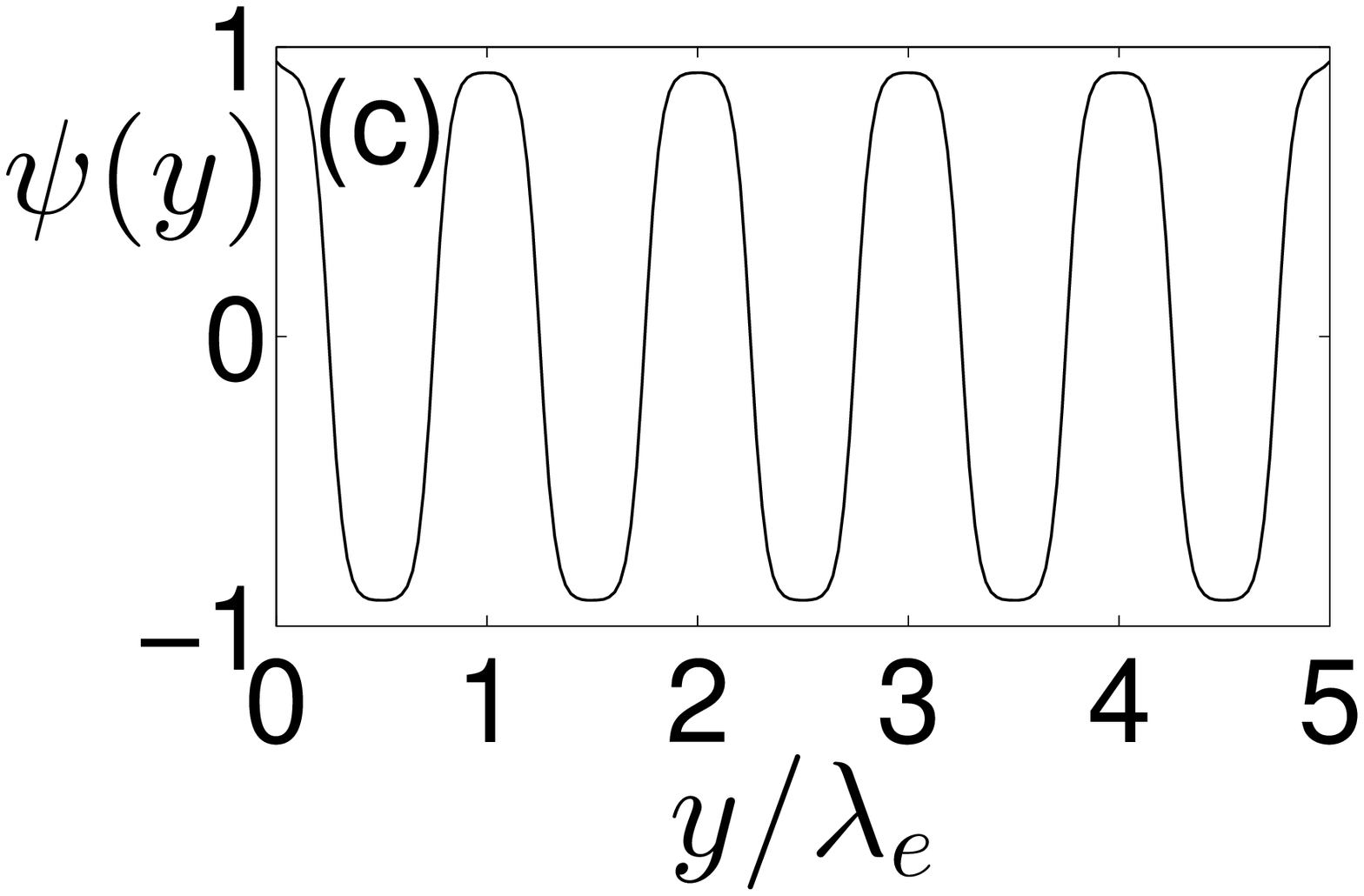} 
\hspace{-0.2cm}
\includegraphics[width=4.2cm]{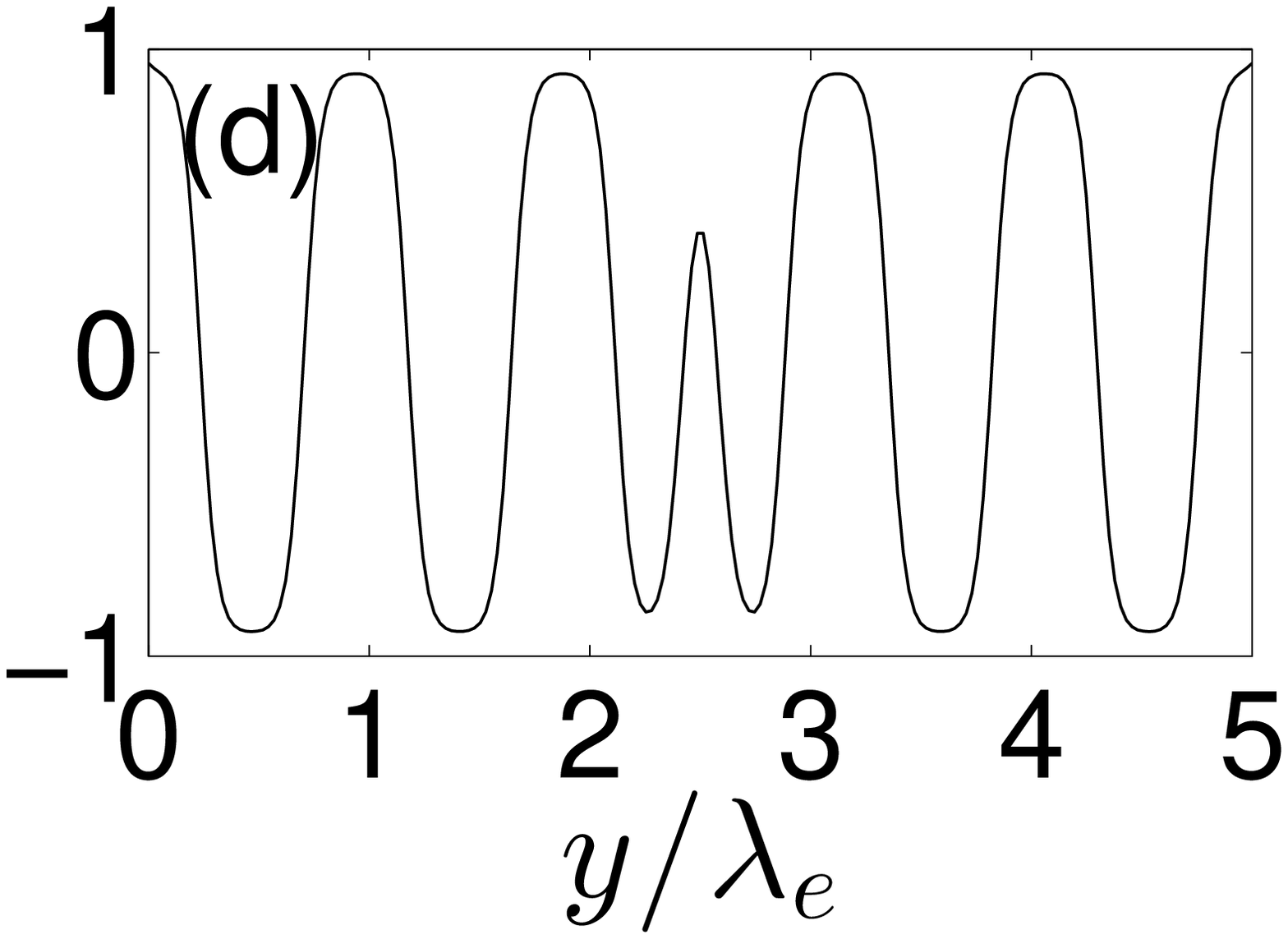}
\end{center}
\caption{\label{saddle2}
Four stationary solutions $\psi(y)$ of Eq.~(\ref{dynglei})
in a film of thickness $L_y=5\lambda_e$ are shown
for symmetric selective boundary conditions
$\psi_0=\psi_{L_y}=1$.
Part (a) and (c) show periodic solutions with
an integer number of  lamellae 
and  part (b) and (d)  so-called (unstable)
saddle-point solutions.
Parameters $\varepsilon =g=1$ and $\alpha =0.015$.
}
\end{figure}
\begin{figure}[ht]
\begin{center}
\includegraphics[width=8.25cm]{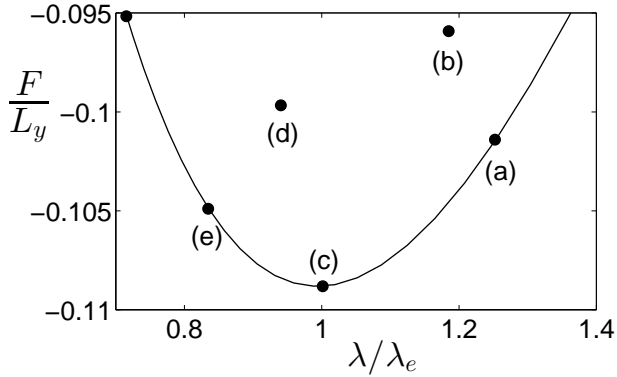} 
\end{center}
\caption{\label{saddle1}
The normalized free energy $F/L_y$
of a copolymer film of thickness $L_y=5\lambda_e$
is shown as a function of the normalized 
width $\lambda / \lambda_e$ (solid line) for lamellae
parallel to selective boundaries with
$\psi_0=\psi_{L_y}=1$ and parameters as in
Fig.~\ref{saddle2}.
(a)-(d) mark the free energy of the corresponding solutions
in Fig.~\ref{saddle2}.
}
\end{figure}

Examples of $\psi(y)$ for lamellae parallel to the boundary are
shown in the case of selective boundary conditions,
$\psi_0=\psi_{L_y}=1$, 
in Fig.~\ref{saddle2}(a) and  Fig.~\ref{saddle2}(c)
for a copolymer film of thickness  $L_y=5 \lambda_e$ 
in the strong segregation regime.
The periodic field $\psi(y)$ in Fig.~\ref{saddle2}(a)
and in Fig.~\ref{saddle2}(c)  
differs in the number of periods
on the interval $[0,L_y]$, corresponding 
to different values of the  wave number $k$.
The  wavelength of the solution with five 
periods in Fig.~\ref{saddle2}(c) 
corresponds to $\lambda=\lambda_e$
at the minimum of the free energy. In Fig.~\ref{saddle2}(a) 
the solution has four periods with a wavelength $\lambda>\lambda_e$
and this stationary solution in an unconfined system is unstable
according to the results  in Fig.~\ref{stab_diagram_komplett}. 
However, dynamical simulations show, that this wavelength is stabilized
in a confined thin film.
The solution in Fig.~\ref{saddle1}(e) has six periods
on the interval $L_y = 5 \lambda_e$
and a wavelength $\lambda$ smaller than  $\lambda_e$. 
This solution is according to the results presented
in  Fig.~\ref{stab_diagram_komplett} expected to be stable
in unconfined systems. 
The free energy $F/L_y$ of the solutions in Fig.~\ref{saddle1}(a) and (e)
is larger than in Fig.~\ref{saddle1}(c) at the minimum of the free energy.

The stationary solutions in 
Fig.~\ref{saddle2}(b) and Fig.~\ref{saddle2}(d) are 
so-called saddle point solutions, which are unstable.
As the characteristic wavelength $\lambda$
of these two solutions we take the distance between two  extrema in the
''undistorted'' range of each solution.   
With this definition the solution in Fig.~\ref{saddle2}(d) 
has a wavelength between the wavelengths of 
the two solutions in Fig.~\ref{saddle2}(e) and (c) and 
the saddle point solution in 
Fig.~\ref{saddle2}(b) has a wavelength between that of the periodic solutions in 
Fig.~\ref{saddle2}(a) and (c).
The free energy of both saddle point solutions is higher
than that of the periodic solutions marked as (a), (c) and (e) in  Fig.~\ref{saddle1}.
The locally strong deformation of the periodic
solutions in Fig.~\ref{saddle2}(b) and (d)
may occur at different locations $y$ in the region $(0,L_y)$,
depending on the initial profile. In order to include or to remove 
one periodicity, as for instance by changing from the solution given in Fig.~\ref{saddle2}(a) to Fig.~\ref{saddle2}(c) 
or reversely, the local maximum of the saddle point solution Fig.~\ref{saddle2}(b) has
to be crossed. Such energy barriers are essentially responsible that states
with a wave number $k \neq k_c$ are stable in BCP films even if the wave number does
not correspond to the minimum of the free energy.

\begin{figure}[ht]
\begin{center}
\includegraphics[width=8.25cm]{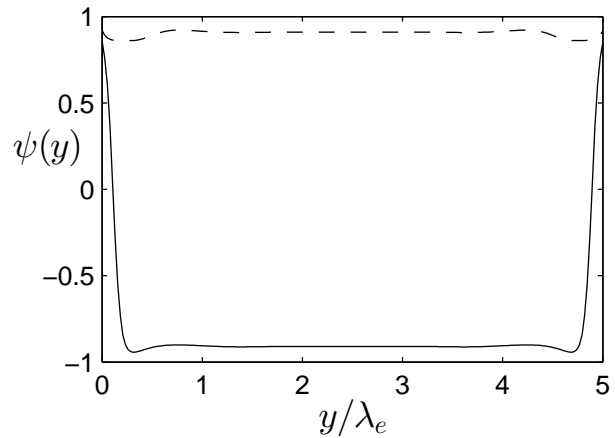} 
\vspace{-2mm}
\end{center}
\caption{\label{persol}
Stationary solutions of Eq.~(\ref{dynglei}) for lamellae oriented
perpendicularly to the boundaries
at a position $x_1$ (dashed), where $\psi(x,y)$
takes its maximum in the bulk  and at $ x_2$
(solid), where $\psi(x,y)$ takes its minimum.
The same parameters as in Fig.~\ref{saddle2}.
}
\end{figure}
\begin{figure}[ht]
\begin{center}
\includegraphics[width=8.25cm]{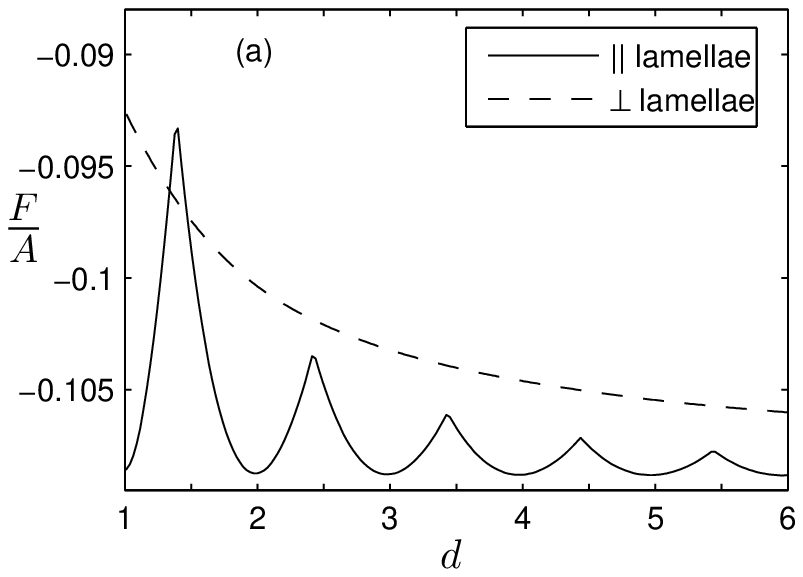}\\
\includegraphics[width=8.25cm]{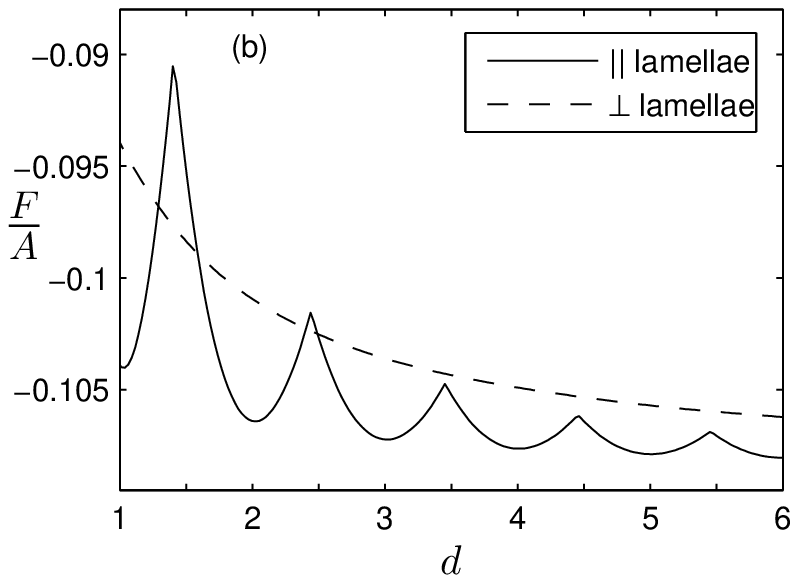}
\end{center}
\caption{\label{selec}
The free energy per unit area $F/A$ as a function of $d$
for lamellae parallel 
or perpendicular to symmetric selective boundaries 
in (a) with $\psi_0=\psi_{L_y}=1$  and
in (b) with  $\psi_0=\psi_{L_y}=0.5$. Parameters  $\alpha=0.015$ and $\varepsilon=g=1$.
}
\end{figure}

The $y$ dependence of the order parameter $\psi$
is rather different in the case of the lamellae perpendicular
to selective boundaries.
In this case $\psi(x,y)$ is a periodic function along the $x$ direction and
selective boundary conditions force 
a finite value  $\psi_{0,L_y}\not = 0$ ($\psi_{0,L_y}>0$) at $y=0,L_y$, 
independent of the phase of the function. 
At positions $x_1$, where
$\psi(x,y)$ takes its maximum 
in the bulk, the order parameter
$\psi(x_1,y)$ is nearly undeformed as a function of $y$, as
can be seen by the dashed line in Fig.~\ref{persol}.
However, the imposed selective boundary condition
requires strong deformations of $\psi(x_2,y)$ along
the $y$ direction at positions $x_2$, where $\psi(x,y)$ takes its minimum in the bulk, as 
indicated by the solid line in Fig.~\ref{persol}.
As a consequence of such strong deformations of $\psi(x_2,y)$ near the  
boundaries, perpendicularly oriented lamellae have for 
selective boundary conditions
a higher free energy than parallel oriented ones
as shown in more detail in  Fig.~\ref{selec}.

The free energy per unit area, $F_\parallel/A$, of a copolymer film with its
lamellae parallel to the boundaries has as a function 
of the film thickness $L_y=d\lambda_e$
local minima at integer values of $d$ as indicated 
by the solid lines in Fig.~\ref{selec}(a) and Fig.~\ref{selec}(b).
For parameters used in Fig.~\ref{selec}(a) the corresponding
minima of the solid line have even an equal height.
The dashed line in Fig.~\ref{selec}(a) shows the normalized 
free energy, $F_\perp/A$,
of lamellae perpendicular to the substrate, which is
for nearly all values of $d$ higher than 
$F_{\parallel}/A$. 
The periodically occurring
strong variation of the order parameter $\psi(x_2,y)$ for perpendicularly
oriented lamella in the case of selective boundary conditions, as indicated at
one position $x_2$ by the solid line in Fig.~\ref{persol},
enhances the free energy compared to the nearly 
undeformed function $\psi(y)$  in Fig.~\ref{saddle2}(a)
for parallel lamellae.  

The decay of $F_\perp(d)/A$ in Fig.~\ref{selec}(a)
indicates that the
weight of the strong deformation
of $\psi(x,y)$ of perpendicularly oriented lamellae near the substrate
becomes smaller with increasing thickness of the copolymer film.
In Fig.~\ref{selec}(a) only for a very thin
film-thickness of about $d\sim 1.5$ the free energy of parallel
oriented lamellae is higher than for perpendicularly oriented ones.

In Fig.~\ref{selec}(b) the normalized free energies 
of parallel and perpendicularly oriented lamella 
are shown in the case of a reduced
selectivity, $\psi_0=\psi_{L_y}=0.5$.
Since the control parameter $\varepsilon$ (resp. $r$)
is unchanged compared to the case in 
Fig.~\ref{selec}(a), 
a reduced value  $\psi_0=\psi_{L_y}=0.5$
requires now a deformation
of the function $\psi(y)$ at the boundaries also for parallel lamellae.
This deformation increases the normalized free energy of parallel
oriented lamellae,
while the normalized free energy 
of perpendicularly oriented ones remains  nearly unchanged,
as can be seen by comparing the dashed lines in 
Fig.~\ref{selec}(a) and (b).
This enhancement of the free energy is stronger
for small values of $d$ than for larger values of $d$, 
because of the decreasing weight of boundary effects
with increasing film thicknesses.

As a consequence of this energy enhancement
in the case of a
reduced preferential adsorption, there are
now two maxima of the free energy of
parallel lamellae
in Fig.~\ref{selec}(b), at about $d\approx 1.5$ and $d \approx 2.5$,
where the free energy is higher than that of lamellae perpendicular to
the substrates. 
Such situations of confined diblock copolymers 
were also studied experimentally 
in thin films in the range $d=1.4-3.2$ by varying the 
selectivity of the substrates \cite{Kellogg:96.1}.
Here, a reduction of the preferential adsorption 
leads at about $d \approx 2.5$
to a frustration and  
lamellae perpendicularly  oriented  to the boundaries. However,
in agreement with our simulations,  
for block copolymer films being less frustrated and also 
for strong preferential adsorption 
the parallel orientation of lamellae 
remains always preferred in
this experiment.

\begin{figure}[ht]
\includegraphics[width=8.25cm]{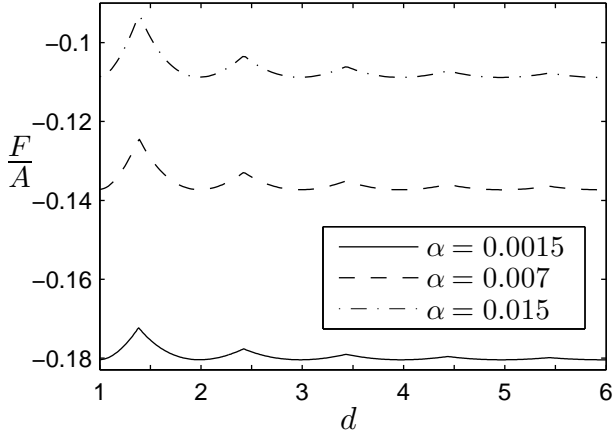} 
\caption{\label{alpabh}The free energy per unit area $F/A$ of 
lamellae  parallel to the substrate as a function of  $d$ for selective boundaries 
$\psi_0=\psi_{L_y}=1$ and different values of $\alpha$ and $\varepsilon=g=1$.}
\end{figure}

The normalized free energy of
lamellae parallel to selective substrates 
becomes smaller with decreasing values of $\alpha$,
as shown in Fig.~\ref{alpabh}. This
trend is similar to the 
$\alpha$-dependence of the 
bulk free energy given by Eq.~(\ref{eq:lmbdeFe})
[see Fig.~\ref{free3}(b)].
Since decreasing values of $\alpha$ 
correspond to increasing values of the thickness of the lamellae,
the weight of surface effects decreases,
that leads to a reduction of the peak height with $\alpha$, as can be
seen in Fig.~\ref{alpabh} too.
\begin{figure}[ht]
\begin{center}
\includegraphics[width=8.25cm]{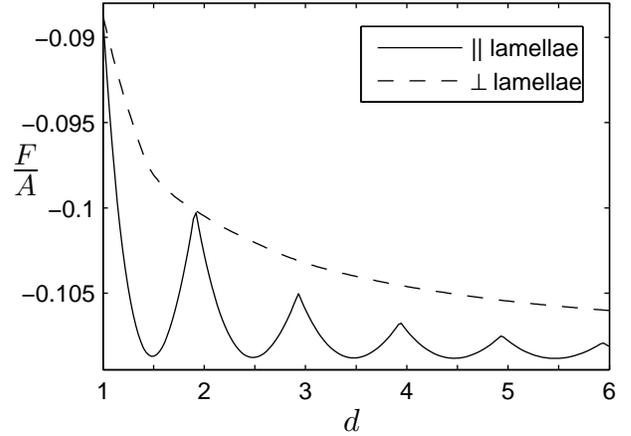}
\vspace{-5mm}
\end{center}
\caption{\label{selec2}
The free energy per unit area $F/A$ of parallel and perpendicularly 
oriented lamellae as a function of 
$d$  for asymmetric selective boundaries:
$\psi_0=1, \psi_{L_y}=-1$. Parameters $\alpha=0.015$ and $\varepsilon=g=1$.}
\end{figure}

For asymmetric selective boundary conditions at the substrates, 
when one of the two substrates 
is preferentially wetted by one block and the other one by 
the second block of the copolymer, the normalized free energy of 
parallel lamellae has local minima at a film thickness 
close to a half-integer multiple of the equilibrium lamellar thickness $\lambda_e$,
i.e. for $d=1.5,2.5,3.5,\dots$, as indicated
by the solid line in Fig.~\ref{selec2}.
A situation with comparable free energies 
for lamella orientations parallel and perpendicular 
to  asymmetric boundaries is only met 
in the range of very thin films  of about $d \approx 2$.
Otherwise the trend, that lamellae parallel to the substrates
have for asymmetric selective boundaries 
a lower free energy, can be explained by the same arguments as 
given above for the case of symmetric selective boundary conditions.

By a reduction of the surface 
interaction strength $g$ (leading to non-interacting 
or quasi-periodic boundary conditions for $g\rightarrow 0$) or by a reduction 
of the preferred difference $\psi_S$ between the concentrations 
of $A$- and $B$ blocks at the boundary
the free energies of both orientations can become comparable
in the range of very thin films like in Fig.~\ref{selec}(b). However,
in the range of thick films a parallel orientation of
lamellae is always preferred in the case of selective boundary conditions.
Detailed studies on that issue can be 
found elsewhere (see, e.g., \cite{Geisinger:99.1,Wang:00.1,Matsen:97.1,Pickett:93.1,Walton:94.1}).
%

\subsection{Neutral boundary conditions}
\label{neutbound}

Neutral boundaries with
$ \psi_0=\psi_{L_y}=0$ correspond to  substrates,
which are neither preferentially
wetted by the $A$- nor by the $B$ block of a copolymer.

A first estimate of the expected
preferred lamellae orientation 
may be gained by considering the effect of neutral boundaries 
in the weak segregation limit
with small values of $r \gtrsim 0$. 
In this range a representation 
of $\psi =A \exp\left(i ~{\bf k}_{\parallel,\perp} \cdot {\bf r}\right)+cc$ 
as in Eq.~(\ref{eq:psi_sp}) is useful, where
${\bf k}_{\perp}=(k,0)$ is the wave vector in the $(x,y)$ plane
of lamellae  perpendicular  
and  ${\bf k}_{\parallel}=(0,k)$
of lamellae parallel to the substrates.
The envelope $A(x,y)$ decays in the case of 
neutral boundaries  from its finite bulk value $A \propto \sqrt{r}$
to the boundary value $A \approx 0 $.
Such a reduction of the envelope causes 
an enhancement of the free energy 
per unit size compared to the case without boundary effects.

The transition layer, in which the  envelope $A(x,y)$ changes
from its bulk value to that at the boundary, 
is for perpendicularly oriented lamellae
according to 
Eq.~(\ref{scalexy})  
proportional to $ \xi_2 \propto r^{-1/4}$,
which is  for small values of $r$ 
smaller than the transition layer 
of parallel oriented lamellae proportional to $\xi_1\propto r^{-1/2}$.
Since the transition range is smaller 
in the case of lamellae perpendicular to the boundaries,
we expect a smaller energy of perpendicularly oriented
lamellae than for
parallel oriented ones.

\begin{figure}[ht]
\begin{center}
\includegraphics[width=8.25cm]{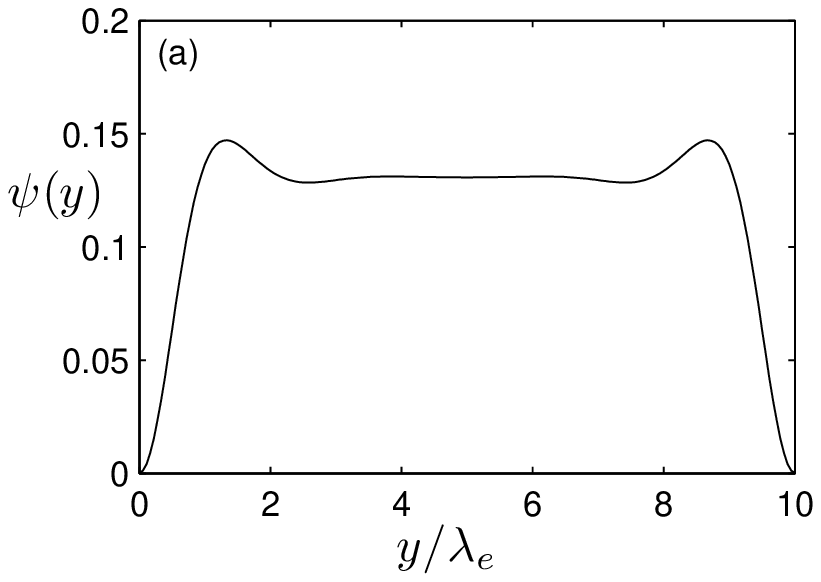} 
\includegraphics[width=8.25cm]{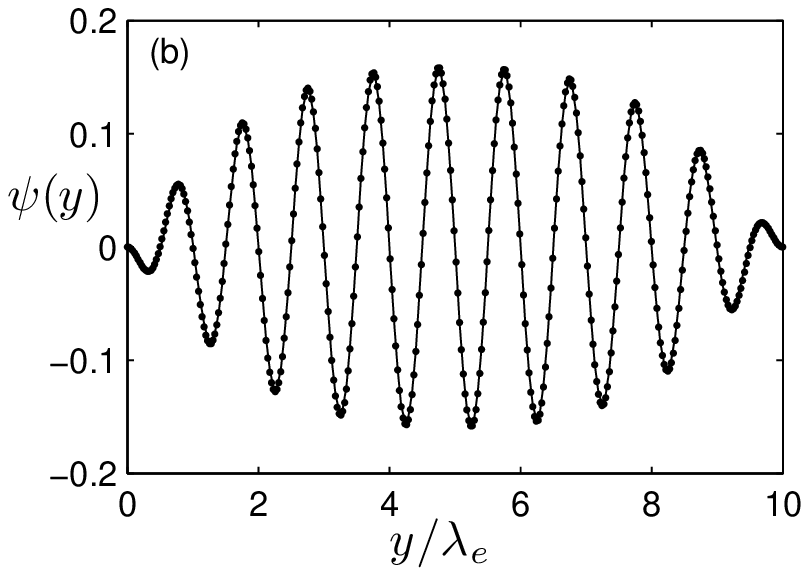}
\vspace{-4mm}
\end{center}
\caption{\label{analyt}
Stationary solutions of Eq.~(\ref{dynglei})
for neutral boundary conditions $\psi_0=\psi_{L_y}=0$: In 
(a) at a position $x_1$ where $\psi(x,y)$ takes its maximum for lamellae 
perpendicular and in (b) parallel to the boundaries.
In (b) the dots mark the analytical approximation as described 
in Appendix \ref{analconf} for the same 
boundary condition.
The parameters are 
$L_y =10\lambda_e$, 
$g=1$, $r =0.021$, $k =0.7$ 
(corresponding to $\varepsilon=1$ and $\alpha=0.24$). 
}
\end{figure}
Full numerical solutions of Eq.~(\ref{dynglei}) by taking into account the
boundary conditions (\ref{boundall}) with 
$\psi_0=\psi_{L_y}=0$ are
shown in Fig.~\ref{analyt} in the 
weak segregation  limit at $r=0.021$ for 
perpendicularly oriented lamellae in part (a) and for
parallel oriented lamellae in part (b). In Fig.~\ref{analyt}(b) we
show also the analytical approximation
of the solution (symbols) for the same
boundary conditions, as described in the Appendix \ref{analconf}.

As indicated by the estimate in the previous paragraph,
the length of the envelope of $\psi(y)$ needed for the
transition from its value in the bulk to that at the boundary
is indeed
larger for parallel oriented lamellae than
for the perpendicularly oriented ones.
A narrower transition
range causes a smaller enhancement of the free energy 
and therefore, in the range of small values of $r \gtrsim 0$ (weak segregation limit)
 lamellae perpendicularly oriented with respect to
the substrates are energetically preferred.
This behavior
extends also to the strong segregation regime with
larger values of $r$,
as we have tested  by further numerical calculations.
\begin{figure}[ht]
\begin{center}
\includegraphics[width=8.25cm]{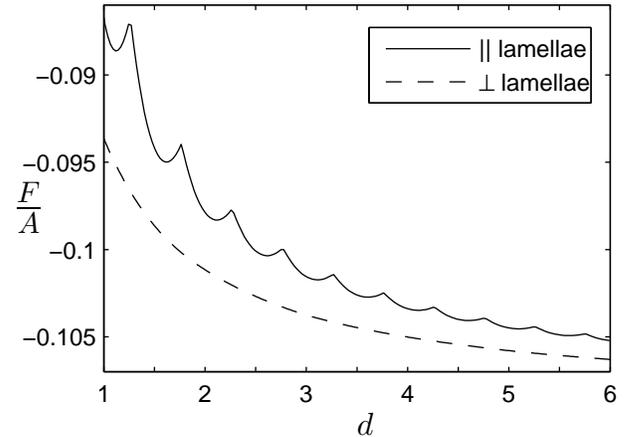}
\end{center}
\caption{\label{neutr}
The free energy per unit area $F/A$ of
parallel and perpendicularly oriented lamellae as a 
function of $d$ for neutral
boundaries $\psi_0=\psi_{L_y}=0$ and the parameters $\alpha=0.015$, $\varepsilon=g=1$.}
\end{figure}

For numerical stationary solutions of
Eq.~(\ref{dynglei}) in the strong segregation regime
the free energy of lamellae, that are perpendicularly
oriented to neutral boundaries, does not differ very much
from the free energy obtained in the case of selective boundaries, as can
be seen by comparing the dashed curves in Fig.~\ref{selec}
and Fig.~\ref{neutr}. On the other hand, the 
decay of the envelope of $\psi(x,y)$ close to
the boundaries, as shown in Fig.~\ref{analyt}(b), 
enhances the free energy of parallel lamellae 
compared to the case of selective 
boundaries, cf. Fig.~\ref{selec}. 
As a consequence of both trends,
in the case of neutral boundaries
perpendicularly oriented lamellae have always a lower
free energy than 
parallel oriented ones, as shown in Fig.~\ref{neutr}.

Note, that the free energy of parallel oriented lamellae
has in the case of neutral boundaries 
local minima as a function of the film thickness 
close to integer and half-integer multiples of $\lambda_e$.
The tendency of lamellae to align perpendicularly to the substrates 
in the case of
neutral boundaries
has been also found in Refs.~\cite{Lee:99.1,Geisinger:99.1}

In the weak segregation limit, i.e. small values of $r$,
approximate analytical solutions of Eq.~(\ref{dynglei})
are derived for lamellae parallel to the substrates, as described in
more detail in Appendix \ref{analconf}. 
Depending on  $\psi_0$ and $\psi_{L_y}$  such an
analytical approximation can be very good as can be seen
for example in Fig.~\ref{analyt}(b).

\subsection{Selective versus neutral boundaries}
It depends on the ratio between 
the extremal values of the amplitude of $\psi$ in 
the bulk and the induced values at the boundaries
whether the boundary conditions act more like
selective or neutral boundary conditions.
This can be recognized for instance by comparing the
difference between the
free energy  of parallel and perpendicularly oriented lamellae
in Fig.~\ref{selec} and Fig.~\ref{neutr}
for the three different values:  $\psi_0=\psi_{L_y}=1$,
 $\psi_0=\psi_{L_y}=0.5$ and $\psi_0=\psi_{L_y}=0$.
While in the case $\psi_0=\psi_{L_y}=1$ the maximum of $\psi(y)$ in
the bulk is 
similar to the imposed value at the boundary, in the other two cases
the maximum in the bulk is 
larger than 
at the boundaries. 

The ratio between the maximum value and
the value at the boundary can also be changed by
changing the quench depth, i.e.
by changing $r$ (respectively $\varepsilon$),
but keeping now the  values $\psi_0=\psi_{L_y}$ fixed.
In this case the maximum bulk value can be either smaller or
larger than the values at boundaries, depending on $r$.
We found that for different values of $\psi_0=\psi_{L_y}$
variations of
the control parameter $r$ do not induce
a reorientation of the lamellae with respect to
the boundaries.

\subsection{Mixed boundary conditions}
The results presented in the 
previous sections indicate that a combination 
of a selective and a neutral boundary condition may
lead to almost equal energies for parallel and perpendicularly 
oriented lamellae over a large range of 
values of the film thickness $L_y$. 
Therefore, we compare in this section for mixed boundary conditions
the free energies of homogeneously oriented lamellae
parallel and perpendicular to the confining substrates.

\begin{figure}[ht]
\begin{center}
\includegraphics[width=8.25cm]{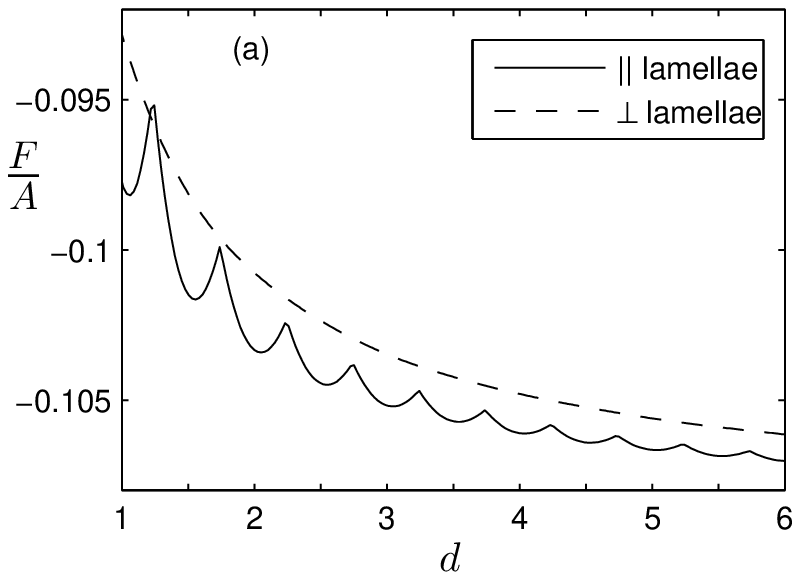}\\
\includegraphics[width=8.25cm]{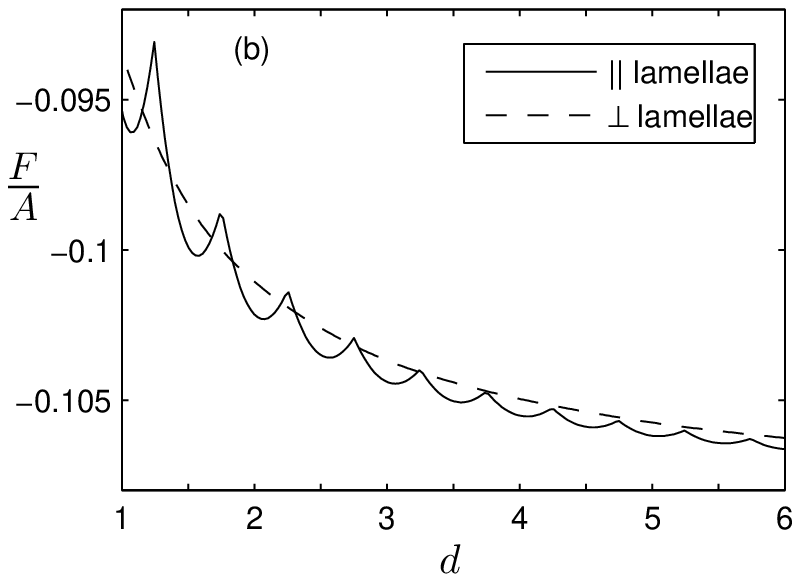}
\end{center}
\caption{\label{mixed}The free energy per unit area $F/A$ of
parallel and perpendicularly oriented lamellae as a function of $d$ for mixed boundary conditions: 
In (a) $\psi_0=0$, $\psi_{L_y}=1$ and in (b)
$\psi_0=0$, $\psi_{L_y}=0.5$. Parameters
$\alpha=0.015$ and $\varepsilon=g=1$.}
\end{figure}
In Fig.~\ref{mixed} the free energy per unit size
is shown as a function of $d$ 
for perpendicularly (dashed lines)
and parallel (solid lines) oriented lamellae
in the case of mixed boundaries with $\psi_0=0$
and either $\psi_{L_y}=1$ or 
$\psi_{L_y}=0.5$.
One may compare these results with those 
given in Fig.~\ref{selec} for symmetric selective
boundary conditions $\psi_0=\psi_{L_y}=1, ~0.5$.
There are two major differences between the
results in both figures. The energy differences between
the two lamellae orientations are smaller for 
mixed boundary conditions in  Fig.~\ref{mixed}
and the free energy of parallel oriented lamellae
now has local minima at integer and half integer values of $d$.

The trend indicated in  Fig.~\ref{mixed} suggests
that by changing the boundary condition at one 
surface from selective to neutral and keeping the other surface neutral, 
the preferred lamellae orientation 
can be changed from parallel to perpendicular.
\begin{figure}[ht]
\begin{center}
\includegraphics[width=8.25cm]{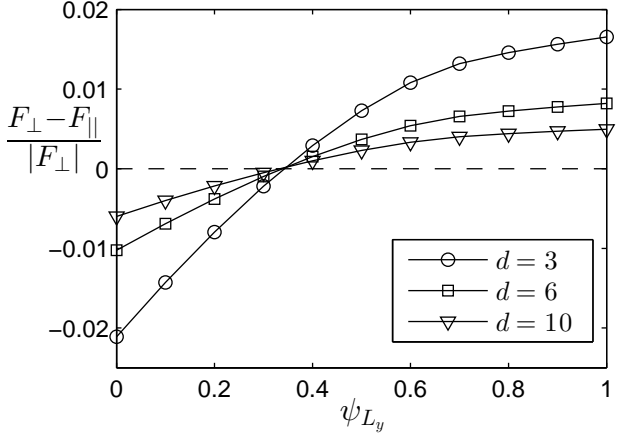}
\end{center}
\caption{\label{ueb}
The relative free energy difference  $(F_{\perp}-F_{\parallel})/|F_{\perp}|$ 
is shown 
for three thicknesses  $L_y=3\lambda_e,~6\lambda_e,
~10 \lambda_e$ of the BCP film
as a function of the 
selectivity $\psi_{L_y}$ at one boundary
and $\psi_0=0$ at the opposite one.
Beyond a critical value $\psi_{L_y}(crit)$ the preferred lamella orientation 
is parallel to the boundaries and perpendicular below. 
Parameters $\alpha=0.015$
and $\varepsilon=g=1$.
}
\end{figure}
This is shown in  Fig.~\ref{ueb}, where the relative
difference of the free energy $(F_{\perp}-F_{\parallel})/|F_{\perp}|$
is plotted as a function of $\psi_{L_y}$
for three different values of the film thickness $L_y=d\lambda_e$.
Fig.~\ref{ueb} shows in addition that the critical 
value  $\psi_{L_y}(crit)$, where both lamellae orientations
have the same free energy, 
is rather independent of the film thickness $d$. This
may be explained as follows. 
For the parameters used in Fig.~\ref{ueb} the two length scales
introduced in Eq.~(\ref{scalexy}) 
are nearly equal and both are small: $\xi_1/\lambda_e \approx 0.13$ and
$\xi_2/\lambda_e \approx 0.1$. I.e. the influence
of the boundary is similar for the three values of the film thickness in Fig.~\ref{ueb},
only the weight of the influence is reduced by
increasing the film thickness. The prior effect leads
to a smaller slope of the curves with larger values of $d$ in Fig.~\ref{ueb}.

\begin{figure}[ht]
\begin{center}
\includegraphics[width=8.25cm]{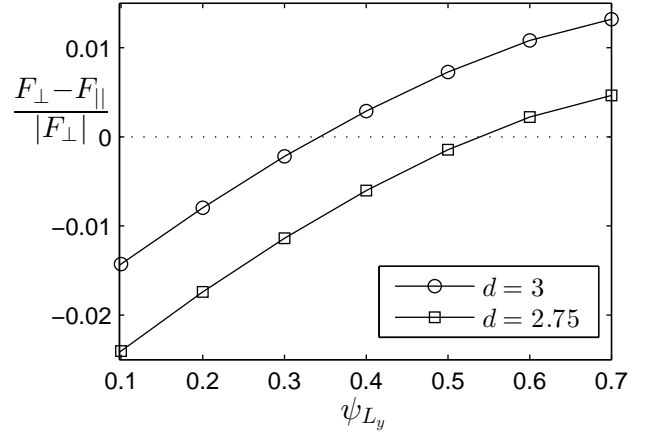}
\end{center}
\caption{\label{uebreich}
The relative free energy difference  $(F_{\perp}-F_{\parallel})/|F_{\perp}|$ 
is shown as a function of the selectivity $\psi_{L_y}$  at one boundary 
with $\psi_0=0$ at the opposite one for two BCP films
with $L_y=2.75 \lambda_e$ (lower curve) and $L_y =3 \lambda_e$
(upper curve).
In both cases the preferred lamella orientation is parallel to the boundaries
beyond a critical value $\psi_{L_y}(crit)$ and perpendicular below. 
Parameters $\alpha=0.015$
and $\varepsilon=g=1$. 
}
\end{figure}

The critical selectivity  $\psi_{L_y}(crit)$
depends weakly on the parameter $\alpha$.
For smaller values of  $\alpha$ and therefore a larger 
lamella period $\lambda_e$, the critical value of $\psi_{L_y} (crit)$ is smaller.
Thus lamellae with a larger period require a smaller 
selectivity of the surface to realign.
Note that in case of relatively thin films the free 
energy of parallel oriented lamellae as a function of thickness 
shows pronounced oscillations between local minima and maxima [see Fig.~\ref{mixed}(a)].
This leads to a weak thickness dependence of the critical 
selectivity when considering the thicknesses that correspond 
to a maximum and a minimum of $F_{\parallel}$ (see Fig.~\ref{uebreich}).
In the case of a maximum of $F_{\parallel}$ the reorientation takes place at higher values of 
$\psi_{L_y}$ than for a minimum.
With increasing film thicknesses this difference is rapidly decreasing.
%

\section{Dynamics of microphase separation} 
\label{Dynphasesep}

The spatio-temporal dynamics of microphase separation in 
copolymers in two spatial dimensions between two parallel boundaries
and the related lamellar (orientational) order
is investigated here.
We also describe typical differences between the evolution
of structures in the {\it strong} and the
{\it weak} segregation regime in BCP films 
confined between different boundaries on the one hand and
in unconfined systems on the other hand.

For numerical simulations of Eq.~(\ref{dynglei}) we use
a central  difference approximation of the spatial 
derivatives with  $\delta x=\delta y=0.5$ 
and an Euler integration of the resulting ordinary 
differential equations with a time step
$\Delta t=10^{-4} -10^{-3}$.
In the unconfined case  periodic boundary conditions
are applied and a system size
 $L_x=L_y=256$ ($\approx 14\lambda_e$ for $\alpha=0.015$) is chosen.
For block copolymer films of thickness $L_y=6 \lambda_e$ 
between two substrates, different combinations of the boundary 
conditions along the $y$ direction are used, cf. Eqs.~(\ref{boundall}), 
and periodic boundary conditions 
along the $x$ direction  with $L_x=4 \lambda_e$, $8 \lambda_e$, or $32 \lambda_e$. 
To mimic a quench we start simulations of Eq.~(\ref{dynglei})
with random initial conditions for $\psi$ of a small amplitude of about $10^{-4}$.
Typical scenarios of the dynamics of microphase separation 
are studied in the {\it strong} segregation regime
at a control parameter $\varepsilon=1$ ($r=3.08$) 
and in the {\it weak} segregation regime 
at $\varepsilon=0.37$ ($r=0.5$). 
Microphase separation can be characterized by the structure 
factor of the evolving patterns,
\begin{eqnarray}
S(\mathbf k, t)&=& |\hat{\psi}(\mathbf k,t)|^2 \;, \;\; \nonumber \\
\mbox{with}~\quad \hat{\psi}(\mathbf k,t) &=& \int e^{i\mathbf k\cdot\mathbf r}\psi(\mathbf r,t)d\mathbf r \;\,.
\end{eqnarray}
Since we expect anisotropy effects 
in BCP films confined between two substrates we introduce
different characteristic lengths along the $x$ and the $y$ direction:
\begin{equation}
l_x(t)=\pi/ \langle k_x\rangle(t) \;, \;\;
l_y(t)=\pi/ \langle k_y \rangle(t) \;.
\end{equation}
The averaged wave numbers $\langle k_i \rangle$ are 
\begin{subequations}
\label{charlae}
\begin{eqnarray}
\langle k_x \rangle(t)=\dfrac{\mathlarger{\int}_{0}^{k_{max}}dk_x\mathlarger{\int}_{-\Delta k}^{\Delta k}dk_y\,\,  S(k_x,k_y,t)k_x}{\mathlarger{\int}_{0}^{k_{max}}dk_x\mathlarger{\int}_{-\Delta k}^{\Delta k}dk_y\,\, S(k_x,k_y,t)} \;,
 \\ [3mm]
\langle k_y \rangle(t)=\dfrac{\mathlarger{\int}_{0}^{k_{max}}dk_y\mathlarger{\int}_{-\Delta k}^{\Delta k}dk_x\,\,  S(k_x,k_y,t)k_y}{\mathlarger{\int}_{0}^{k_{max}}dk_y\mathlarger{\int}_{-\Delta k}^{\Delta k}dk_x\,\, S(k_x,k_y,t)} \;,
\end{eqnarray}
\end{subequations}
where $\Delta k$ is the half-width of the corresponding peak of the structure 
factor along $k_x$ and $k_y$, respectively.
%

\subsection{Unconfined systems}
\begin{figure}[ht]
\begin{center}
\includegraphics[width=4cm]{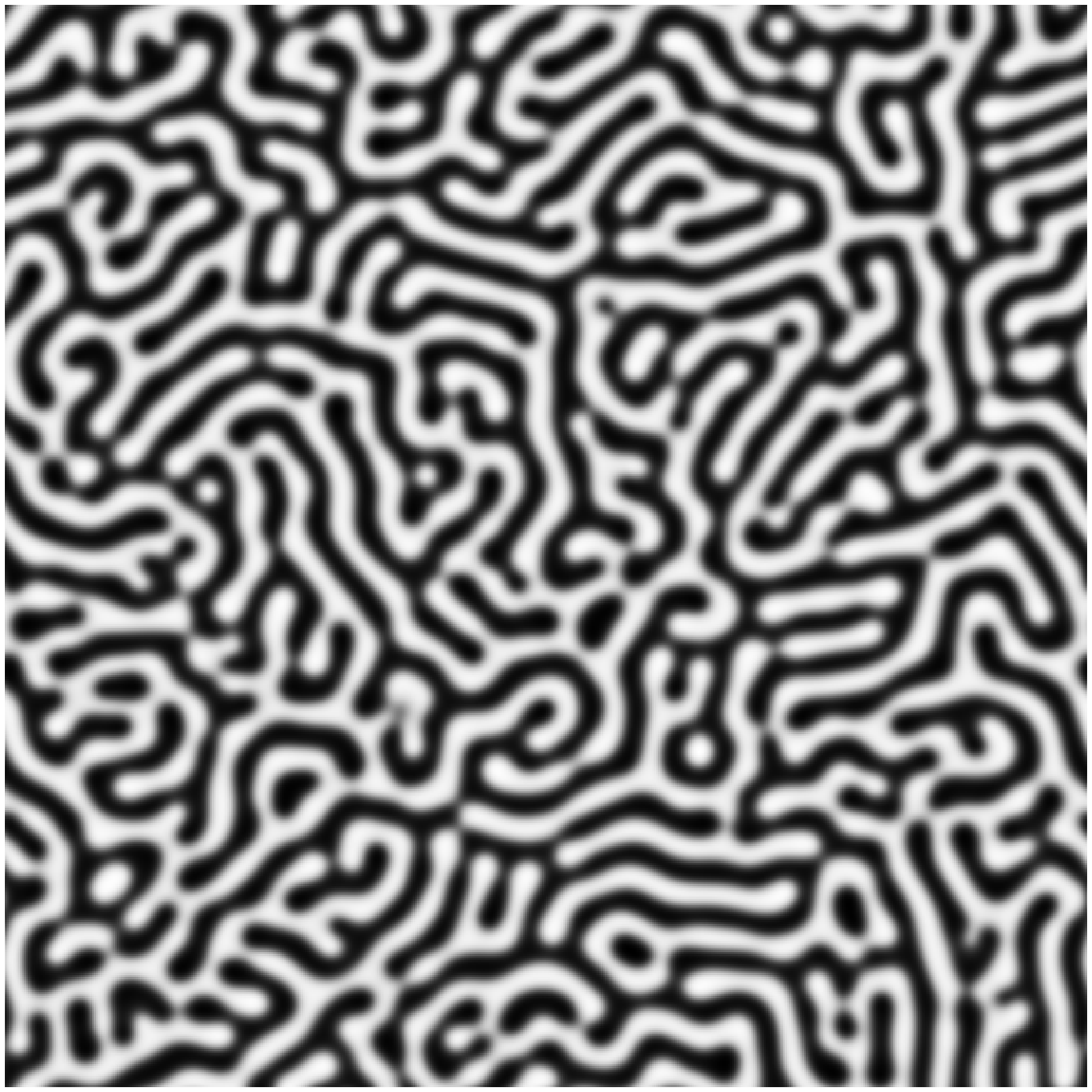}\hspace{0.1cm} 
\includegraphics[width=4cm]{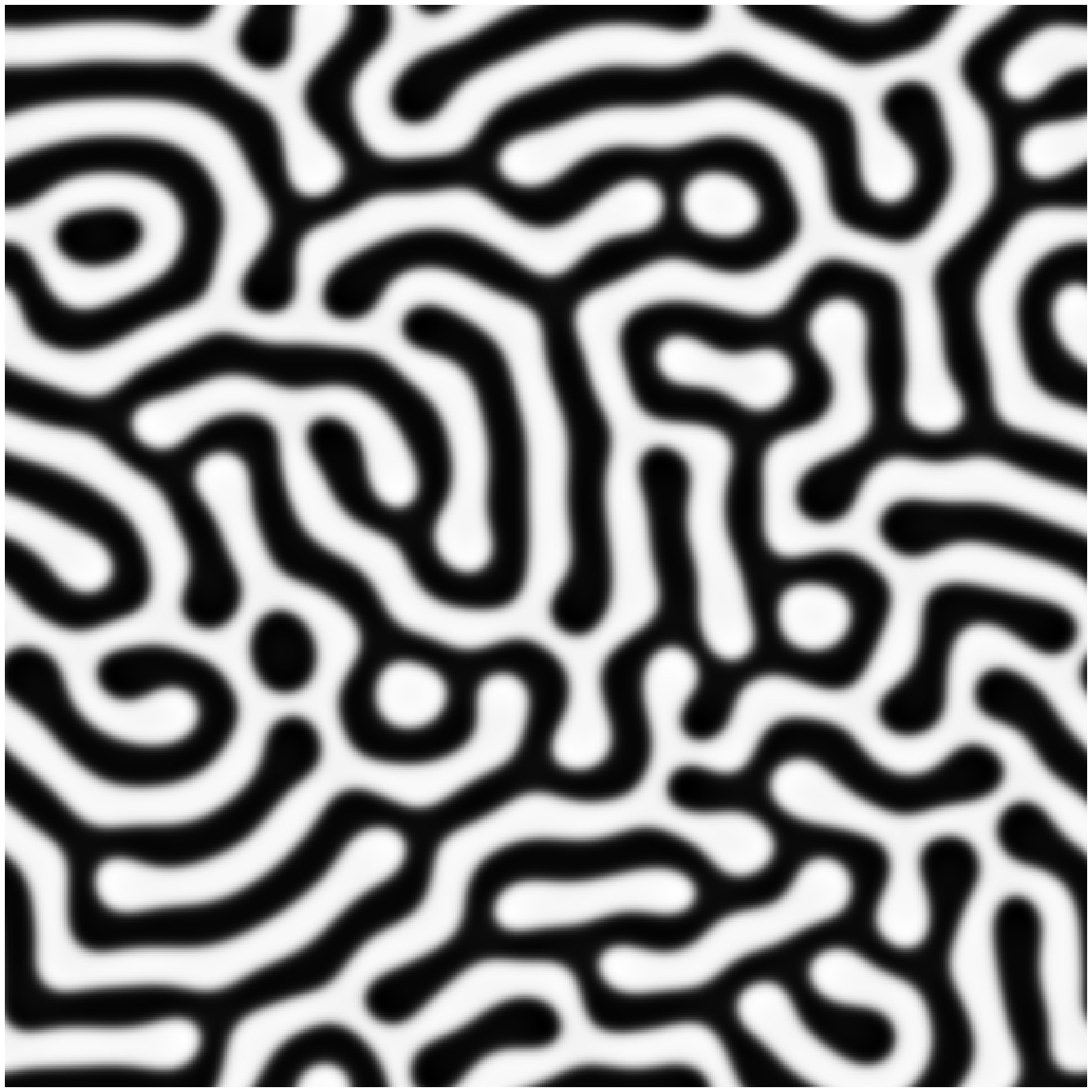}\\ 
(a)\hspace{3.8cm}(b)\\ \vspace{0.1cm}
\includegraphics[width=4cm]{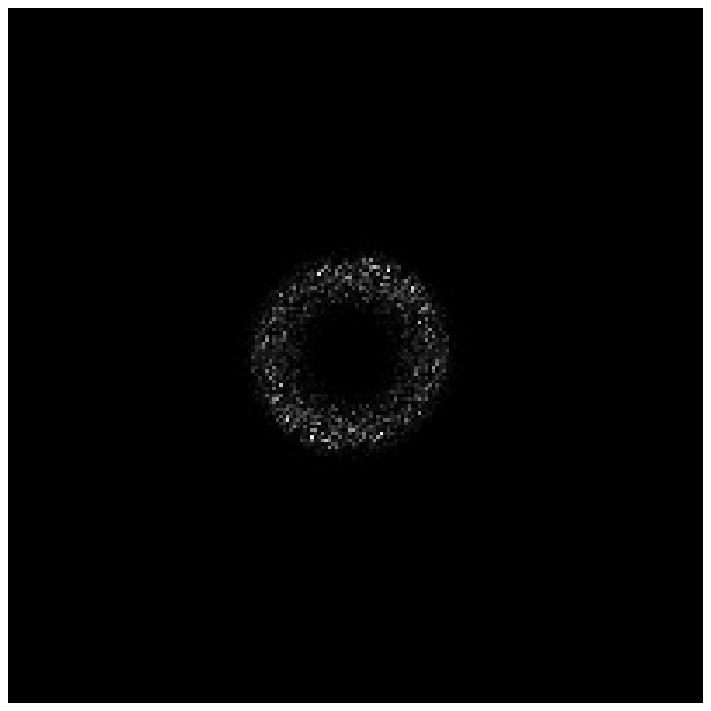}\hspace{0.1cm} 
\includegraphics[width=4cm]{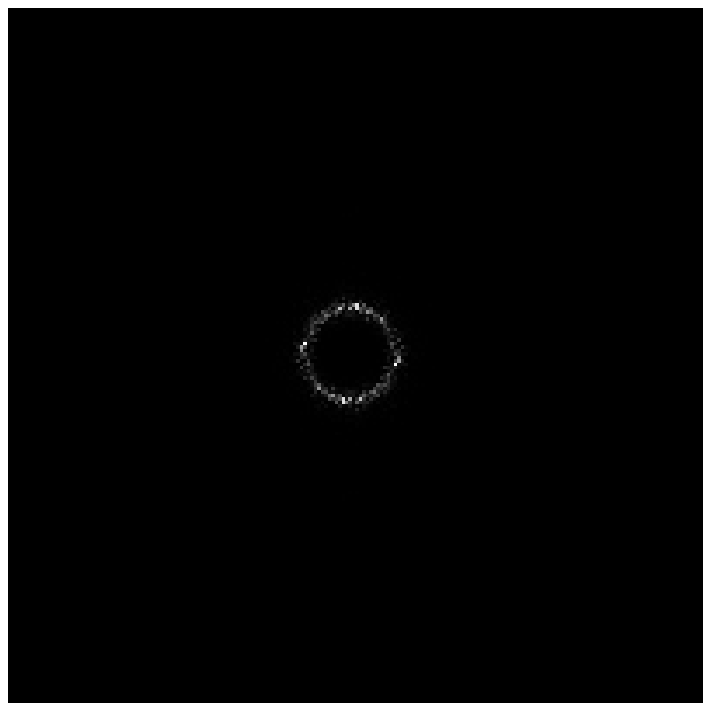}\\
(c)\hspace{3.8cm}(d) 
\end{center}
\caption{\label{2dper}
Microphase separation in the strong segregation regime 
for the parameters $\varepsilon=1$ ($r = 3.08$) and $\alpha=0.015$
is shown in a two-dimensional system with
$L_x=L_y=256 \approx 14 \lambda_e$ ($\lambda_e=19.3$)
and  periodic boundary conditions: In (a) at
the time $t=10^2$ with the average wavelength 
of about $\lambda \approx 11.0$
and in (b) at $t=10^4$ with  $\lambda \approx 17.5$. 
Dark and bright
regions in the top part correspond to 
$A$- and $B$ block rich phases,
respectively. The bottom part (c) and (d) 
show the corresponding structure factors, where the bright regions
correspond to large values of $S(\mathbf k, t)$.
}
\end{figure}

During microphase separation in diblock copolymers
the most unstable perturbation  with respect 
to the homogeneous basic state  $\psi=0$
has the wave number $k_m=\sqrt{\varepsilon/2}$, cf. Eq.~(\ref{maxgrowth}), 
similar as in binary mixtures \cite{Langer:71.1}. 
The coarsening regime in diblock copolymers
below $T_c$
is at early stages similar as in polymer blends,
as indicated by two 
snapshots of a simulation in Fig.~\ref{2dper}.
In diblock copolymers  
the coarsening process of phase separation is limited
by the chemical bond between an $A$- and $B$ block,
which limits the domain size of phase separation 
to the order of the chain length of diblock copolymers.

The pattern shown in Fig.~\ref{2dper}(b)
has an average wavelength still 
below the optimal wavelength $\lambda_e$ 
at the minimum of the free energy.
With a further progress of time the mean wavelength 
approaches only very slowly towards $\lambda_e$, 
because the system has to get rid 
of lamellar imperfections by diffusion processes.

\begin{figure}[ht]
\begin{center}
\includegraphics[width=8.25cm]{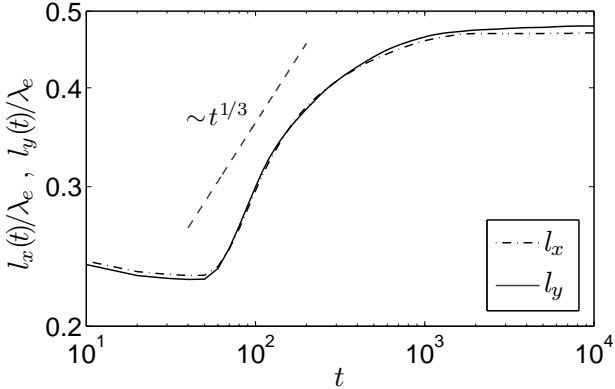}
\end{center}
\caption{\label{dyn_period}
The temporal evolution of the characteristic
length scales $l_x(t)$ and $l_y(t)$ 
(averaged over 10 independent runs) after a quench is shown 
in the {\it strong} segregation
regime for the same parameters as in Fig.~\ref{2dper}
and periodic boundary conditions.}
\end{figure}

\begin{figure}[ht]
\begin{center}
\includegraphics[width=8.25cm]{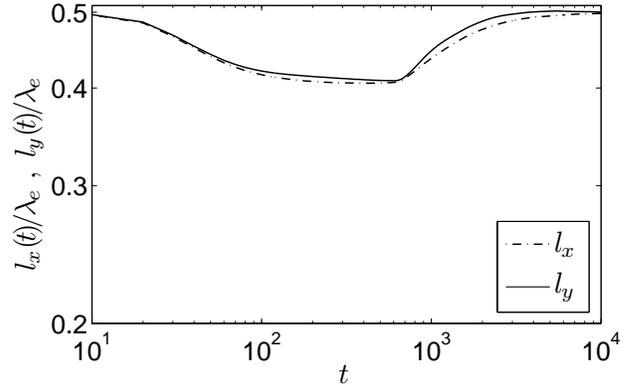}
\end{center}
\caption{\label{dyn_period2}
The temporal evolution of the
characteristic length scales $l_x(t)$ and $l_y(t)$ 
(averaged over 10 independent runs) after a quench 
is shown in the {\it weak}
segregation regime at $\varepsilon=0.37$ ($r= 0.5$)
for a system with periodic
boundary conditions and otherwise the same parameters as in 
Fig.~\ref{2dper} and in Fig.~\ref{dyn_period}.
 }
\end{figure}

In an unconfined system the two length scales $l_x(t)$ and $l_y(t)$ evolve
in a similar way and this isotropic behavior of the block copolymer melt
is also reflected by the rotational symmetry
of the structure 
factor shown by the parts (c) and (d) of Fig.~\ref{2dper}.
During the intermediate regime between 
the early stage of phase separation with
a dominating growth of the perturbation of wave number 
$k_m$ and the late stage of coarsening 
with an average domain size, $l_x \approx l_y \approx \lambda_e/2$, one observes
the scaling $l_x\sim l_y\sim t^{1/3}$  as shown in Fig.~\ref{dyn_period},
which is common for polymer blends.
Such a scenario is typical for a deep quench into the 
strong segregation regime at about $\varepsilon \sim 1$ and beyond. 

In the weak segregation 
regime ($\varepsilon=0.37$ or smaller)
the wave number of the fastest growing mode  
during the early stage of phase separation, $k_m=\sqrt{\varepsilon/2}$, 
is already closer to 
$2\pi/\lambda_e$ and therefore, one observes during pattern evolution
only a slight variation
of the scales $l_x(t)$ and $l_y(t)$, as shown in Fig.~\ref{dyn_period2}.
Typical snapshots 
of phase separation during the early stage 
in the weak segregation regime are similar to patterns
shown in Fig.~\ref{2dper}(b).

\subsection{Confined systems}
\label{Dynconfine}
\begin{figure}[ht]
 \begin{center}
\includegraphics[width=2.5cm]{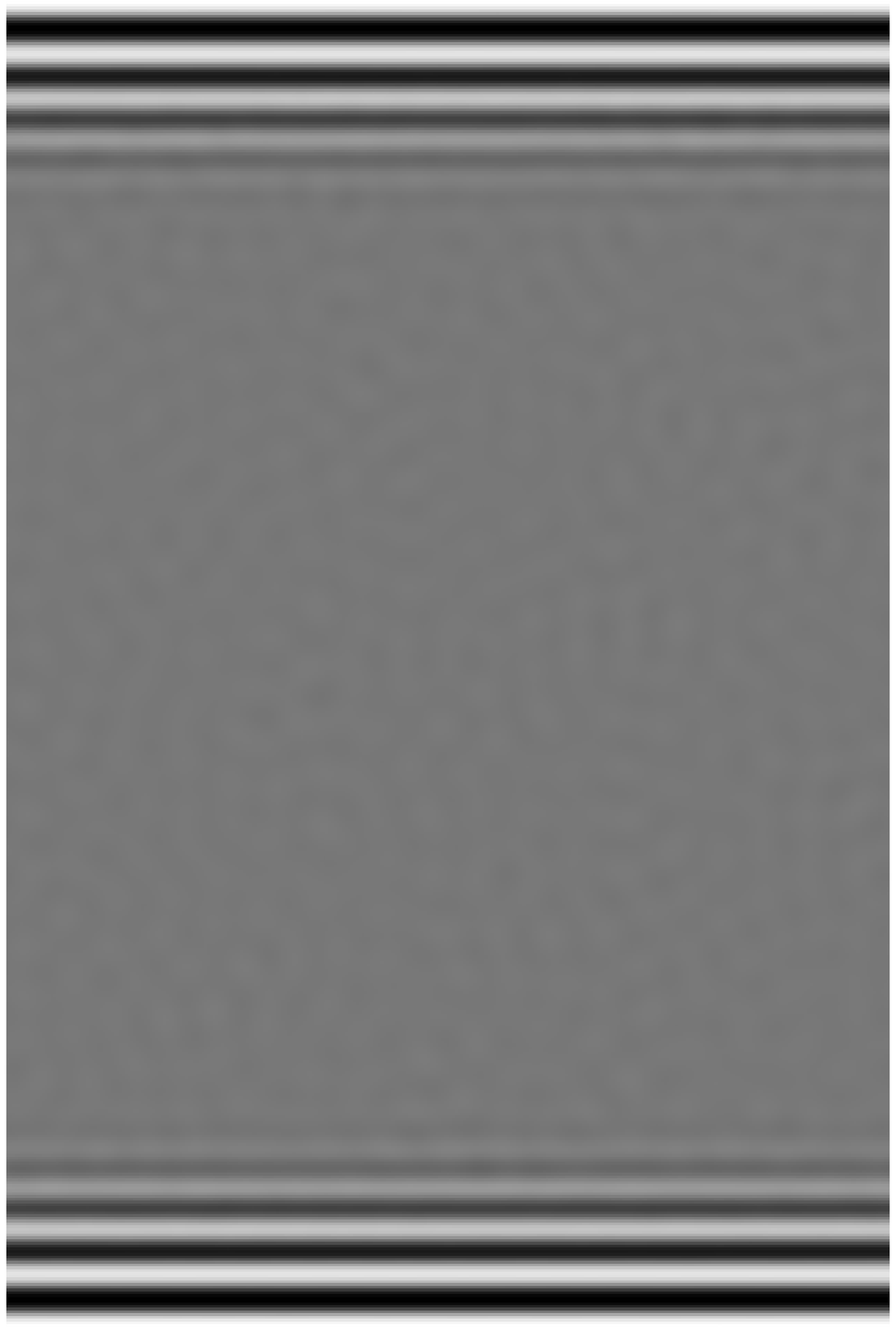}\hspace{0.3cm}
\includegraphics[width=2.5cm]{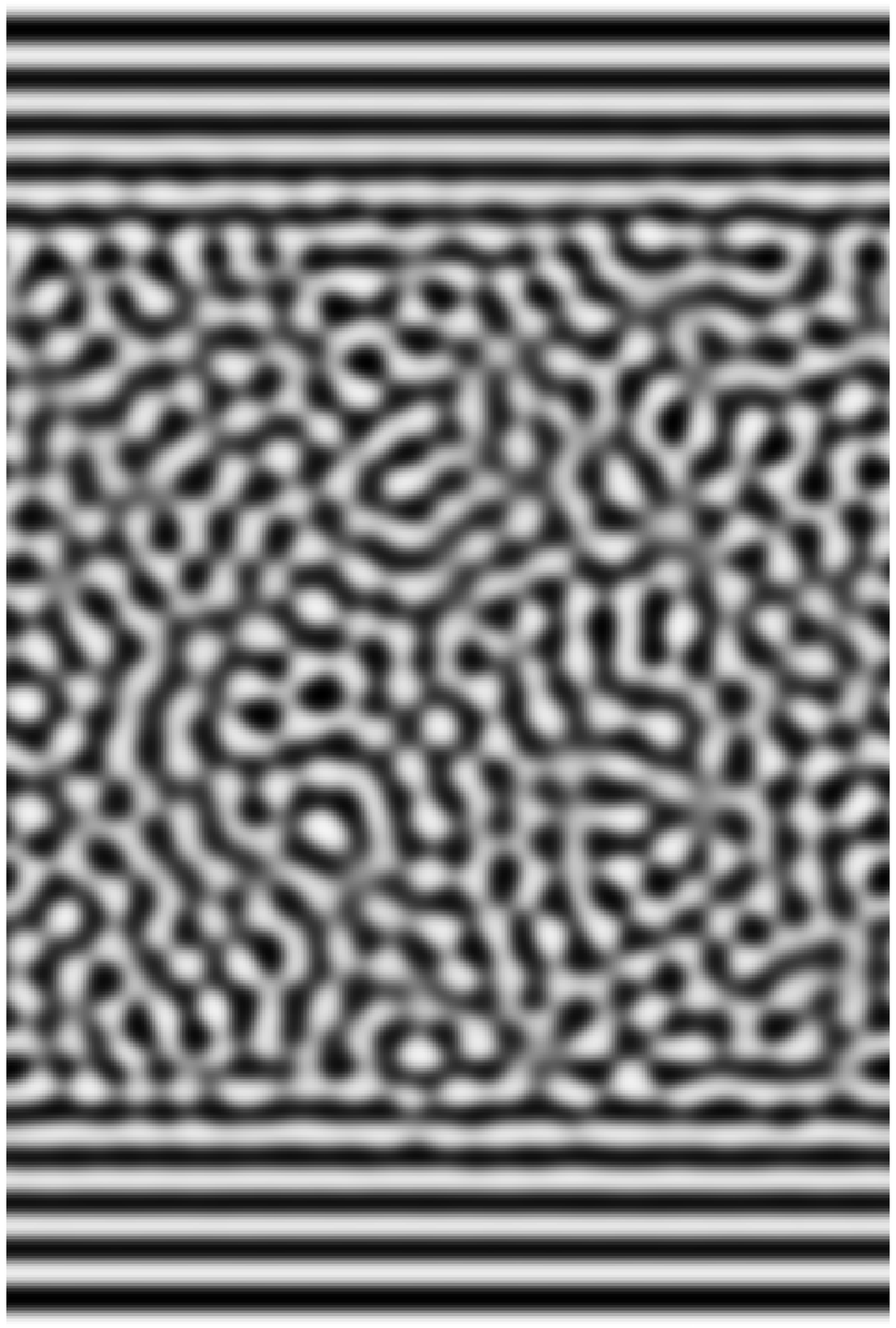}\hspace{0.3cm}
\includegraphics[width=2.5cm]{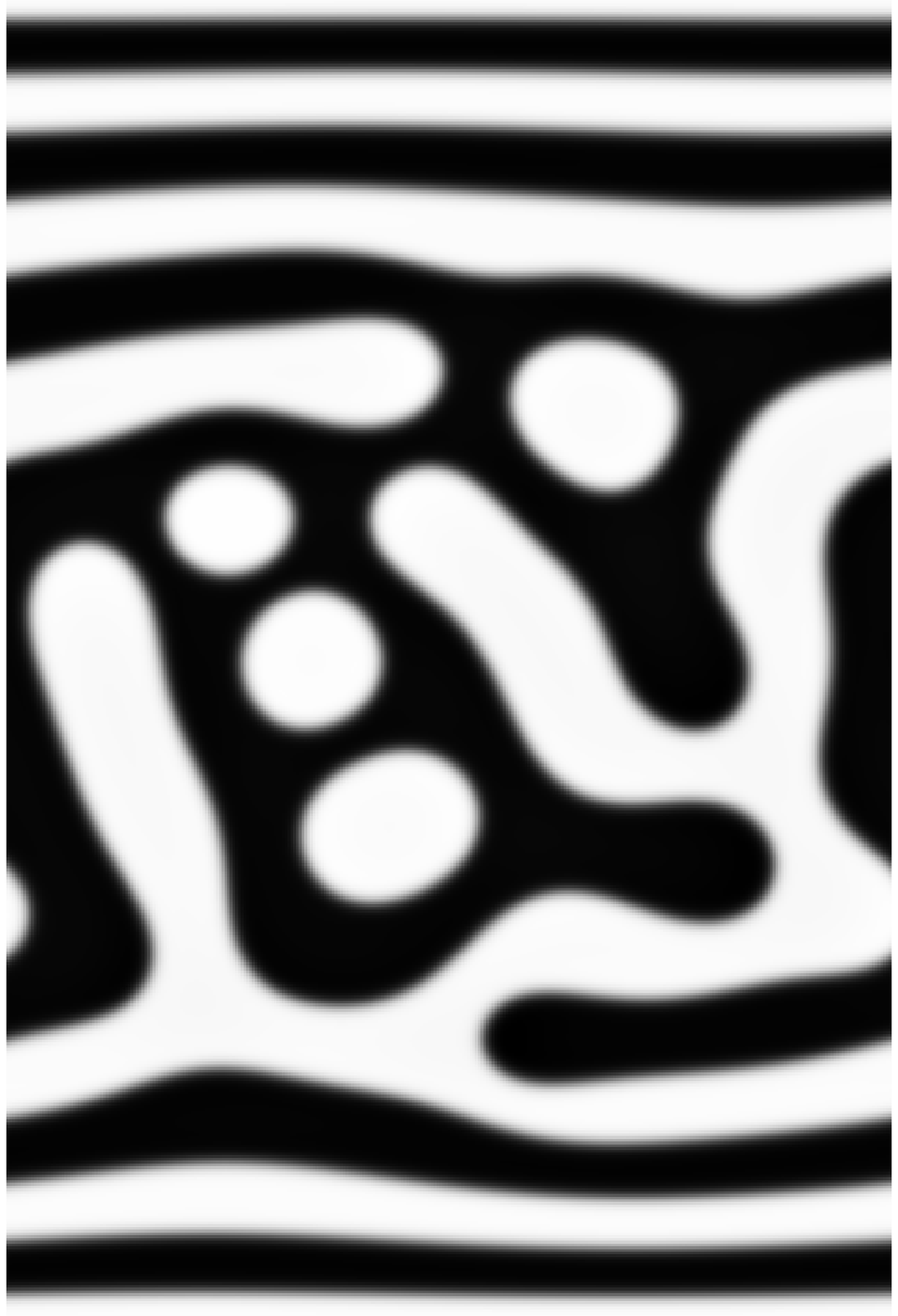}\\
(a)\hspace{2.5cm}(b)\hspace{2.5cm}(c)
\end{center}
\caption{\label{conf}
Microphase separation in a BCP film between 
two selective boundaries with $\psi_0=\psi_{L_y}=1$ 
is shown in the strong segregation regime 
at $\varepsilon=1$ ($r= 3.08$) in (a)
at time  $t=20$ 
after the quench, in (b) at $t=40$, and in (c) at $t=10^4$ close
to the final state. 
Parameters $L_x=4\lambda_e$, $L_y=6\lambda_e$,
$\alpha=0.0015$  and $g=1$.
}
\end{figure}

For a block copolymer film confined between
two selective boundaries with $\psi_0=\psi_{L_y}=1$,
three snapshots 
during microphase separation in the strong
segregation regime at $\varepsilon=1$ ($r =3.08$) are shown in Fig.~\ref{conf},
which will be compared later 
with results in the weak segregation limit.

After a deep quench at $t=0$
far below threshold $\varepsilon_c=2\sqrt{\alpha}$
(for $\alpha=0.0015$ one has $\varepsilon_c=0.077$)   
the selective boundary conditions trigger close to the
substrates immediately a large value of
the order parameter $\psi$ 
and in this case lamellae orient parallel
to the substrates. One should note, that 
for the parameters in Fig.~\ref{conf}
the wavelength of the fastest growing 
mode $\lambda_m=2\pi/k_m= 2\pi \sqrt{2/\varepsilon}
= 2\pi \sqrt{2}\approx 8.9$
is much smaller than the 
wavelength $\lambda_e= 4 (2\varepsilon)^{1/6}\alpha^{-1/3} \approx 39$ 
at the minimum of the free energy of a lamellar structure.

In the strong segregation regime
the correlation lengths 
in the $x$ and $y$ direction are rather small 
and therefore a surface induced orientational 
order of the lamellae occurs only within short ranges
near the boundaries as indicated by parts (a) and (b) in Fig.~\ref{conf}.
Further away from the substrates
the orientation of lamellae is only weakly
influenced by boundary conditions and the lamellae 
are disordered, whereby this disorder resembles very much 
to that in the unconfined case,
as shown in Fig.~\ref{2dper} and has also been observed in experiments
on confined thin films \cite{Park:07.1,Nealey:2010.1}.
The average wavelength of the structure tends 
in the long time limit to 
$\lambda_e$.

\begin{figure}[ht]
\begin{center}
(a)\includegraphics[width=8.25cm]{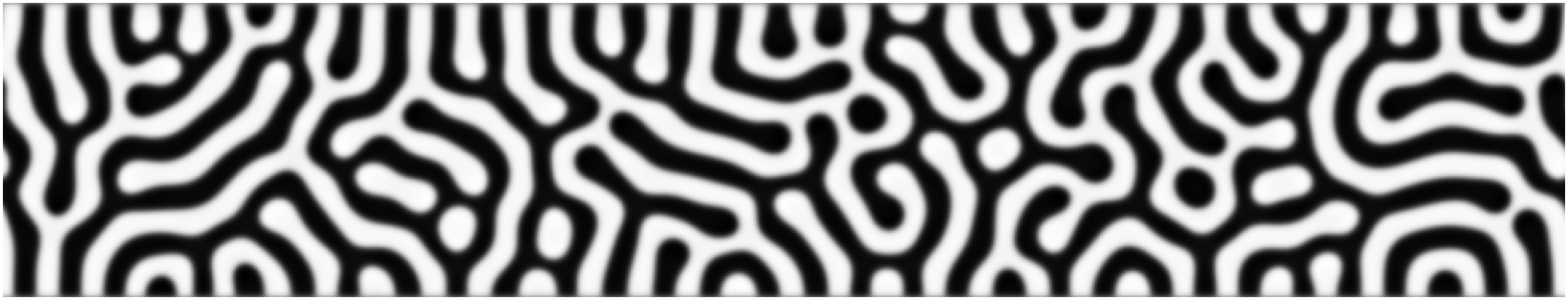}\\
\vspace{3mm} 
(b)\includegraphics[width=8.25cm]{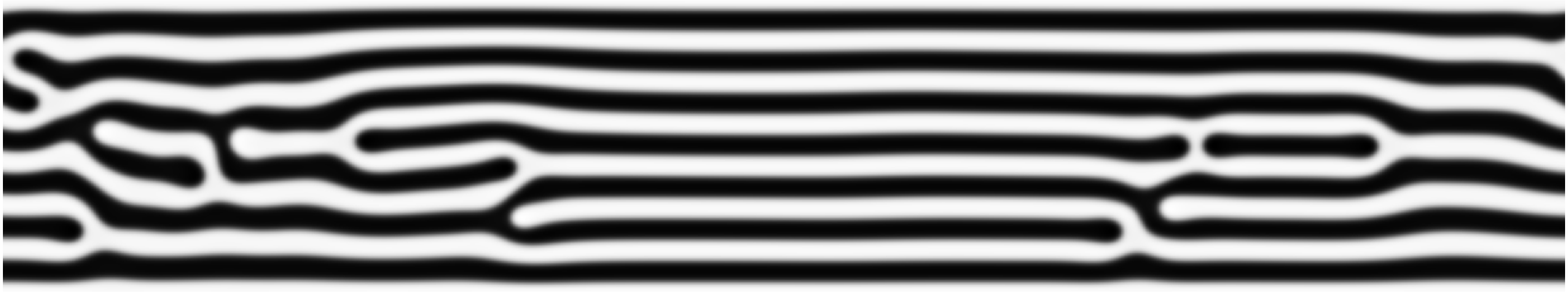}\\ 
\vspace{3mm}
(c)\includegraphics[width=8.25cm]{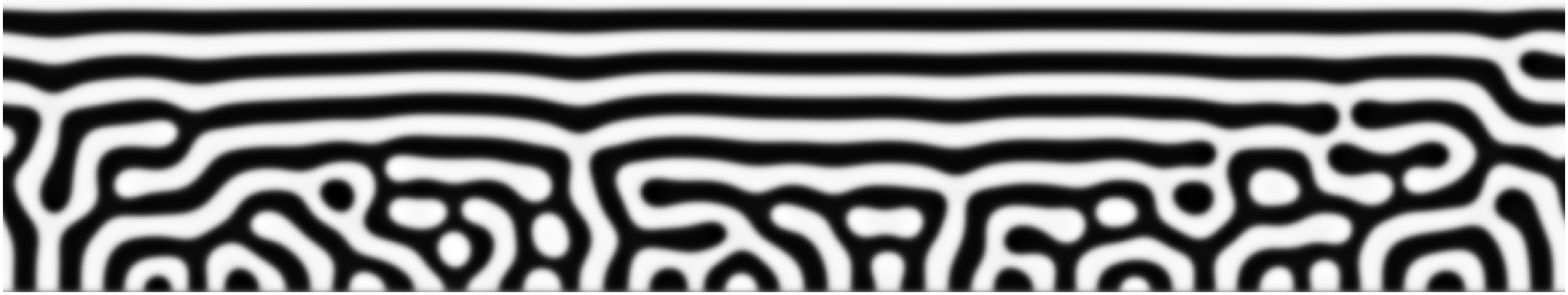} 
\end{center}
\caption{\label{profiles} 
Microphase separation is shown in the strong segregation regime 
in a confined system at $t=10^4$ 
after a quench: In (a) for neutral, $\psi_0=\psi_{L_y}=0$,
in (b) for
symmetric selective, $\psi_0=\psi_{L_y}=1$,
and in (c) 
for mixed boundary conditions, $\psi_0=0$, $\psi_{L_y}=1$.
Parameters $L_x=32 \lambda_e$, $L_y=6 \lambda_e$, $\alpha=0.015$, $\varepsilon=1$
 ($r=3.08$) and
$g=1$.
}
\end{figure}

Typical lamellar structures at the late stage of
microphase separation in the strong segregation 
regime are shown in  Fig.~\ref{profiles}
for three types of boundary conditions.
Neutral boundary conditions $\psi_0=\psi_{L_y}=0$
are used in Fig.~\ref{profiles}(a),  
symmetric selective boundaries $\psi_0=\psi_{L_y}=1$
in Fig.~\ref{profiles}(b) and in Fig.~\ref{profiles}(c) mixed boundary conditions, $\psi_0=0$, $\psi_{L_y}=1$. 
The simulations of Eq.~(\ref{dynglei}) were started 
with random initial conditions.

In the case of symmetric selective boundaries in Fig.~\ref{profiles}(b)
lamellae parallel to the substrates
have the lower free energy as 
shown for the defect free lamellar order 
in Fig.~\ref{selec}.
For neutral boundaries  in Fig.~\ref{profiles}(a)
an orthogonal 
lamellae orientation close to a boundary is favored, 
which is in agreement with the results shown in Fig.~\ref{neutr}.
In the case of mixed boundary conditions in Fig.~\ref{profiles}(c)
lamellae are 
oriented parallel close to the selective (upper) boundary and
perpendicular to the neutral (lower) boundary. 

The pattern away from the boundaries 
in the bulk shows for neutral boundaries [Fig.~\ref{profiles}(a)] a stronger disorder 
compared to the case of selective boundaries [Fig.~\ref{profiles}(b)].
In the strong segregation regime 
the coherence lengths are small for both cases,
but in the case of selective 
boundary conditions larger values of $\psi$ 
close to the boundary  are induced and this
causes a more regular
lamellae orientation in the bulk.

The pattern Fig.~\ref{profiles}(b)
consists of seven periods parallel to the horizontal
$x$-axis around the 
center of the image and six periods with $\lambda=\lambda_e$
close to the right end, whereby
both regions are connected by a  pattern including defects.
The wave numbers corresponding to seven and six lamellae
between the boundaries 
lie  both in the range $ k/k_c\ge 1$ of the stability
diagram in Fig.~\ref{stab_diagram_komplett},
where a straight and defect free lamellar order is
linearly stable with respect to small perturbations.

In simulations started with random initial
conditions, patterns with $k_m> k_c$ grow
with the largest rate and therefore a lamellar
order with  small wavelengths is preferred during the
early stage of microphase separation.  
Since one has in Fig.~\ref{stab_diagram_komplett}
a wide wave number range $k> k_c$ 
of stable straight lamellae, a relaxation of a pattern
like in  Fig.~\ref{profiles}(b) to the homogeneous state with
six lamellae, which has the lowest free energy, is a long lasting process.

We showed in the previous section in Fig.~\ref{mixed}(a) 
that in the case of mixed boundary conditions and 
parameters as in Fig.~\ref{profiles}(c) a defect-free order of lamellae
parallel to the substrates has a lower free energy 
than perpendicularly oriented ones.
In simulations of extended systems
with mixed boundaries,
where patterns with defects may occur,
neither a parallel nor a perpendicular orientation of
lamellae is preferred in the bulk. 
Moreover, the pattern in Fig.~\ref{profiles}(c)
shows, that for mixed boundary conditions
an orientational transition 
across the block-copolymer film can be expected, from parallel oriented
lamellae at the selective (upper)
boundary to a perpendicular lamellae orientation at the neutral (lower) boundary.
The free energy of the pattern in Fig.~\ref{profiles}(c)
is higher than the free energy of parallel oriented
lamellae as shown Fig.~\ref{mixed}(a)
and lower than that of perpendicularly oriented ones.

\begin{figure}[ht]
\begin{center} 
\includegraphics[width=8.25cm]{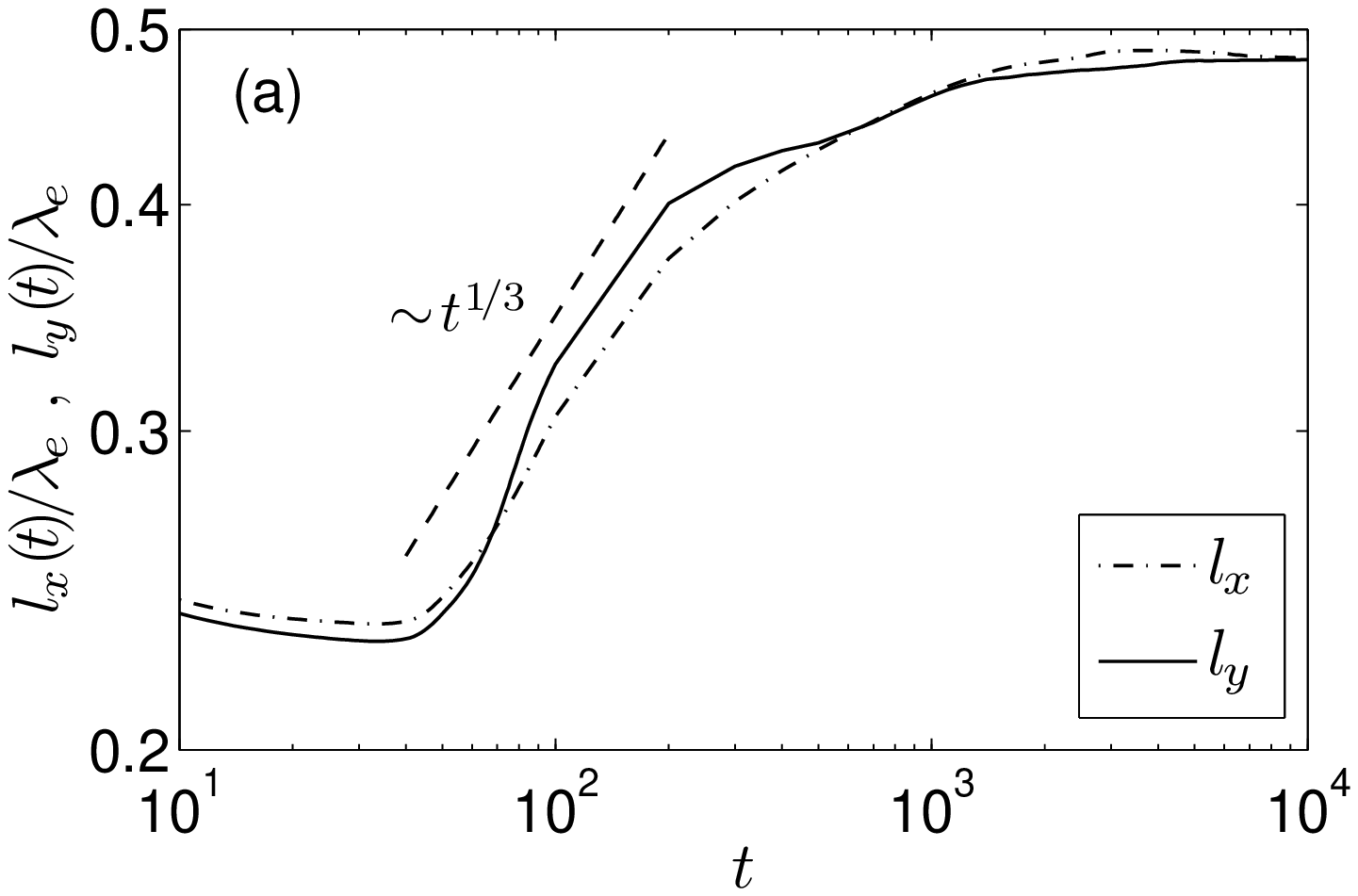}\\ 
\includegraphics[width=8.25cm]{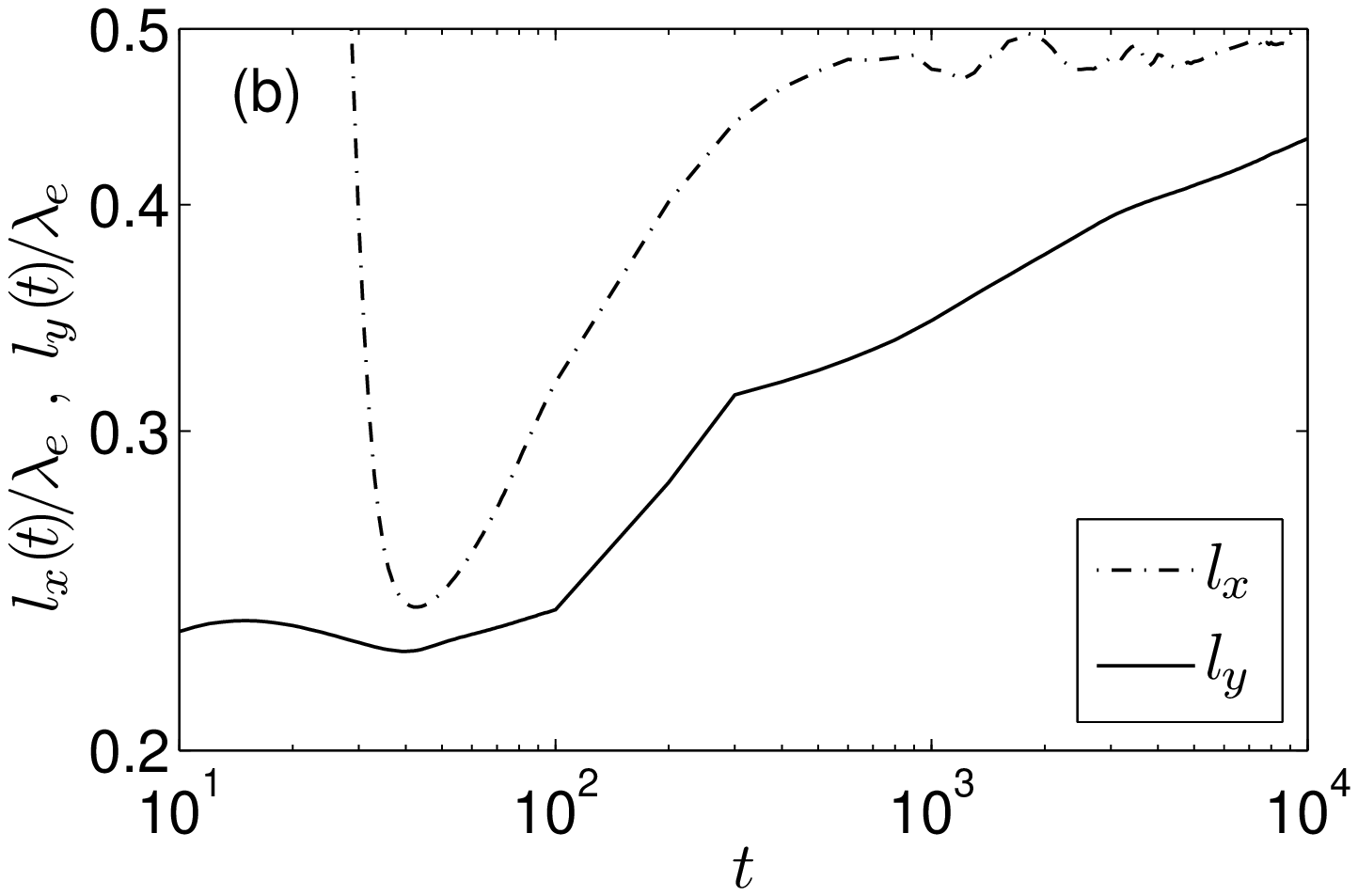}
\end{center}
\caption{\label{dummy}
The time dependence of the characteristic length scales 
$l_x(t)$ and $l_y(t)$ (averaged over three runs) is shown
in the strong segregation regime 
for the same parameters as in Fig.~\ref{profiles}:
(a) neutral boundaries, $\psi_0=\psi_{L_y}=0$, and (b) 
selective boundaries, $\psi_0=\psi_{L_y}=1$.
}
\end{figure}

The temporal evolution of 
the lengths $l_x(t)$ and $l_y(t)$ 
in the case 
of neutral boundary conditions, as
shown in  Fig.~\ref{dummy}(a),
is rather similar to the unconfined case shown in Fig.~\ref{dyn_period}.
During the early stage of phase separation
the dominating length 
scale is again that of the fastest
growing mode, which is followed by the intermediate coarsening regime
with $l_x \sim l_y \propto t^{1/3}$, before
$l_x(t)$ and $l_y(t)$ terminate again at the typical length scale
$\lambda_e/2$ of a diblock copolymer.

In the case of symmetric selective boundary conditions, 
$\psi_0=\psi_{L_y}=1$,  
the length scales 
$l_x$ and $l_y$ 
exhibit a different behavior during the initial stage of phase separation
and especially the behavior of $l_x(t)$ is changed significantly,
as shown in Fig.~\ref{dummy}(b).
A comparison of Fig.~\ref{conf}(a) and Fig.~\ref{conf}(b)
reveals, that during the early stage of microphase separation
compositional waves are induced by the selective 
boundaries and they propagate into the copolymer film.
These induced composition waves near the boundaries 
have a quasi-infinite wavelength $l_x$ 
along the $x$ direction, while the wavelength $l_y$ along the $y$ direction 
behaves similar as in Fig.~\ref{dummy}(a)
for neutral boundaries.
Far away from the selective boundaries one 
finds a random lamellae orientation
and therefore $l_x$ behaves in the
bulk at an intermediate and late stage of 
microphase separation similar
as in the case of neutral boundaries.
\begin{figure}[ht]
\begin{center}
(a)\includegraphics[width=8.25cm]{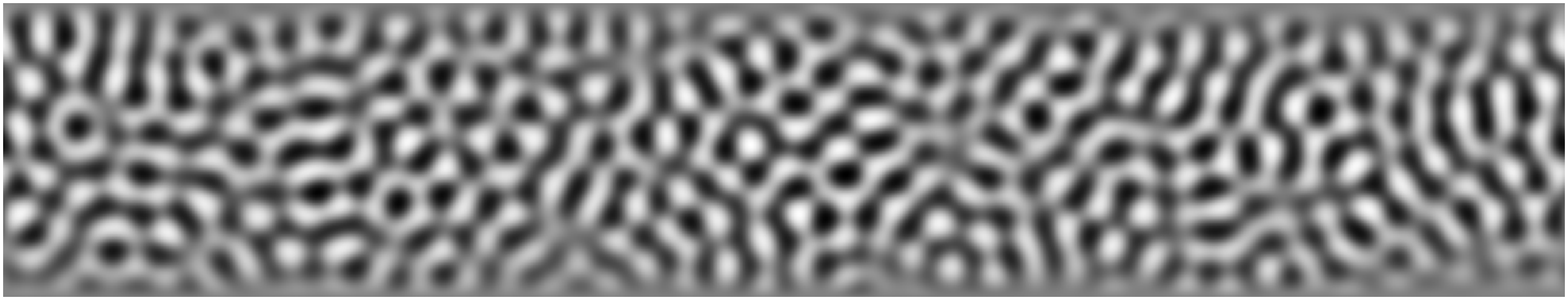}\\
\vspace{3mm} 
(b)\includegraphics[width=8.25cm]{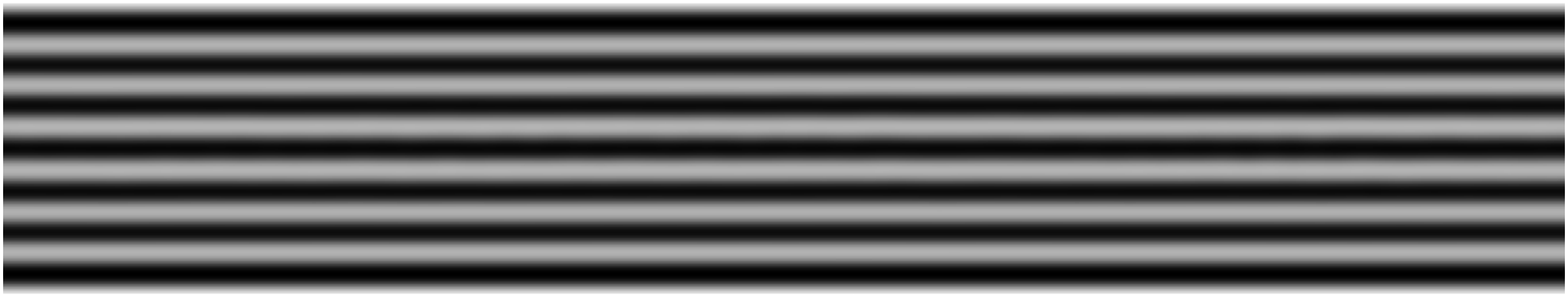}\\ 
\vspace{3mm}
(c)\includegraphics[width=8.25cm]{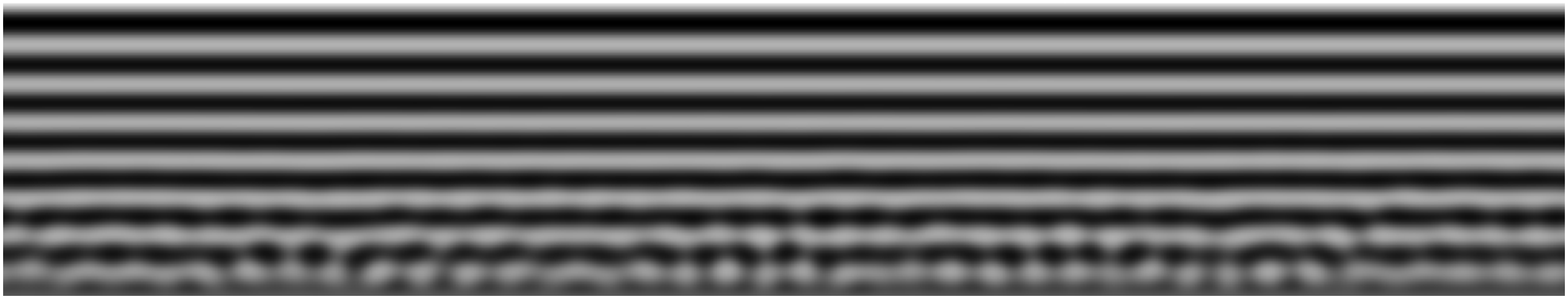} 
\end{center}
\caption{\label{profiles037} 
Microphase separation is shown in a confined system
in the {\it weak} segregation
regime with $\varepsilon=0.37$ ($r=0.5$)
at the time $t=500$ after a quench and
for three boundary conditions: (a) neutral, (b) symmetric selective 
and mixed in (c).
Other parameters as in Fig.~\ref{profiles}.
}
\end{figure}

For comparison, we show in Fig.~\ref{profiles037}
late stage patterns in
the {\it weak} segregation regime at $\varepsilon= 0.37$ ($r=0.5$)
for the same confined systems as in Fig.~\ref{profiles}.
These patterns show at a time $t=500$ already a similar
order as in the strong segregation limit 
at $t=10^4$ (Fig.~\ref{profiles}) in spite of the
fact that the dynamics is slower for smaller
values of the control parameter $r$.
However, by a reduction of the control parameter
from $r = 3.08$ in the strong segregation limit
to $r= 0.5$ one has 
an enhancement of the length scales 
from $\xi_1\approx 0.13$ to 
$\xi_1 \approx 0.32$ and from $\xi_2 \approx 0.10$ to $\xi_2 \approx 0.16$.
These higher coherence lengths increase 
simultaneously the action range of the boundaries and
both effects cause a higher lamellar  order in a thin copolymer film 
within a shorter time (Fig.~\ref{profiles037}).

Similar as in the strong segregation regime one finds
in the case of selective boundaries [Fig.~\ref{profiles037}(b)]
again a higher lamellar order  
than in the case of neutral boundaries [Fig.~\ref{profiles037}(a)].
This is in agreement with the observation, that
$\xi_1$ is for $r = 0.5$ roughly by a factor of $2$
larger than $\xi_2$.
This reasoning also confirms the results for the
case of mixed boundaries [Fig.~\ref{profiles037}(c)], where the
action range of the neutral (lower) boundary is smaller than
that of the selective (upper) boundary.

The characteristic lengths $l_x(t)$ and $l_y(t)$ develop for $r= 0.5$ and
neutral boundaries again very similar as in the
unconfined case in Fig.~\ref{dyn_period2}. In the
case of selective boundaries [Fig.~\ref{profiles037}(b)]
the two scales $l_x(t)$ and $l_y(t)$ show a similar
behavior as in the strong segregation limit, only the
saturation of
$l_y(t)$ takes place already at $t \sim 10^2$.

In summary selective boundary 
conditions are more efficient for controlling 
the  orientation of lamellae in copolymer films than
neutral ones. A comparison
of the results in Fig.~\ref{profiles}
and in Fig.~\ref{profiles037} suggests in addition
that a quench to a small 
value of $\varepsilon \gtrsim \varepsilon_c$, followed 
by a further enhancement of $\varepsilon$ into the
strong segregation regime,  
favors a coherent order of the lamellae.
In the case of mixed boundary conditions 
one obtains ''mixed'' lamellar structures
as shown in Fig.~\ref{profiles}(c), which can
be also interpreted as a coexistence of two different
boundary induced lamellae orientations.

\begin{figure}[ht]
\begin{center} 
\includegraphics[width=8.25cm]{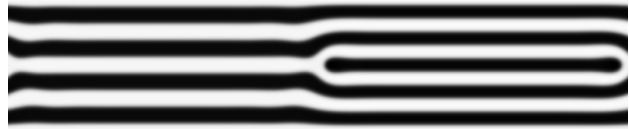}
\end{center}
\caption{\label{coex_par}
The spatial coexistence of structures with
four and five lamellae parallel to the boundaries
at a distance $L_y= 4.4373 \lambda_e$ 
is shown in the case of selective boundary 
conditions $\psi_0=\psi_{L_y}=1$.
Parameters $\varepsilon=g=1$ and $\alpha=0.015$. 
}
\end{figure}

As described in Sec.~\ref{confined}, 
different numbers of parallel oriented lamella 
between selective substrates can have at certain values
of the distance $L_y$ between the boundaries
equal free energies. For example at $L_y=4.4373$
and for parameters as given in Fig.~\ref{coex_par} 
solutions with five and four lamellae parallel
to the substrates have the same free energy.
Such a coexistence is shown in Fig.~\ref{coex_par}
where the interface between both solutions
does not move. This coexistence has a strong similarity for
instance with observations presented in Fig.~3(a) in Ref.\cite{Park:07.1}.
This example
indicates that in the case of a film thickness, which
is not an integer multiple of $\lambda_e$, one may
observe a spatial variation of the number of lamellae
in a block copolymer film.

\begin{figure}[ht]
\begin{center} 
\includegraphics[width=8.25cm]{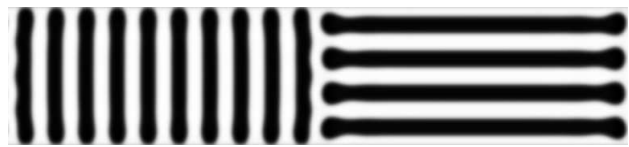}
\end{center}
\caption{\label{coex_perp}
The spatial coexistence of structures with
lamellae parallel and perpendicular to the boundaries
at a distance $L_y= 4.3854 \lambda_e$ 
is shown in the case of selective boundary 
conditions $\psi_0=\psi_{L_y}=0.4$.
Parameters $\varepsilon=g=1$ and $\alpha=0.015$. 
}
\end{figure}

Also the free energy of parallel and perpendicularly oriented
lamellae can be equal for certain boundary conditions
and film thicknesses, as discussed in Sec.~\ref{confined}.
In Fig.~\ref{coex_perp} we show an example for parameters, where
parallel and perpendicularly oriented 
lamellae have the same free energy. According to 
the interface between both orientations the free energy of
the structure in Fig.~\ref{coex_perp} is slightly higher than
that of the pure parallel or perpendicular orientation.
Nevertheless, as the interface between coexisting lamellae orientations in Fig.~\ref{coex_perp}
is not moving the coexisting pattern is long lasting.

\subsection{Dynamics of orientational ordering}

In diblock copolymers confined between boundaries
the rotational symmetry is broken. The
structure factor $S({\bf k},t)$ as well as the 
characteristic length scales $l_x$ and $l_y$, as introduced
above, do not provide a sufficient quantitative characterization of the
orientational order of lamellae.
Besides the so-called Euler characteristics \cite{Mecke:96.1} or a complex 
demodulation method \cite{Kassner:01.1} the lamellar morphology may be 
described in the framework of a 
network analysis \cite{Campbell:12.1,Pujari:12.1},
as applied in this section.

A basic element of the following 
analysis is image processing and the open
source library OpenCV \cite{Opencv:00.1} is used
for
the detection of the interfaces between
the $A$- and $B$-rich regions in the 
2-dimensional binarized images of the field $\psi(x,y)$.
The curves 
along these interfaces are approximated by polygonal chains
resulting into a set of segments of lengths $l_i$ with
the corresponding segment orientation angle $\theta_i$ relative 
to the $x$-axis.
These data allow the calculation of an average interface segment length
\begin{eqnarray}
\label{segmlength}
\langle l_s \rangle = \frac{1}{N} \sum_{i=1}^N l_i \;,
\end{eqnarray}
over which neighboring
lamellae are parallel to each other,
as well as the calculation of the average orientation of segments and the number 
of segments.
These criteria offer an improved distinction 
between patterns of different morphology.

\begin{figure}[H]
\begin{center}
\includegraphics[width=8.25cm]{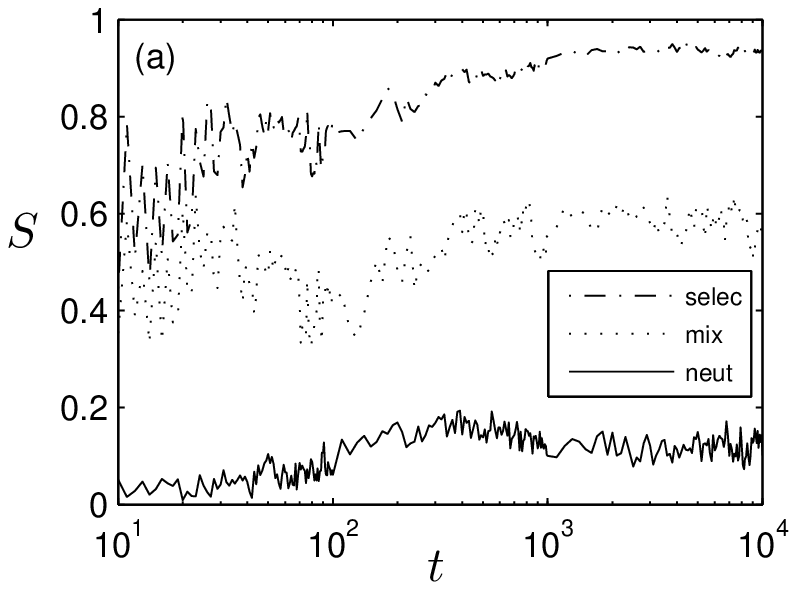}
\includegraphics[width=8.25cm]{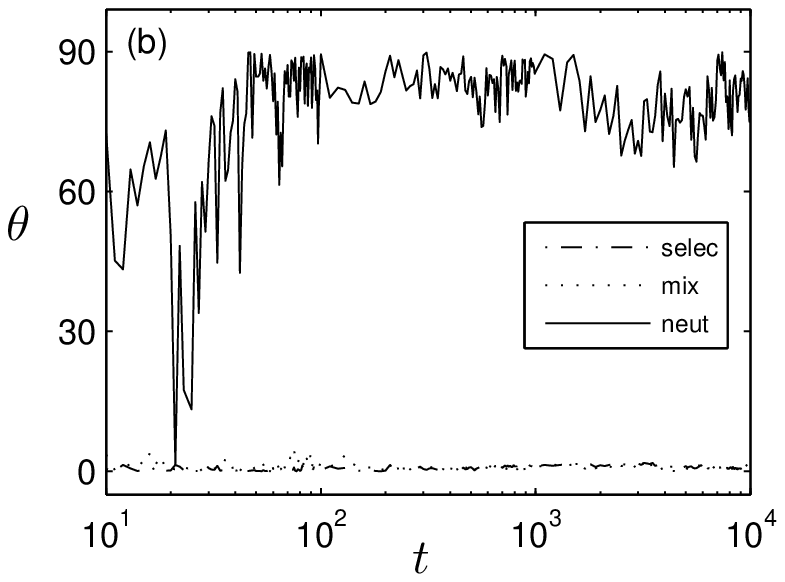}
\includegraphics[width=8.25cm]{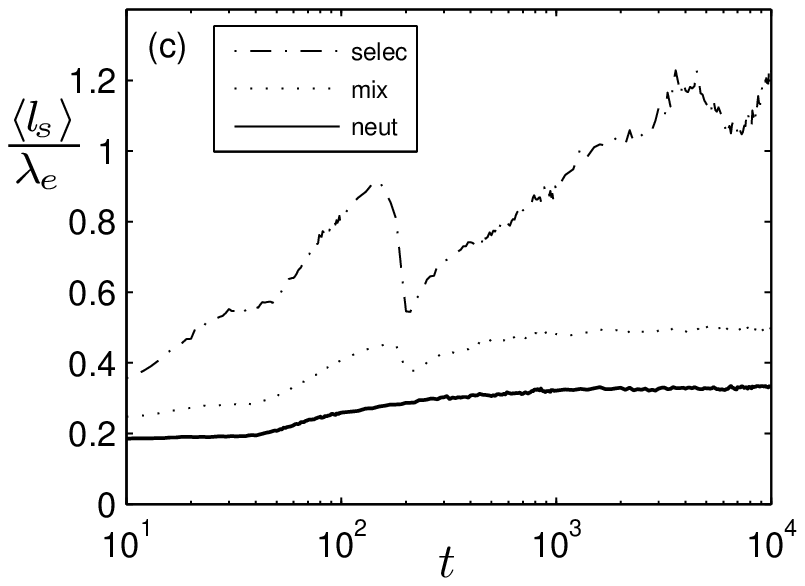} 
\end{center}
\caption{\label{fig:avrstheta} 
In part (a) the
scalar order parameter $S$, in (b) the lamellae orientation 
angle $\theta$ [cf. Eq.~(\ref{eq:Stheta})], and in (c) the
averaged segment length $\langle l_s \rangle$ 
[cf. Eq.~(\ref{segmlength})] is shown
in the strong segregation regime
as a function 
of time for the same parameters 
as in Fig.~\ref{profiles} and Fig.~\ref{dummy}
for the boundary conditions symmetric selective ($\psi_0=\psi_{L_y}=1$),
symmetric neutral 
($\psi_0=\psi_{L_y}=0$) and mixed ($\psi_0=0$, $\psi_{L_y}=1$).
}
\end{figure}

An order parameter of the segment distribution, similar to the order 
parameter in nematic liquid crystals \cite{deGennes:93.1}, is an appropriate 
quantity for a characterization of the lamellar patterns
during microphase separation.
Since the orientation angles $\theta_i$ and $\theta_i + \pi$ 
are equivalent, the order parameter is given by a symmetric second rank  
and traceless tensor
\begin{eqnarray}
\label{eq:top}
&& \hat{Q} = 
\begin{pmatrix} Q_{xx} & Q_{xy} \\ Q_{xy} & -Q_{xx} \end{pmatrix} \;,
\nonumber \\
&& Q_{xx} = \frac{\sum_{i=1}^N l_i \cos(2 \theta_i)}{\sum_{i=1}^N l_i} \;,
\nonumber \\
&& Q_{xy} = \frac{\sum_{i=1}^N l_i \sin(2 \theta_i)}{\sum_{i=1}^N l_i} \;.
\end{eqnarray}
With the scalar order parameter $S$ and the 
averaged orientation angle $\theta$ with respect to the $x$-axis,
\begin{eqnarray}
\label{eq:Stheta}
S = \sqrt{Q_{xx}^2 + Q_{xy}^2} \;, \;\;
\theta = \frac{1}{2} \textrm{arccos}(Q_{xx}/S) \;.
\end{eqnarray}
With the unit vector (the director) $\mathbf{n}=(n_x,n_y)= (\cos\theta, \sin\theta)$
the tensor order parameter can also be written in 
the following form:
\begin{eqnarray}
\label{eq:top2}
Q_{i j} = S (2 n_i n_j - \delta_{i j}) \;,
\end{eqnarray}
where $S$ has for perfectly ordered
segments the value $S =1$ and for an isotropic orientational
distribution of the segments one has $S=0$.
For the same parameters as used in 
Fig.~\ref{profiles} and Fig.~\ref{dummy}  now 
the quantities $S$, $\theta$ and
$\langle l_s \rangle$ are calculated as a function of time 
and the results are shown in 
Fig.~\ref{fig:avrstheta}.
The snapshots in Fig.~\ref{profiles} indicate
a significantly higher orientational order of the lamellae
in the case of  
selective  boundary conditions compared to neutral ones. This difference
in the orientational order can now be quantified by comparing 
$S(t)$ for the two boundary conditions, as can be seen in 
Fig.~\ref{fig:avrstheta}(a).

The temporal evolution
of the average boundary segment length $\langle l_s \rangle$ after a 
deep 
quench is shown in Fig.~\ref{fig:avrstheta}(c) and the 
mean orientation $\theta$ of segments in Fig.~\ref{fig:avrstheta}(b)
for the three
different boundary conditions: symmetric selective, neutral
and mixed. 

As can be already seen in Fig.~\ref{profiles}, 
the average segment length of straight lamellae
without defects takes in the thin film geometry its smallest value 
in the case of neutral boundary conditions. 
Simultaneously, one observes for neutral boundary conditions
the
smallest values of the scalar order parameter $S(t)$
on the time scale shown in  Fig.~\ref{fig:avrstheta}(a)
as well the strongest fluctuations of $\theta(t)$. 
Since the
action length of the boundaries in the case of neutral boundary 
conditions is small and the removal of defects is a slow
process, $\langle l_s \rangle$ increases only slowly
as function of time ($\langle l_s \rangle \to L_y$ in the long time limit).

\begin{figure}[ht]
\begin{center}
(a)\includegraphics[width=8.25cm]{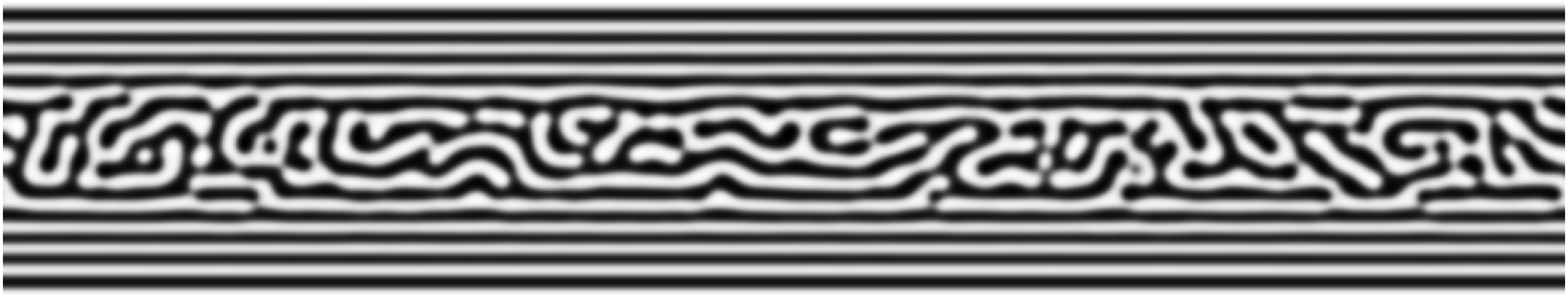}\\ 
\vspace{3mm}
(b)\includegraphics[width=8.25cm]{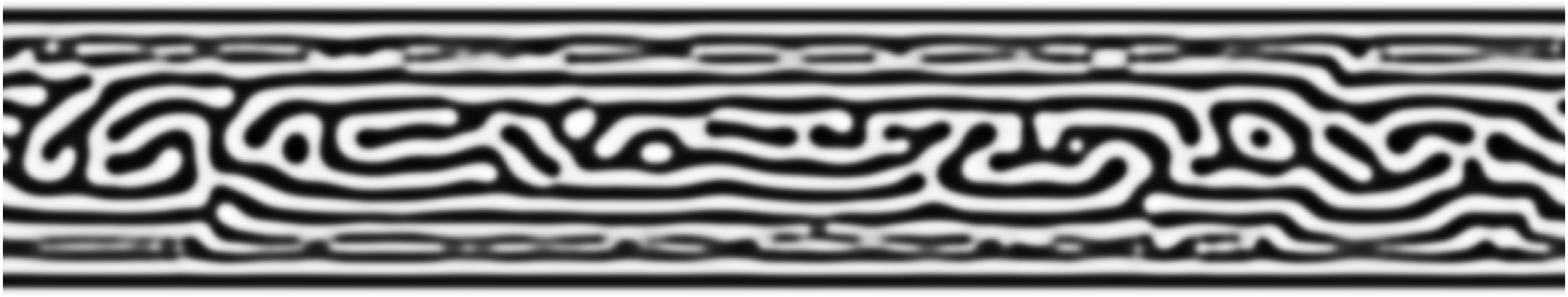}\\
\vspace{3mm}
(c)\includegraphics[width=8.25cm]{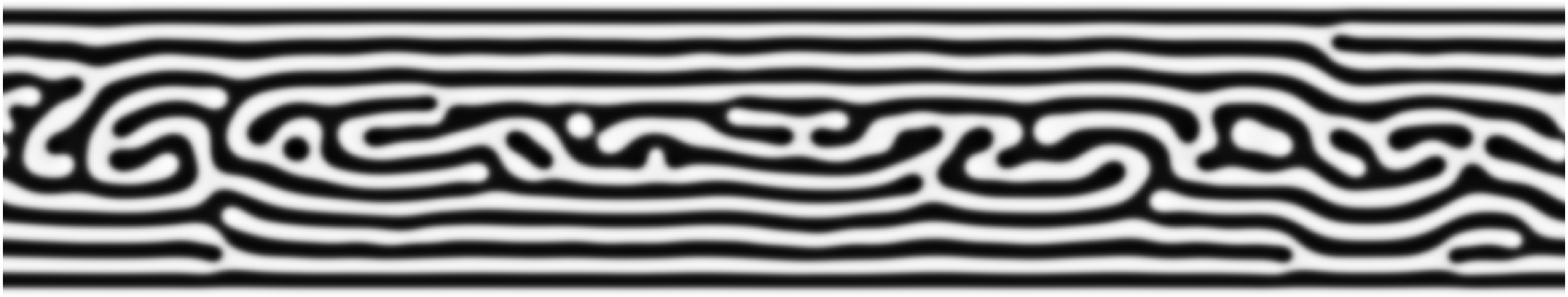}
\end{center}
\caption{\label{fig:erklaer} 
Snapshots of $\psi(x,y,t)$
at $t=100$ (a), $t=200$ (b) and $t=300$ (c)
are shown for symmetric selective
boundary conditions
and the same parameters 
as in Fig.~\ref{profiles}.}
\end{figure}

In the case of selective boundaries the action length of
the substrates is larger and therefore
oriented lamellae are formed much earlier. Consequently, one observes
higher values  of $\langle l_s \rangle$
and of the order parameter $S$ much earlier. The fact, that $S(t)$
has still not reached the value $S=1$
at about $t =10^4$ in  Fig.~\ref{fig:avrstheta}(a)
is related to the few defects left, as can
be seen for instance in Fig.~\ref{profiles}(b).
As the regular 
structure is represented by lamellae parallel to the 
substrates, $\langle l_s \rangle \to L_x$ for long time dynamics.
The reduction of $\langle l_s \rangle$ for selective
boundaries at about $t = 200$ in Fig.~\ref{fig:avrstheta}(c)
is related to intermediate structures that occur during
coarsening as indicated by the transition
from the pattern in Fig.~\ref{fig:erklaer}(a) to the pattern in Fig.~\ref{fig:erklaer}(b).

\begin{figure}[H]
\begin{center}
\includegraphics[width=8.25cm]{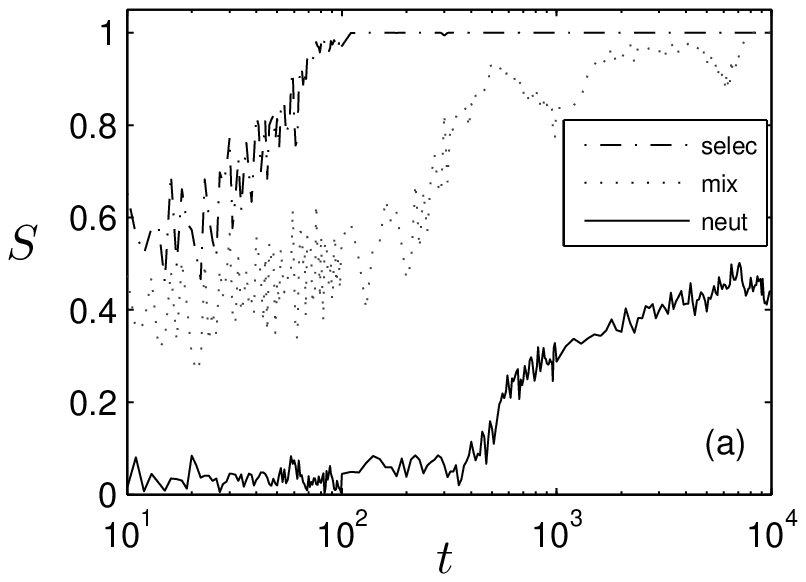}
\includegraphics[width=8.25cm]{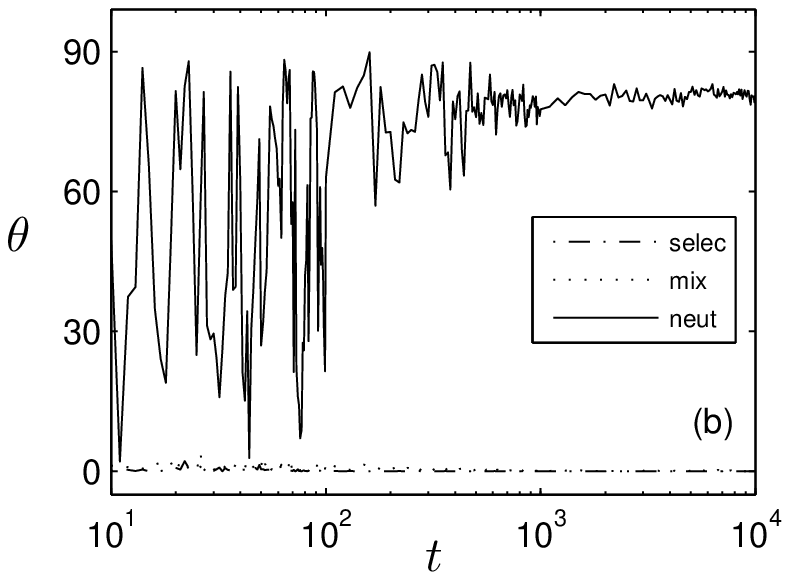}
\includegraphics[width=8.25cm]{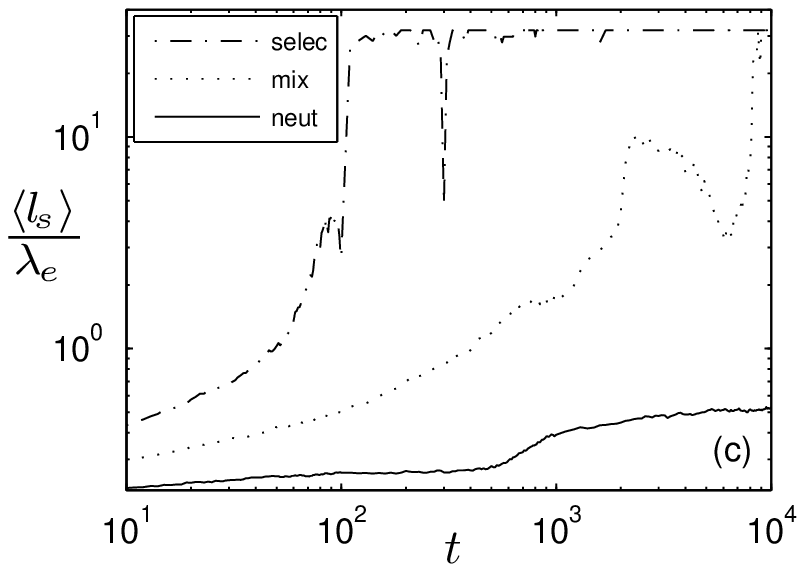} 
\end{center}
\caption{\label{fig:avrstheta037} 
In part (a) the
scalar order parameter $S$ and in (b) the lamellae orientation 
angle $\theta$ [cf. Eq.~(\ref{eq:Stheta})], and
in (c) the averaged segment length $\langle l_s \rangle$ 
[cf. Eq.~(\ref{segmlength})] is shown
in the weak
segregation regime as a function 
of time for the  same parameters 
as in Fig.~\ref{profiles037}
for the boundary conditions, symmetric selective ($\psi_0=\psi_{L_y}=1$),
symmetric neutral 
($\psi_0=\psi_{L_y}=0$) and mixed ($\psi_0=0$, $\psi_{L_y}=1$).
}
\end{figure}

The rather early achieved orientational order
for selective boundaries is also indicated by
the behavior of $\theta(t)$, which approaches zero
quite early in
Fig.~\ref{fig:avrstheta}(b). 
The large fluctuations of $\theta(t)$ in Fig.~\ref{fig:avrstheta}(b)
for the neutral boundary conditions
reflect the coarsening and the related removal 
of defects on the route to a higher orientational order.
In this case the boundary segments are preferentially oriented perpendicular to the 
substrates resulting into the orientation angle $\theta \approx \pi/2$
for long-time evolution.
In case of mixed boundaries, one obtains
a mixing of both trends with respect to the orientation of lamellae.
The selective (upper) surface triggers lamellae oriented parallel 
to the substrate whereas the neutral (lower) surface initiates lamellae 
oriented perpendicular to the substrate with less defects as 
for the two neutral boundaries [see Fig.~\ref{profiles}(b)], and accordingly 
the results for $S$ and $\langle l_s \rangle$, represented by 
dotted lines in Fig.~\ref{fig:avrstheta}(a) and Fig.~\ref{fig:avrstheta}(c), lie between the two
symmetric cases.
For comparison we show the temporal evolution of 
$S$, $\theta$ and  $\langle l_s\rangle$
in Fig.~\ref{fig:avrstheta037}
in the weak segregation 
regime  with $\varepsilon=0.37$.
Both, the behavior of $S(t)$ and 
of $\langle l_s\rangle$ confirm, that in the
weak segregation regime an orientational order
is reached on a smaller time scale as
in the case of the strong segregation regime.

The composition waves with almost equilibrium 
wavelength result into $\langle l_s \rangle \approx L_x$ 
for the selective and mixed boundary conditions.
The selective boundaries provide the fastest formation of 
regular parallel lamellae and therefore the fastest saturation of $\langle l_s \rangle$.
In case of neutral boundaries $\langle l_s \rangle$ is 
increased faster in time compared to the deep quench 
[Fig.~\ref{fig:avrstheta}(c)] with the tendency $\langle l_s \rangle \to L_y$.
In case of a not too deep quench the scalar order 
parameter $S$ shown in Fig.~\ref{fig:avrstheta037}(a) 
grows much faster in time compared to the deep quench.
The boundary segments are again preferentially oriented 
perpendicular to the boundaries for the neutral boundary 
conditions and parallel to the boundaries for the selective 
and mixed boundary conditions [Fig.~\ref{fig:avrstheta037}(b)].
Thus even this quite simple analysis of lamellar patterns 
provides quantitative characteristics of the dynamics and 
influence of the boundary conditions on the resulting patterns, 
that are complimentary to the standard analysis of the structure factor.

%

\section{Summary and conclusions}
\label{conclusion}

The formation and stability of lamellae
in block copolymers has been investigated in terms of a
mean-field model. 
A method for the determination of the stable
wave number band has been introduced in Sec.~\ref{statstab}
and the shape of this band provides the basis for a deeper understanding
of stable lamellae conformations which are locked
in experiments on BCP films to different wavelengths when using spatially periodic 
chemical nano patterning of substrates \cite{Kim:03.1}.

We have found, similar as in previous calculations in terms of
self-consistent mean field theories or phenomenological
free energy models \cite{Matsen:97.1,Walton:94.1}, that selective boundaries induce 
lamellae orientations parallel 
to the substrates and in the case of neutral boundaries lamellae orient  
perpendicular to surfaces. We present also estimates in terms of
the different length scales parallel and perpendicular to
the lamellae, whether
lamellae orient parallel or perpendicular to confining boundaries.
Some of the
lamellae conformations calculated within this work resemble very much
the lamellae orientations observed experimentally in thin BCP films 
at neutral substrates in Refs.~\cite{Park:07.1,Nealey:2010.1}. We derive
also analytical expressions
in the case of lamellae parallel to substrates
for the concentration modulation 
perpendicular to the boundaries, which can be useful for
qualitative considerations in further works.

In the case of mixed boundary conditions, i.e. 
selective at one boundary and neutral at the opposite boundary, 
we find a critical value ${\psi_S}(crit)$
of the selectivity below which the energetically preferred, homogeneous lamellae 
orientation changes from parallel to perpendicular with 
respect to the confining boundaries for any film thickness.

\begin{figure}[ht]
\begin{center}
\includegraphics[width=8.25cm]{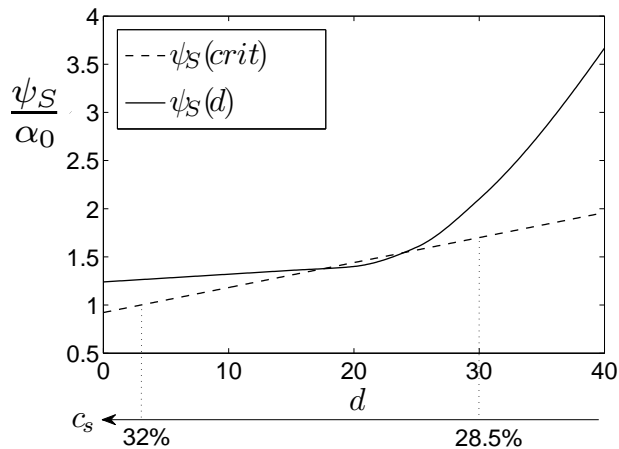} 
\end{center}
\caption{\label{reorient}
Critical value $\psi_S(crit)$ of the surface selectivity 
and  $\psi_S$
rescaled by $\alpha_0$
 as a function  of $d$.
The additional axis at the bottom indicates the solvent concentration $c_s$.}
\end{figure}

The results obtained are interesting also with regard to 
a recently used strategy to control the long range lamellae order 
in diblock copolymer films, where a 
thickness-dependent orientation of lamellae
has been found \cite{Olszowka:09.1}.
While the lamellae oriented parallel to the substrate in the ranges of film thicknesses, 
$d<19$ and $d>23$, a perpendicular
orientation in the range $19\leq d\leq 22$ was observed \cite{Olszowka:09.1}.
At a first sight, this 
experimental observation seems to be 
in contradiction to our results. However, these
experiments were done in the presence of a solvent that changed the degree of
swelling of the BCP film.
Although our mean-field model does not contain explicitly the effects of a solvent
on the lamellae formation and its interaction with surfaces,  
we suggest  to use  our results for an interpretation of the mentioned 
experiment.
It has been found that in the presence of a solvent the degree of 
swelling $\phi=\lambda_e/\lambda_s$ (where $\lambda_s$ stands for 
the lamellae period in the swollen state) depends on the film thickness \cite{Olszowka:09.1}.
For very thin films ($d \approx 3$) the degree of swelling is 
around $\phi \approx 0.68$ whereas for thicker films ($d \approx 30$) 
it is about $\phi \approx 0.715$, which means,  thicker films swell about $4\%$ 
less than thinner films and the concentration of the solvent $c_s$ is 
decreasing with increasing the film thickness.
Accordingly, for thinner films in the swollen state we calculate 
an ``effective value'' of the model parameter $\alpha\sim \lambda^{-4}:$ $\alpha_s=(0.68)^4\alpha_0$ ($\alpha_0$ 
indicates the interaction parameter in the absence of a solvent).
It is about $18\%$ smaller than for thick films where $\alpha_s=(0.715)^4\alpha_0$.
Thus to model the influence of the solvent we can assume that the parameter $\alpha$ 
is effectively increased by increasing the film thickness.
In addition our simulations reveal that the critical absorption at the surface, 
${\psi_S}(crit)$,  increases with $\alpha$.
Assuming a linear behavior of ${\psi_S}(crit)$ as a function of $\alpha$
\begin{equation}\label{gl1}
 {\psi_S}(crit)=a\cdot\alpha+b,
\end{equation}
in a small range around $\alpha_s(d=30)/\alpha_s(d=3) \approx 1.2$,
we calculate the slope $a \approx 14.8$.
Therefore, in the range $d=3 \cdots 30$,
\begin{equation}
 {\psi_S}(crit)[d=30]-{\psi_S}(crit)[d=3] \approx 0.7 \alpha_0 \;,
\end{equation}
and ${\psi_S}(crit)$ increases slightly with the film thickness.
On the other hand the solvent may reduce the selectivity of the confining surface (see, e.g., \cite{Cavicchi:05.1}).
Thus, ${\psi_S}$ is decreasing with an increasing solvent concentration.
According to the swelling behavior it means that ${\psi_S}$ is increasing 
with increasing the film thickness $d$ as in thicker films the solvent concentration
is lower.
The exact form of the curve ${\psi_S}(d)$ can be determined 
experimentally by measuring the BCP-substrate interfacial tension in the swollen state for 
various film thicknesses.
Combining now the two effects, the addition of a solvent has, we end up 
with two curves that, depending on their relative position, offer the 
possibility for a reorientation effect of the lamellae as a function of 
the film thickness (see Fig.~\ref{reorient}).
The region where ${\psi_S}(d)<{\psi_S}(crit)$ indicates a possible transition region.

Of course the curve for ${\psi_S}(d)$ in Fig.~\ref{reorient} is to some extend hypothetical and the shape 
has to be determined experimentally via measuring the interfacial tension.
Nevertheless, our results indicate a route for 
further experiments on the thickness-dependent lamellae reorientation.
Especially for diblock copolymers with a pronounced change of the 
degree of swelling as a function of the film 
thickness (like, e.g., in \cite{Zettl:10.1}, where the increase of the 
solvent uptake with decreasing film thickness is more than $10$\%) 
a considerably wide reorientation range may be realized by a suitable 
tuning of the wetting properties of the confining surfaces.

In addition to the energetic considerations of the influence
of boundaries on the homogeneous lamellae orientations, we 
also investigated the dynamical evolution of lamellae structures
between boundaries. In the case of mixed boundaries, 
one also finds complex lamellae conformations, even if they
have a higher free energy than a homogenous lamellar order
either parallel or perpendicular to the confining parallel boundaries.
Simulations of the 
time-dependent mean field model show, 
that the type of boundary condition determines strongly  
the evolution of the orientational order
as well as the number of defects in BCP films 
parallel and perpendicular to the boundaries, which has been
quantified by using an order parameter for the
characterization of the lamellae orientation.
The consideration of different quench depths
in combination with various boundary conditions
provides a strategy for the experimental preparation
of defect-free oriented lamellae.

\begin{acknowledgement}
We are grateful for inspiring discussions with W. Baumgarten, M. Hauser,
M. M\"uller, W. Pesch and L. Tsarkova and for useful hints given by M. Khazimullin on using OpenCV.
This work was supported by the German Science Foundation via the Research 
Center SFB 840 and the Research Unit FOR 608.
\end{acknowledgement}

\begin{appendix}
%
\section{Numerics of nonlinear solutions}
\label{app_numstab}
Here we describe the numerical determination of stationary periodic solutions
of Eq.~(\ref{dynglei}) for $\beta =0$
as well as their linear stability. 
Since Eq.~(\ref{dynglei}) is isotropic one can choose the $x$ direction 
parallel to the wave vector of the periodic solution.
A Fourier expansion of the periodic solution $\psi_k(x)$, as
given by 
Eq.~(\ref{Fourieransatz}), 
leads together with Eq.~(\ref{dynglei}) after projection
onto the $j$-th Fourier mode $e^{i k x j}$ to a set of
nonlinear equations 
for the coefficients $A_j$:
\begin{eqnarray}
\label{statlsg}
\hspace*{-0.7cm}&& [ \varepsilon (j k)^2-(j k)^4-\alpha]A_j 
- (j k)^2 \sum\limits_{l,m} A_l A_m A_{j-l-m} =0 \;,
\nonumber \\
\hspace*{-0.7cm}&& j=-M \ldots M \;.
\end{eqnarray}
For $M>1$ this system of nonlinear 
equations is solved numerically by Newton's iteration method and $M$ is adjusted to keep the relative error smaller than $10^{-6}$.
For $\varepsilon=1$ this accuracy can be maintained
in the case of $ \alpha=0.001$ with a very steep density profile 
by choosing $M=120$ modes and in the case of smoother density variations by $M=15$ modes.
For larger values of $\alpha$
a smaller number of modes is required
as the solution becomes more harmonic.
The linear stability of 
periodic solutions $\psi_k(x)$ of Eq.~(\ref{dynglei}) 
with respect to small perturbations $ \psi_1(x,y,t)$ is
investigated as follows. One starts with the
ansatz $\psi(x)=\psi_k(x) + \psi_1(x,y,t)$ and a
linearization of the basic equation~(\ref{dynglei}) 
with respect to $\psi_1(x,y,t)$ gives
\begin{eqnarray}
\label{linstab}
\partial_t \psi_1(x,y,t) = \nabla^2 \left( -\varepsilon + 3 \psi_k^2 
- \nabla^2 \right) \psi_1 - \alpha \psi_1 \;,
\end{eqnarray}
wherein the spatially periodic function $\psi_k$
enters parametrically.
For a solution of this linear equation (\ref{linstab}) 
with
periodic coefficients one uses a 
Floquet ansatz,
\begin{eqnarray}\label{ansatz2}
\psi_1(x,y,t) = e^{\sigma t} e^{i s (x \cos\theta + y \sin\theta)} \phi_F(x) 
+ \textrm{c.c.} \;,
\end{eqnarray}
with the Floquet parameter $s$, a $2\pi/k$-periodic 
function $\phi_F(x)$ and the angle $\theta$ enclosed between the wave 
vector of the perturbation $\psi_1$ and the wave vector of the basic periodic solution.
This periodic function $\phi_F(x)$ can be represented by a Fourier expansion
\begin{eqnarray}\label{phiF}
\phi_F(x) = \sum_{n=-M}^M D_n ~e^{i k x n} \;.
\end{eqnarray}
Taking into account Eq.~(\ref{Fourieransatz}) 
for $\psi_k$ the linear partial differential equation Eq.~(\ref{linstab}) 
is transformed after projection into an eigenvalue problem
\begin{eqnarray}
\label{ew}
\hspace*{-0.7cm}&& \sigma D_n = \left\{ \varepsilon C_n - C_n^2 - \alpha \right\} D_n
- 3 C_n \sum_{l,m} A_l A_m D_{n-l-m} \;,
\nonumber \\
\hspace*{-0.7cm}&& C_n = (k n + s \cos\theta)^2 + s^2 \sin^2\theta \;,
\nonumber \\
\hspace*{-0.7cm}&& n=-M \ldots M \;,
\end{eqnarray}
where the coefficients $A_l$ are determined by Eq.~(\ref{statlsg}).
We are interested in the growth rate $\sigma(\varepsilon,k,s,\theta)$, i.e., in the eigenvalue $\sigma$ with the largest real part.
The condition $Re[\sigma(\varepsilon,k,s,\theta)] =0$ yields the stability boundaries $\varepsilon = \varepsilon_0(k,s,\theta)$.
For $Re(\sigma) = 0$ and $\theta =0$ one finds $\varepsilon = \varepsilon_E(k)$ that determines the Eckhaus boundary.
In the case of $Re(\sigma)= 0$ and $\theta =\pi/2$ 
the corresponding $\varepsilon = \varepsilon_{ZZ}(k)$ gives the zig-zag line.

\section{Stability of weakly nonlinear solutions}
\label{app_stab}

As described in Sec.~\ref{stab_func} a periodic solution 
in a two-dimensional isotropic system may be
destabilized by modulations along the wave vector (Eckhaus instability), 
undulations perpendicular to it (zig-zag instability), 
and a combination of both types (skewed varicose) \cite{Cross:93.1}.
Two of the instability branches are given in 
Fig.~\ref{stab_diagram_komplett} and they may be determined
analytically near threshold by 
analyzing the stability of the solution given by Eq.~(\ref{eq:A0}) for $P=0$
with respect to 
small perturbations $\delta A$:
\begin{eqnarray}
\label{eq:dA}
&& A = A_0 e^{i Q x} + \delta A \;.
\end{eqnarray}
The analytical form of the perturbation is as follows:
\begin{eqnarray}
&& \delta A = e^{\sigma t} e^{i Q x} \{ a_1 \exp[i s (x \cos\theta + y \sin\theta)]
\nonumber \\
&& \qquad\qquad + a_2 \exp[-i s (x \cos\theta + y \sin\theta)] \} \;. \qquad
\end{eqnarray}
A linearization of Eq.~(\ref{eq:A}) with respect to small perturbations
$\delta A$ leads for the growth rate $\sigma$
to a $2 \times 2$ eigenvalue problem  (see e.g. \cite{Cross:93.1,Cross:09.1}).
In the special case $\theta=0$ and under the condition
of neutral growth $\sigma=0$
the following expression for the control parameter $r_E$
at the Eckhaus stability boundary ($\theta =0$, longitudinal instability)
follows:
\begin{eqnarray}
\label{eq:r_EA}
r_E = \frac{6 Q^2}{k_c^2} = 6 (\tilde{k} - 1)^2 \;.
\end{eqnarray}
A comparison with the neutral curve in Eq.~(\ref{eq:r_NA})
shows that the width of curve $r_E(Q)$ is narrower than  $r_N(Q)$
by the famous factor 
\begin{align}
  \frac{r_E(Q)}{ r_N(Q)}= \frac{1}{\sqrt{3}}
\end{align}
for Eckhaus stability boundary \cite{Eckhaus:65.1,Kramer:85.1,Cross:93.1,Cross:09.1}.

The zig-zag stability boundary (transversal instability) 
results for the case $\theta =\pi/2$ and $\sigma=0$:
\begin{eqnarray}
\label{eq:r_ZZA}
Q = 0 \; \textrm{, i.e.,} \;\; \tilde{k}_{ZZ} = 1 \;.
\end{eqnarray}
Hence, the stationary weakly nonlinear solution given by  Eq.~(\ref{eq:A0}) 
is linearly stable in the region $Q \ge 0$ ($\tilde{k} \ge 1$) 
between the zig-zag line in Eq.~(\ref{eq:r_ZZA}) and the Eckhaus boundary
 $Q_E= \sqrt{r k_c^2/6~}$.
The results for the stability boundaries found so far in this appendix
in the framework of 
the amplitude equation Eq.~(\ref{eq:A}) are typically valid only 
in the vicinity of the critical point ($r \gtrsim 0$, $|k - k_c| \ll 1$).
An essential improvement of the analytical results for 
the stability diagram 
can be obtained by the Galerkin approach in a one-mode approximation,
as described in the following.

Inserting the ansatz given by Eq.~(\ref{eq:psi_sp}) into Eq.~(\ref{eq:alt}) 
and projecting on the critical mode $e^{-i k_c x}$ one gets in the leading order
\begin{eqnarray}
\label{eq:ampl}
\hspace*{-0.7cm}&&\partial_t A = \nabla^2_{k_c} [ -2 k_c^2 (r+1) A + 3|A|^2 A - \nabla^2_{k_c} A ] 
- k_c^4 A \;, \nonumber 
\\
\hspace*{-0.7cm}&&\mbox{with} \qquad \nabla^2_{k_c} \equiv (\partial_x + i k_c)^2 + \partial_y^2 \;.
\end{eqnarray}
Above threshold one may derive for the amplitude $A_0$ 
with the ansatz Eq.~(\ref{eq:A0}) via  Eq.~(\ref{eq:ampl})
the following expression:
\begin{eqnarray}
\label{eq:amplA0_2}
&& A_0^2 = \frac{2 k_c^2}{3} \left[ r - \frac{\big[ (Q+k_c)^2 - k_c^2 \big]^2}{2 k_c^2 (Q+k_c)^2} \right] 
\nonumber \\
&& \qquad = \frac{2 k_c^2}{3} \left[ r - \frac{(\tilde{k}^2 - 1)^2}{2 \tilde{k}^2} \right] \;.
\end{eqnarray}
One can easily see that 
this amplitude $A_0$ vanishes at the 
neutral curve Eq.~(\ref{eq:r_N}) for arbitrary values of $r$.
Inserting perturbation Eq.~(\ref{eq:dA}) with the amplitude 
given by (\ref{eq:amplA0_2}) into Eq.~(\ref{eq:ampl}) 
the growth rate $\sigma$ is again calculated from a $2 \times 2$ eigenvalue problem.
The various stability boundaries are determined via
the neutral stability condition
$\sigma(r,Q,\theta) =0$ in terms of the control parameter $r(Q,\theta)$ by keeping simultaneously only the leading terms in $s$.
Minimization of $r(Q,\theta)$ with respect to $\theta$ gives in the range $Q>0$ the angle $\theta=0$ 
and therefore an Eckhaus stability boundary
\begin{eqnarray}
\label{eq:r_Eckhaus}
r_E = 
\frac{3 \tilde{k}^8 - 2 \tilde{k}^6 - 6 \tilde{k}^2 + 5}{2 \tilde{k}^2 (\tilde{k}^4 + 3)} \;,
\end{eqnarray}
that coincides with the result obtained in the framework of the free 
energy considerations [see Eq.~(\ref{eq:r_E})].
For $\tilde{k}$ between the neutral curve given by Eq.~(\ref{eq:r_N}) 
and the Eckhaus boundary in Eq.~(\ref{eq:r_Eckhaus}) the periodic solution
in Eq.~(\ref{eq:A0}) with $A_0$ from Eq.~(\ref{eq:amplA0_2}) 
is unstable with respect to long-wavelength perturbations along the wave 
vector, i.e. $s \to 0$ and $\theta =0$.

In the vicinity of the band center $\tilde{k}=1$ one has in the 
leading order $r_E = 6 (\tilde{k} - 1)^2 + \cdots$ in agreement 
with the result derived via the standard 
amplitude equation [see Eq.~(\ref{eq:r_EA})].
A minimization of $r(Q,\theta)$ in the range $Q<0$ gives the angle $\theta=\pi/2$ 
for the zig-zag instability line
\begin{eqnarray}
\label{eq:r_Zigzag}
Q_{ZZ} = 0 \; \textrm{, i.e.,} \;\; \tilde{k}_{ZZ} = 1 \;,
\end{eqnarray}
in agreement with Eq.~(\ref{eq:zz_line}).
For the perturbations Eq.~(\ref{eq:dA}) with $s \to 0$, $\theta =\pi/2$ 
the growth rate $\sigma$ is negative for $\tilde{k}$ on the right hand side of 
the zig-zag line Eq.~(\ref{eq:r_Zigzag}) up to the Eckhaus boundary Eq.~(\ref{eq:r_Eckhaus}).
A skew varicose instability with $0< \theta < \pi/2$ does not occur.

\section{Weakly nonlinear solution under confinement \label{analconf}}

Close to the
onset of microphase separation ($r\gtrsim 0$) 
the dynamics of the amplitude of the periodic order parameter field $\psi(\mathbf r)$
is governed by the 
Newell-Whitehead-Segel amplitude equation (\ref{eq:A}),
as described in Sec.~\ref{weaknon}. This equation has spatially periodic
solutions in extended systems as described in  Sec.~\ref{weaknon},
but it may also be solved in the presence of boundaries.

Here we take into account boundary conditions
for the case of lamellae oriented parallel to 
the substrates. With the ansatz
\begin{equation}
\label{apppsiB}
 \psi(y)=\sqrt{\frac{2k_c^2r}{3}}~B(y)e^{ik_c(y-y_0)}+\mbox{c.c.}\,.
\end{equation}
one gets, starting from Eq.~(\ref{eq:alt}), the
following equation of the envelope $B(y)$:
\begin{equation}
\label{eq:B}
 \frac{2}{r k_c^2}\partial_y^2 B+\left(1-\mid B \mid^2\right)B=0\,.
\end{equation}
This equation has constant solutions
of the form
\begin{equation}
 B_0(y)=B_0=\pm 1,
\end{equation}
corresponding to a spatially periodic field $\psi(y)$
of constant amplitude.

Eq.~(\ref{eq:B}) also has the solution
\begin{equation}
\label{saddleapp}
 B(y)=B_0 \tanh\left[\frac{k_c\sqrt{r}}{2}(y-\tilde y)\right]\,
\end{equation}
and this may be used to construct approximate solutions
for lamellae parallel to the two boundaries. The decomposition 
(\ref{apppsiB}) is based on the assumption, that
$B(y)$ varies slowly on the scale $2\pi/k_c$ and close to
threshold one may simplify the boundary conditions 
(\ref{boundall}) to the following
conditions,
\begin{subequations}
\label{BCapp}
\begin{align}
&\partial_y\psi|_{y=0,L_y}=0\,,\\
&\psi|_{y=0}=\psi_0=c\cdot\psi_b\,,\\
&\psi|_{y=L_y}=\psi_{L_y}=c\cdot\psi_b\,,
\end{align}
\end{subequations}
by neglecting higher order derivatives of $\psi$.
The constant $c$ is used to switch between different types of
boundary conditions: $c=0$ corresponds to neutral boundaries, and
$c\neq 0$ to selective (symmetric) boundaries.

In order to fulfill the boundary conditions (\ref{BCapp})
at $y=0$ and $y=L_y$, we use a  
linear superposition of the solution
(\ref{saddleapp}) 
of the form 
\begin{equation}
 B(y)=v_1+v_2Y_1(y)+v_3Y_2(y)\,,
\end{equation}
where the constants $v_1,v_2,v_3$ indicate the
possible signs $\pm 1$ and with
\begin{subequations}
\begin{eqnarray}
 Y_1(y)=\tanh\left[\frac{k_c\sqrt{r}}{2}(y-y_1)\right]\,,\\
 Y_2(y)=\tanh\left[\frac{k_c\sqrt{r}}{2}(y-y_2)\right]\,.
\end{eqnarray}
\end{subequations}
The constants $y_1$ and $y_2$ are determined by
the boundary conditions.

With the bulk value
\begin{align}
 \psi_b= \sqrt{\frac{8k_c^2r}{3}}\,
\end{align}
the order parameter $\psi(y)$ takes the form
\begin{equation}
\label{soluBB}
 \psi(y)=\psi_b[v_1+v_2Y_1(y)+v_3Y_2(y)]\cos\left[k_c(y-y_0)\right]\,.
\end{equation}

Inserting Eq.~(\ref{soluBB}) into the boundary conditions
(\ref{BCapp}) one has the
following equations
\begin{subequations}
 \begin{align}\nonumber
 &\left[v_2(1-Y_1^2(0))\frac{\sqrt{r}}{2}\right]\cos(k_cy_0)+\\
 &[v_1+v_2Y_1(0)-v_3]\sin(k_cy_0)=0\,,\\ \nonumber
 &\left[v_3(1-Y_2^2(L_y))\frac{\sqrt{r}}{2}\right]\cos(k_cy_0)+\\
 &[v_1+v_2+v_3Y_2(L_y)]\sin(k_cy_0)=0\,,\\
 &[v_1+v_2Y_1(0)-v_3]\cos(k_cy_0)=c\,,\\
 &[v_1+v_2+v_3Y_2(L_y)]\cos(k_cy_0)=c\,
\end{align}
\end{subequations}
for the determination of the constants of the ansatz.
Here we assumed a system size $L_y=n\cdot (2\pi)/k_c$
and exploited the approximation that the boundaries do not influence each other 
so that $Y_1(L_y)=1$ and $Y_2(0)=-1$. From this one may directly deduce $v_1=v_3=-v_2$.

\begin{figure}[H]
\begin{center}
\includegraphics[width=8.25cm]{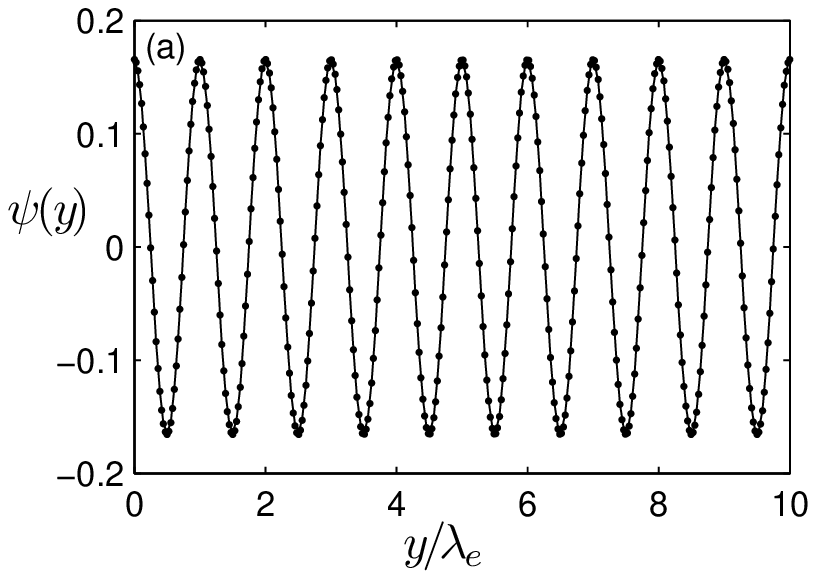} 
\includegraphics[width=8.25cm]{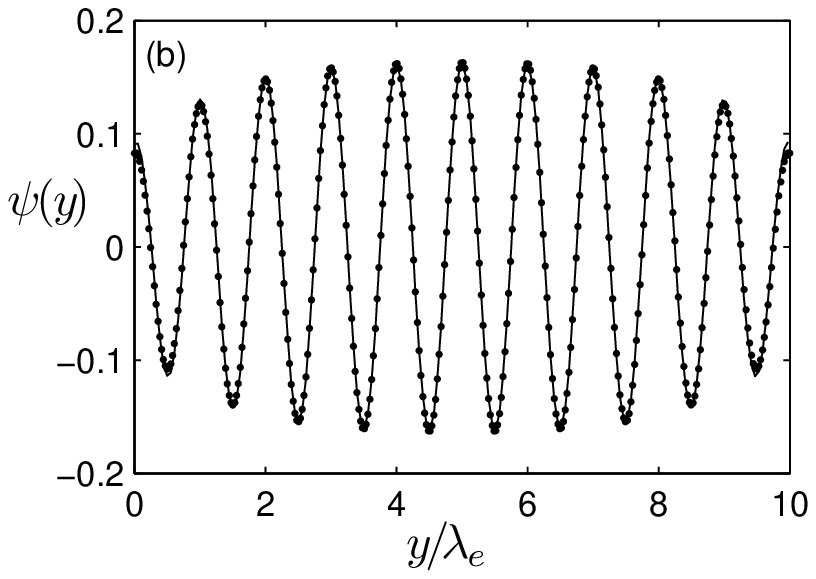}   
\end{center}
\caption{\label{analyt_neu}
The numerical solution $\psi(y)$  of Eq.~(\ref{dynglei})
is shown for various boundary conditions for lamellae
parallel to the boundary: In part
(a) for selective boundaries $c=1$ and
in (b) for selective boundaries with a reduced selectivity $c=0.5$.
The solid lines mark the numerical
solution and the analytical approximation given by
Eq.~(\ref{soluBB}) is displayed by the dots.
The parameters are 
$L_y =10\lambda_e$,
$g=1$, $r =0.021$, $k_c =0.7$ 
(corresponding to $\varepsilon=1$ and $\alpha=0.24$).}
\end{figure}

In the following some special cases are considered to illustrate the high quality of this approximation.
The cases of selective ($c=1$) or neutral ($c=0$) boundaries may be solved explicitly. 
For the case of neutral boundaries we obtain $k_c y_0=\pm \pi/2$, $y_1=0$ and $y_2=L_y$ as a solution [see Fig.~\ref{analyt} (b)].
$v_1$ remains arbitrary in this case.
If $c=1$ the value at the boundary corresponds to the bulk value and this case is conform
with the periodic one where $B(y)=\pm 1=\mbox{const.}$ and $y_0=0$ [see Fig.~\ref{analyt_neu} (a)]. 
For $c<1$ the selectivity is reduced
[see Fig.~\ref{analyt_neu} (b)]. In this case an explicit solution is not possible
but one may use the approximation $\sin(k_cy_0)\approx0$ if $c$ is not too small
what corresponds to $y_0\approx0$. With this approximation we find $Y_1(0)\approx c$ 
and $Y_2(L_y)\approx -c$ and $v_1=-1$ as a solution. The example shown in Fig.~\ref{analyt_neu} (b)
is for $c=0.5$ what results in $y_1\approx-10.83$ and $y_2\approx L_y+10.83$. Although
this is only an approximation it still fits very well the full numerical solution. Of course this case
may be studied more accurately by taking into account the wave number change. In this
case one has more free parameters to adjust reasonably. Furthermore in the case
of mixed boundaries this approximation becomes more complex and
higher order derivatives have to be taken into account. 
The solution of the resulting equations becomes involved
and the advantage of the approximations gets lost in comparison
with the full numerical solution of the problem.

\end{appendix}


\providecommand*\mcitethebibliography{\thebibliography}
\csname @ifundefined\endcsname{endmcitethebibliography}
  {\let\endmcitethebibliography\endthebibliography}{}


\begin{tocentry}
\includegraphics[width=8.9cm]{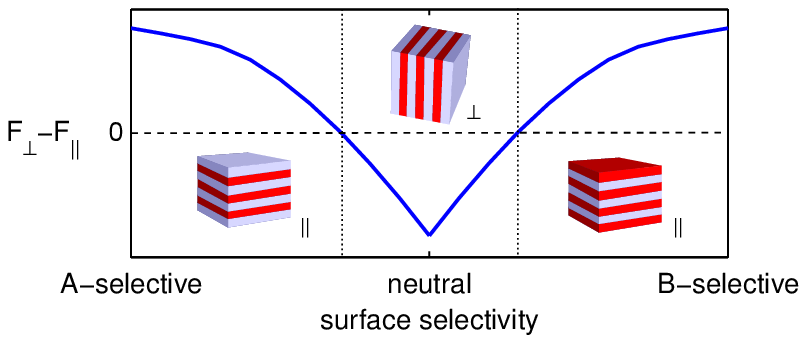}
\end{tocentry}

\end{document}